\documentclass[10pt, a4paper, onecolumn]{article} 
\usepackage[title]{appendix}
\usepackage{lscape}
\usepackage{hyperref}
\usepackage{amsfonts,amssymb}
\usepackage{graphicx}
\usepackage{float}
\usepackage{amsmath}
\usepackage{cuted}
\usepackage{geometry}
\usepackage{rotating}
\usepackage{pdflscape}
\usepackage{blindtext}

\newcommand{\ket}[1]{\left|#1\right\rangle}
\newcommand{\bra}[1]{\left\langle#1\right|}

 \usepackage{braket}
\usepackage{chngcntr}
 \numberwithin{equation}{section}
\title{Fundamental description of Wannier qubits of any topology \newline in semiconductor by analytical and numerical computations}
\author{ Krzysztof Pomorski$^{1,2} $ \\ \\ 
		\newline\newline 
     1: Cracow University of Technology,\\ Faculty of Computer Science and Telecommunications, \\ Department of Computer Science, Poland 
     \\ \\
    2: Quantum Hardware Systems (\texttt{www.quantumhardwaresystems.com}) 
}
\providecommand{\keywords}[1]
{
  \small	
  \textbf{\textit{Keywords---}} #1
}

\begin{document}
\maketitle

\begin{abstract}
Justification of tight-binding model from Schroedinger formalism for various topologies of position-based semiconductor qubits is presented in this work. Simplistic tight-binding model allows for description of single-electron devices at large integration scale. However it is due to the fact that tight-binding model omits the integro-differential equations that arise from electron-electron interaction in Schroedinger model. Two approaches are given in derivation of tight-binding model from Schroedinger equation. First approach is conducted by usage of Green functions obtained from Schroedinger equation. Second approach is given by usage of Taylor expansion applied to Schroedinger equation. The obtained results can be extended for the case of many Wannier qubits with more than one electron and can be applied to 2 and 3 dimensional model. Furthermore various correlation functions are proposed in Schroedinger formalism that can account for static and time-dependent electric and magnetic field polarizing given Wannier qubit system. One of the central results of presented work relies on the emergence of dissipation processes during smooth bending of semiconductor nanowires both in the case of classical and quantum picture. Presented results give the base for physical description of electrostatic Q-Swap gate of any topology using open loop nanowires. We observe strong localization of wavepacket due to nanowire bending.
\end{abstract}

\tableofcontents

\section{Philosophy behind charged based classical and quantum logic}
\begin{figure}
\centering
\includegraphics[scale=0.6]{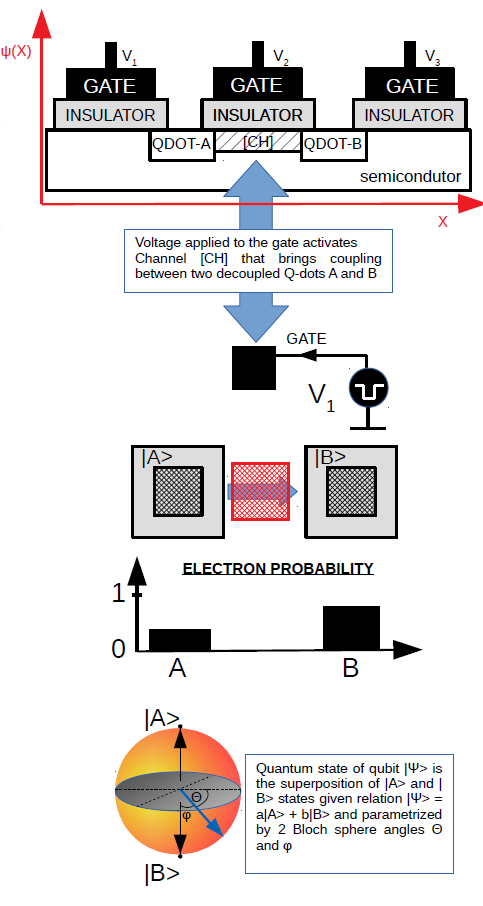}\includegraphics[scale=0.6]{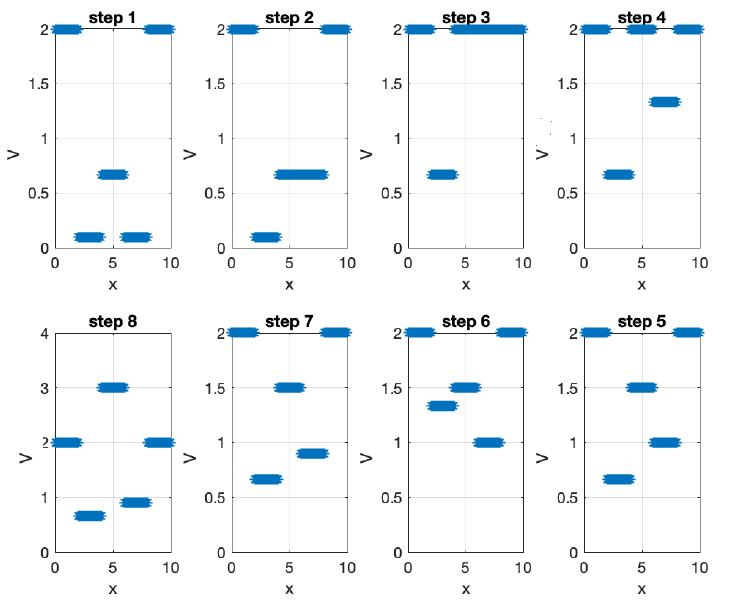}
\caption{Left:Position based qubit also known as Wannier qubit in CMOS circuit implementation as by \cite{SEL}, \cite{Cryogenics}, \cite{Nbodies},\cite{qchip}, Right: Effective potentials V(x) for single electron under different voltage biasing circumstances \cite{Noninvasive}.}
\label{fig:qubit1}
\includegraphics[scale=0.6]{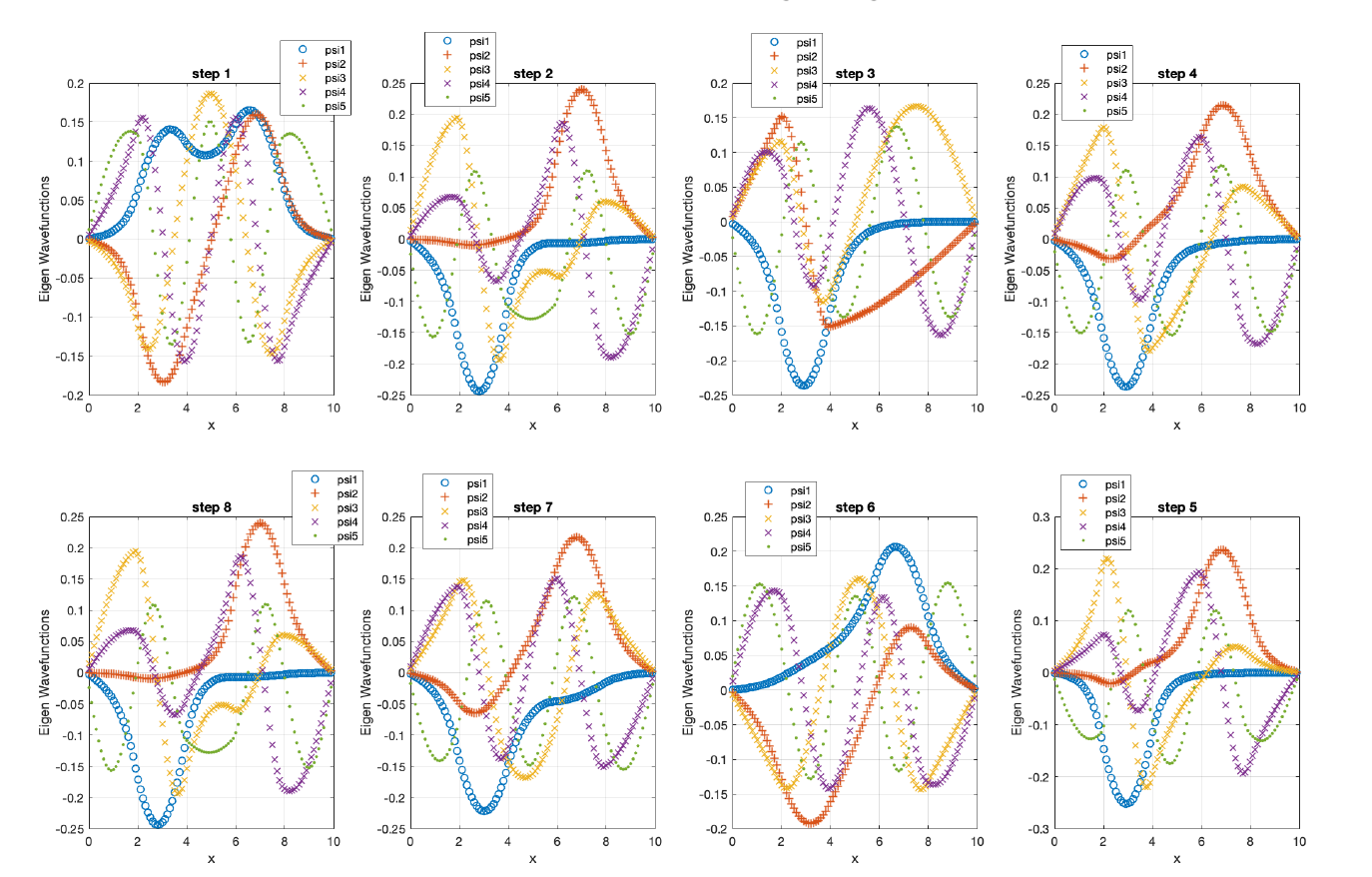}
\caption{Electron wavefunctions for subsequency eiengenergies \cite{Noninvasive} obtained by Wannier qubit. One can spot maximum localized functions for various qubit electrostatic biasing potentials expressed by effective potential. }
\label{fig:qubit2}
\end{figure}

Single electron devices in semiconductor quantum dots \cite{Likharev} becomes quite promising way of implementation of qubit and quantum computation as well as quantum communication and it was being confirmed experimentally by \cite{Fujisawa}, \cite{Petta}.
This is particularly attractive perspective in the framework of CMOS technology \cite{Dirk}. In case of small field effect transistor the source and drain are playing the role of quantum dots whose connectivity is regulated by voltage applied on the top gate as depicted in Fig.\ref{fig:qubit1} and in Fig.\ref{fig:qubit2}.

The most simple model describing semiconductor Wannier position based qubit is given by simplistic tight-binding model. We have dynamics of quantum state with time given by
\begin{eqnarray}
\hat{H}_t \ket{\psi(t)}=
\begin{pmatrix}
E_{p1}(t) & t_{s12}(t) \\
t_{s12}^{*}(t) & E_{p2}(t)
\end{pmatrix}
\begin{pmatrix}
\alpha(t) \\
\beta(t)
\end{pmatrix}
=i\hbar \frac{d}{dt}
\begin{pmatrix}
\alpha(t) \\
\beta(t)
\end{pmatrix}=i\hbar \frac{d}{dt}\ket{\psi(t)}, |\alpha(t)|^2+|\beta(t)|^2=1, \nonumber \\
\end{eqnarray}
where $E_{p1}$ denotes maximum localized energy due to presence of electron on the left quantum dot and
$E_{p2}$ denotes maximum localized energy due to presence of electron on the right quantum dot and $|t_{s12}|$ is hopping energy due to electron movement from left to right quantum dot. We can refer tight-binding model to Schrodinger equation
\begin{eqnarray}
\psi(x,t)=\alpha(t)w_L(x)+\beta(t)w_R(x), \int_{-\infty}^{+\infty} dx|w_{L(R)}(x)|^2=1, \int_{-\infty}^{+\infty} dx w_{R}^{*}(x) w_{L}(x)=\ket{w_R}\bra{w_L}=0, \nonumber \\
\alpha(t)=\int_{-\infty}^{+\infty}dxw_L^{*}(x)\psi(x,t)=\bra{w_L}\ket{\psi},\beta(t)=\int_{-\infty}^{+\infty}dxw_R^{*}(x)\psi(x,t)=\bra{w_L}\ket{\psi},
\end{eqnarray}
where $w_L(x)$ and $w_R(x)$ are maximum localized orthonormal wavefunctions (Wannier functions) of single electron on left and right quantum dot.
They are defined by wavefunction distribution implementing maximum occupancy of electron on left or right side.
Prescription for tight-binding model in terms of Wannier functions is given by
\begin{eqnarray}
E_{p1(p2)}=\int_{-\infty}^{+\infty}dxw_{L(R)}(x)^{*}(\frac{-\hbar^2}{2m}\frac{d^2}{dx^2}+V(x))w_{L(R)}(x)=\int_{-\infty}^{+\infty}dxw_{L(R)}(x)^{*}\hat{H}(x)w_{L(R)}(x), \nonumber \\
t_{s12(s21)}=\int_{-\infty}^{+\infty}dxw_{L(R)}(x)^{*}(\frac{-\hbar^2}{2m}\frac{d^2}{dx^2}+V(x))w_{R(L)}(x)=\int_{-\infty}^{+\infty}dxw_{L(R)}(x)^{*}\hat{H}(x)w_{R(L)}(x),
\end{eqnarray}
where $H$ is Hamiltonian of 2 quantum dot system. In this work we justify the formulas for $E_{p1}$, $E_{p2}$, $t_{s12}$ and for $t_{s21}$ for most general case in terms of
Green functions and Taylor wavefunction expansion for any physical situation.
Physics is quite much relying on mass and electric charge conservation principles and both quantities are quasi-continuous on macroscale and on nanoscale become integer multiplicity of elementary values. The charge flows in such a way as the energy of the electric/magnetic field tends to be minimized. One of its consequences is the repulsion of two charges of the same sign and attraction of two charges of opposite signs that is commonly known as Coulomb law. The electric currents flow also tends to minimize the energy of magnetic field so the most equilibrium state of isolated capacitor is discharged device. Due to electron and hole mobility charge can be used for information or energy transfer across metallic or semiconductor nanowires. One can use the electric and magnetic fields as parameters controlling the evolution of the given physical system with time so desired final state can be achieved upon previous setting the system with initial configuration that is formally expressed by circuit theory both in classical and in quantum regime. Furthermore the simple rules of dynamics of charged billiard balls confined in boxes can lead to a simple scheme for implementation of logical operations as
logical inverter or controllable inverter (CNOT gate) as depicted in Fig.1 and by \cite{Likharev}, \cite{Dirk}, \cite{Panos}, \cite{Pomorski_spie}, \cite{SEL}, \cite{qchip}. However the electric charge is confirmed to be quantized by experiments (except fractional quantum Hall effect where fractionation of electric charge is being observed) and expressed by electron, proton or hole charge in condensed matter systems. The quantization and control of single electron flow by distinct integer values can be achieved in nanotechnological experiments as in the chain of coupled quantum dots that can have particularly small diameters in semiconductors and in most recent CMOS technology, so even size of 3nm can be achieved for very highly integrated circuits. In such types of structures the usage of magnetic fields is less practical since waveguides and solenoids are very hardly scalable. Therefore it is favourable to use only electric fields as a controlling factor so it promotes Wannier qubits that are also known as position-based qubits. Wannier qubits are using maximum localized wavefunctions as present in two coupled quantum dots in order to encode quantum information in qubit what makes such qubit from eigenergy based qubit using two eigenergies to span the qubit state. However it shall be underlined that even in cryogenic conditions the semiconductors are having the intrinsic noise that is significantly higher than in case of superconductors. 
\section{From Schroedinger to Wannier functions}
Let us consider the system of 2 coupled quantum dots and let us assign the occupancy of left quantum dot by electron as wavepacket presence in $x \in (-\infty,0)$ and wavepacket occupancy of right quantum dot as wavepacket presence in $x \in (0,+\infty)$. We can assume that Wannier wavefunctions are linear transformation of system eigenenergy wavefunctions.

We propose maximum localized orthonormal Wannier functions of the form
\begin{eqnarray}
w_L(x)=(+\alpha \psi_{E1}(x)+\beta \psi_{E2}(x))=w_{1,1}\psi_{E1}(x)+w_{1,2}\psi_{E2}(x), \nonumber \\
w_R(x)=(-\beta \psi_{E1}(x)+\alpha\psi_{E2}(x))=w_{2,1}\psi_{E1}(x)+w_{2,2}\psi_{E2}(x), \nonumber \\
\end{eqnarray}
and formally we have
\begin{eqnarray}
\begin{pmatrix}
w_L(x), \\
w_R(x)
\end{pmatrix}= \hat{W}
\begin{pmatrix}
\psi_{E1}(x), \\
\psi_{E2}(x)
\end{pmatrix}
=
\begin{pmatrix}
w_{1,1} & w_{1,2} \\
w_{2,1} & w_{2,2}
\end{pmatrix}
\begin{pmatrix}
\psi_{E1}(x), \\
\psi_{E2}(x)
\end{pmatrix},
 \hat{W}^{-1}=\frac{1}{Det(\hat{W})}
 \begin{pmatrix}
w_{2,2} & -w_{1,2} \\
-w_{2,1} & w_{1,1}
\end{pmatrix}
\end{eqnarray}

We have 4 conditions to be fullfilled:
\begin{eqnarray} \label{eqnset}
1=\int dx w_L^{*}(x)w_L(x)=\int dx (+\alpha^{\dag} \psi_{E1}(x)^{\dag}+\beta^{\dag} \psi_{E2}(x)^{\dag})(\alpha\psi_{E1}(x)+\beta \psi_{E2}(x)), \nonumber \\
1=\int dx w_R^{*}(x)w_R(x)=\int dx (-\beta^{\dag} \psi_{E1}(x)+\alpha^{\dag}\psi_{E2}(x)^{\dag})(-\beta\psi_{E1}(x)+\alpha \psi_{E2}(x)), \nonumber \\
0=\int dx w_R^{*}(x)w_L(x)=\int dx (-\beta^{\dag} \psi_{E1}(x)^{\dag}+\alpha^{\dag} \psi_{E2}(x)^{\dag})(\alpha\psi_{E1}(x)+\beta \psi_{E2}(x)), \nonumber \\
0=\int dx w_L^{*}(x)w_R(x)=\int dx (\alpha^{\dag}\psi_{E1}(x)^{\dag}+\beta^{\dag} \psi_{E2}(x)^{\dag})(-\beta \psi_{E1}(x)+\alpha \psi_{E2}(x)). 
\end{eqnarray}
Due to orthogonality of 2 wavefunctions $\psi_{E1}$ and $\psi_{E2}$ we have from first two equations $|\alpha|^2+|\beta|^2=1$, so $|\beta|^2=1-|\alpha|^2$.
From 3rd and 4th equation we have $\alpha^{\dag}\beta=\alpha\beta^{\dag}$ that is fullfilled when $\alpha=|\alpha|e^{i\delta}, \beta=|\beta|e^{-i\delta}=\sqrt{1-|\alpha|^2}e^{-i\delta}$.

Therefore we have

We propose maximum localized orthonormal Wannier functions of the form
\begin{eqnarray}
w_L(x)=(+|\alpha|e^{i\delta} \psi_{E1}(x)+\sqrt{1-|\alpha|^2}e^{-i\delta} \psi_{E2}(x)), \nonumber \\
w_R(x)=(-\sqrt{1-|\alpha|^2}e^{-i\delta} \psi_{E1}(x)+|\alpha|e^{i\delta}\psi_{E2}(x)), \nonumber \\
\end{eqnarray}
or
\begin{eqnarray}
w_L(x)=(+|\alpha|e^{i\delta} \psi_{E1}(x)+\sqrt{1-|\alpha|^2}e^{-i\delta} \psi_{E2}(x)), \nonumber \\
w_R(x)=(+\sqrt{1-|\alpha|^2}e^{-i\delta} \psi_{E1}(x)-|\alpha|e^{i\delta}\psi_{E2}(x)), \nonumber \\
\end{eqnarray}

The last criteria to be matched is that $w_L(x)$ is maximum localized on the left quantum dot that geometric position is given by $x \in (-\infty,0)$ and that $w_R(x)$ is maximum localized on the right quantum dot that is denoted by $x \in (0,+\infty)$. Formally we can define
\begin{eqnarray}
S_L(\alpha)[\psi_{E1}(x),\psi_{E2}(x)]=\int_{-\infty}^{0}w_L(x)^{*}w_L(x)dx= \nonumber \\
=\int_{-\infty}^{0}dx[+|\alpha|e^{-i\delta} \psi_{E1}^{\dag}(x)+\sqrt{1-|\alpha|^2}e^{+i\delta} \psi_{E2}^{\dag}(x))][+|\alpha|e^{i\delta} \psi_{E1}(x)+\sqrt{1-|\alpha|^2}e^{-i\delta} \psi_{E2}(x))]= \nonumber \\
=\int_{-\infty}^{0}dx [ (1-|\alpha|^2)|\psi_{E2}|^2+|\alpha|^2|\psi_{E1}(x)|^2+|\alpha|\sqrt{1-|\alpha|^2}(e^{i\delta}\psi_{E1}\psi_{E2}^{\dag}(x)+e^{-i\delta}\psi_{E1}^{\dag}\psi_{E2}(x)) ]
\end{eqnarray}
Since $S_L(\alpha)[\psi_{E1}(x),\psi_{E2}(x)]$ reaches maximum with respect to $|\alpha|$ and $\delta$ we have $\frac{d}{d|\alpha|}S_L(\alpha)[\psi_{E1}(x),\psi_{E2}(x)]=0$ and $\frac{d}{d|\delta|}S_L(\alpha)[\psi_{E1}(x),\psi_{E2}(x)]=0$.
We have
\begin{eqnarray}
\frac{d}{d|\alpha|}S_L(\alpha)[\psi_{E1}(x),\psi_{E2}(x)]=[ -2|\alpha|\int_{-\infty}^{0}dx|\psi_{E2}(x)|^2+2|\alpha|\int_{-\infty}^{0}dx|\psi_{E1}(x)|^2+ \nonumber \\
+[\sqrt{1-|\alpha|^2}-\frac{|\alpha|^2}{\sqrt{1-|\alpha|^2}}](e^{i\delta}\int_{-\infty}^{0}dx\psi_{E1}(x)\psi_{E2}(x)^{\dag}(x)+e^{-i\delta}\int_{-\infty}^{0}dx\psi_{E1}^{\dag}(x)\psi_{E2}(x)) ]=0.
\end{eqnarray}
Having only $\psi_{E1}(x), \psi_{E2}(x) \in R$ we have $\psi_{E1}(x)\psi_{E2}(x)^{\dag}=\psi_{E1}(x)^{\dag}\psi_{E2}(x)=\psi_{E1}(x)\psi_{E2}(x)$ and thus from $\frac{d}{d|\alpha|}S_L(\alpha)=0$ we have
\begin{eqnarray}
[ -\int_{-\infty}^{0}dx|\psi_{E2}(x)|^2+\int_{-\infty}^{0}dx|\psi_{E1}(x)|^2 
+[\frac{\sqrt{1-|\alpha|^2}}{|\alpha|}-\frac{|\alpha|}{\sqrt{1-|\alpha|^2}}]cos(\delta)(\int_{-\infty}^{0}dx\psi_{E1}(x)\psi_{E2}(x)(x))=0.
\end{eqnarray}
as well as
\begin{eqnarray}
|\alpha|\sqrt{1-|\alpha|^2}(ie^{i\delta}\int_{-\infty}^{0}dx\psi_{E1}(x)\psi_{E2}(x)^{\dag}(x)-ie^{-i\delta}\int_{-\infty}^{0}dx\psi_{E1}^{\dag}(x)\psi_{E2}(x)) ]=0.
\end{eqnarray}
what simply brings $e^{i\delta}=e^{-i\delta}$ and thus $e^{2i\delta}=1$ what implies $\delta=k\Pi$, so $e^{i\delta}=e^{-i\delta}=1$.
In case we obtain
\begin{eqnarray}
[ -\int_{-\infty}^{0}dx|\psi_{E2}(x)|^2+\int_{-\infty}^{0}dx|\psi_{E1}(x)|^2 
+[\frac{\sqrt{1-|\alpha|^2}}{|\alpha|}-\frac{|\alpha|}{\sqrt{1-|\alpha|^2}}](\int_{-\infty}^{0}dx\psi_{E1}(x)\psi_{E2}(x))=0.
\end{eqnarray}
and thus
\begin{eqnarray}
[ -c_2+c_1 
+[\frac{\sqrt{1-|\alpha|^2}}{|\alpha|}-\frac{|\alpha|}{\sqrt{1-|\alpha|^2}}]c_0=0.
\end{eqnarray}
that is equivalent to
\begin{eqnarray}
+\frac{1-2|\alpha|^2}{|\alpha|\sqrt{1-|\alpha|^2}}=\frac{c_2-c_1}{c_0}=r=\frac{\int_{-\infty}^{0}dx(|\psi_{E2}(x)|^2-|\psi_{E1}(x)|^2)}{\int_{-\infty}^{0}dx\psi_{E1}(x)\psi_{E2}(x)}.
\end{eqnarray}
and hence
\begin{eqnarray}
1+4|\alpha|^4-4|\alpha|^2=r^2|\alpha|^2-r^4|\alpha|^4.
\end{eqnarray}
that can be summarized
\begin{eqnarray}
(r^4+4)|\alpha|^4-(4+r^2)|\alpha|^2+1=0.
\end{eqnarray}
and $\Delta=(4+r^2)^2-4(r^4+4)=16+r^4+8r^2-4r^4-16=-3r^4+8r^2=4r^2(2-\frac{3}{4}r^2)$.
Consequently
\begin{eqnarray}
|\alpha|^2=\frac{(4+r^2)\pm 2r\sqrt{2-\frac{3}{4}r^2}}{2(r^4+4)}.
\end{eqnarray}
In order to have $|\alpha|>=0$ we have the condition $2>\frac{3}{4}r^2$ that is $2.666(6)=\sqrt{\frac{8}{3}}>|r|$.
Finally one can write

\begin{eqnarray}
\begin{pmatrix}
w_L(x) \\
w_R(x)
\end{pmatrix}
=
\begin{pmatrix}
+\sqrt{\frac{(4+r^2)\pm 2r\sqrt{2-\frac{3}{4}r^2}}{2(r^4+4)}} & +\sqrt{1-\frac{(4+r^2)\pm 2r\sqrt{2-\frac{3}{4}r^2}}{2(r^4+4)}} \\
-\sqrt{1-\frac{(4+r^2)\pm 2r\sqrt{2-\frac{3}{4}r^2}}{2(r^4+4)}} & +\sqrt{\frac{(4+r^2)\pm 2r\sqrt{2-\frac{3}{4}r^2}}{2(r^4+4)}}
\end{pmatrix}
\begin{pmatrix}
\psi_{E1}(x) \\
\psi_{E2}(x)
\end{pmatrix},
\end{eqnarray}
 with $r=\frac{\int_{-\infty}^{0}dx(|\psi_{E2}(x)|^2-|\psi_{E1}(x)|^2)}{\int_{-\infty}^{0}dx\psi_{E1}(x)\psi_{E2}(x)}$. Such reasoning can be conducted for any 2 different energy levels as well as for N different energetic levels.
 If quantum state is given as
 \begin{eqnarray}
 |\psi>=e^{\frac{E_1 (t-t_0)}{i \hbar}}e^{i\gamma_{E1}}\sqrt{p_{E1}}|E_1>+e^{\frac{E_2 (t-t_0)}{i \hbar}}e^{i\gamma_{E2}}\sqrt{p_{E2}}\sqrt{p_{E2}}|E_2>
 \end{eqnarray}
 then

\begin{eqnarray}
\begin{pmatrix}
\alpha(t) w_L(x) \\
\beta(t) w_R(x)
\end{pmatrix}
=
\begin{pmatrix}
+\sqrt{\frac{(4+r^2)\pm 2r\sqrt{2-\frac{3}{4}r^2}}{2(r^4+4)}} & +\sqrt{1-\frac{(4+r^2)\pm 2r\sqrt{2-\frac{3}{4}r^2}}{2(r^4+4)}} \\
-\sqrt{1-\frac{(4+r^2)\pm 2r\sqrt{2-\frac{3}{4}r^2}}{2(r^4+4)}} & +\sqrt{\frac{(4+r^2)\pm 2r\sqrt{2-\frac{3}{4}r^2}}{2(r^4+4)}}
\end{pmatrix}
\begin{pmatrix}
e^{\frac{E_1 (t-t_0)}{i \hbar}}e^{i\gamma_{E1}}\sqrt{p_{E1}} \psi_{E1}(x) \\
e^{\frac{E_2 (t-t_0)}{i \hbar}}e^{i\gamma_{E2}}\sqrt{p_{E2}} \psi_{E2}(x)
\end{pmatrix},
\end{eqnarray}
The last implies that
\begin{eqnarray}
\alpha(t)= \int_{-\infty}^{+\infty}dx
\begin{pmatrix}
[\sqrt{\frac{(4+r^2)\pm 2r\sqrt{2-\frac{3}{4}r^2}}{2(r^4+4)}}]^{*}\psi_{E1}^{\dag}(x), & [\sqrt{1-\frac{(4+r^2)\pm 2r\sqrt{2-\frac{3}{4}r^2}}{2(r^4+4)}}]\psi_{E2}^{\dag}(x)
\end{pmatrix} \times \nonumber \\
\begin{pmatrix}
+\sqrt{\frac{(4+r^2)\pm 2r\sqrt{2-\frac{3}{4}r^2}}{2(r^4+4)}}, & +\sqrt{1-\frac{(4+r^2)\pm 2r\sqrt{2-\frac{3}{4}r^2}}{2(r^4+4)}} \\
-\sqrt{1-\frac{(4+r^2)\pm 2r\sqrt{2-\frac{3}{4}r^2}}{2(r^4+4)}}, & +\sqrt{\frac{(4+r^2)\pm 2r\sqrt{2-\frac{3}{4}r^2}}{2(r^4+4)}}
\end{pmatrix}
\begin{pmatrix}
e^{\frac{E_1 (t-t_0)}{i \hbar}}e^{i\gamma_{E1}}\sqrt{p_{E1}} \psi_{E1}(x) \\
e^{\frac{E_2 (t-t_0)}{i \hbar}}e^{i\gamma_{E2}}\sqrt{p_{E2}} \psi_{E2}(x)
\end{pmatrix}= \nonumber \\
=\int_{-\infty}^{+\infty}dx
\begin{pmatrix}
w_{1,1}^{*}\psi_{E1}^{\dag}(x), & w_{1,2}^{*}\psi_{E2}^{\dag}(x)
\end{pmatrix} \times 
\begin{pmatrix}
w_{1,1} & w_{1,2} \\
w_{2,1} & w_{2,2}
\end{pmatrix}
\begin{pmatrix}
e^{\frac{E_1 (t-t_0)}{i \hbar}}e^{i\gamma_{E1}}\sqrt{p_{E1}} \psi_{E1}(x) \\
e^{\frac{E_2 (t-t_0)}{i \hbar}}e^{i\gamma_{E2}}\sqrt{p_{E2}} \psi_{E2}(x)
\end{pmatrix}= \nonumber \\
=\int_{-\infty}^{+\infty}dx
\begin{pmatrix}
w_{1,1}^{*}\psi_{E1}^{\dag}(x), & w_{1,2}^{*}\psi_{E2}^{\dag}(x)
\end{pmatrix}
\begin{pmatrix}
w_{1,1}e^{\frac{E_1 (t-t_0)}{i \hbar}}e^{i\gamma_{E1}}\sqrt{p_{E1}} \psi_{E1}(x)+w_{1,2}e^{\frac{E_2 (t-t_0)}{i \hbar}}e^{i\gamma_{E2}}\sqrt{p_{E2}} \psi_{E2}(x) \\
w_{2,1}e^{\frac{E_1 (t-t_0)}{i \hbar}}e^{i\gamma_{E1}}\sqrt{p_{E1}} \psi_{E1}(x)+w_{2,2}e^{\frac{E_2 (t-t_0)}{i \hbar}}e^{i\gamma_{E2}}\sqrt{p_{E2}} \psi_{E2}(x)
\end{pmatrix}= \nonumber \\
=w_{1,1}^{*}w_{1,1}e^{\frac{E_1 (t-t_0)}{i \hbar}}e^{i\gamma_{E1}}\sqrt{p_{E1}} +w_{1,2}^{*}w_{2,2}e^{\frac{E_2 (t-t_0)}{i \hbar}}e^{i\gamma_{E2}}\sqrt{p_{E2}}= \nonumber \\
=[+\sqrt{\frac{(4+r^2)\pm 2r\sqrt{2-\frac{3}{4}r^2}}{2(r^4+4)}}]^{\dag}[+\sqrt{\frac{(4+r^2)\pm 2r\sqrt{2-\frac{3}{4}r^2}}{2(r^4+4)}}]e^{\frac{E_1 (t-t_0)}{i \hbar}}e^{i\gamma_{E1}}\sqrt{p_{E1}} + \nonumber \\
\Bigg[ +\sqrt{1-\frac{(4+r^2)\pm 2r\sqrt{2-\frac{3}{4}r^2}}{2(r^4+4)}} \Bigg]^{\dag}
\Bigg[+\sqrt{\frac{(4+r^2)\pm 2r\sqrt{2-\frac{3}{4}r^2}}{2(r^4+4)}} \Bigg] e^{\frac{E_2 (t-t_0)}{i \hbar}}e^{i\gamma_{E2}}\sqrt{p_{E2}}=\alpha(t). \nonumber \\
\end{eqnarray}
and
 \begin{eqnarray}
\beta(t)= \int_{-\infty}^{+\infty}dx
\begin{pmatrix}
 -[\sqrt{1-\frac{(4+r^2)\pm 2r\sqrt{2-\frac{3}{4}r^2}}{2(r^4+4)}}]\psi_{E1}^{\dag}(x), & [\sqrt{\frac{(4+r^2)\pm 2r\sqrt{2-\frac{3}{4}r^2}}{2(r^4+4)}}]^{*}\psi_{E2}^{\dag}(x)
\end{pmatrix} \times \nonumber \\
\begin{pmatrix}
+\sqrt{\frac{(4+r^2)\pm 2r\sqrt{2-\frac{3}{4}r^2}}{2(r^4+4)}}, & +\sqrt{1-\frac{(4+r^2)\pm 2r\sqrt{2-\frac{3}{4}r^2}}{2(r^4+4)}} \\
-\sqrt{1-\frac{(4+r^2)\pm 2r\sqrt{2-\frac{3}{4}r^2}}{2(r^4+4)}}, & +\sqrt{\frac{(4+r^2)\pm 2r\sqrt{2-\frac{3}{4}r^2}}{2(r^4+4)}}
\end{pmatrix}
\begin{pmatrix}
e^{\frac{E_1 (t-t_0)}{i \hbar}}e^{i\gamma_{E1}}\sqrt{p_{E1}} \psi_{E1}(x) \\
e^{\frac{E_2 (t-t_0)}{i \hbar}}e^{i\gamma_{E2}}\sqrt{p_{E2}} \psi_{E2}(x)
\end{pmatrix}= \nonumber \\
=\int_{-\infty}^{+\infty}dx
\begin{pmatrix}
w_{2,1}^{*}\psi_{E1}^{\dag}(x), & w_{2,2}^{*}\psi_{E2}^{\dag}(x)
\end{pmatrix} \times 
\begin{pmatrix}
w_{1,1} & w_{1,2} \\
w_{2,1} & w_{2,2}
\end{pmatrix}
\begin{pmatrix}
e^{\frac{E_1 (t-t_0)}{i \hbar}}e^{i\gamma_{E1}}\sqrt{p_{E1}} \psi_{E1}(x) \\
e^{\frac{E_2 (t-t_0)}{i \hbar}}e^{i\gamma_{E2}}\sqrt{p_{E2}} \psi_{E2}(x)
\end{pmatrix}= \nonumber \\
=\int_{-\infty}^{+\infty}dx
\begin{pmatrix}
w_{2,1}^{*}\psi_{E1}^{\dag}(x), & w_{2,2}^{*}\psi_{E2}^{\dag}(x)
\end{pmatrix}
\begin{pmatrix}
w_{1,1}e^{\frac{E_1 (t-t_0)}{i \hbar}}e^{i\gamma_{E1}}\sqrt{p_{E1}} \psi_{E1}(x)+w_{1,2}e^{\frac{E_2 (t-t_0)}{i \hbar}}e^{i\gamma_{E2}}\sqrt{p_{E2}} \psi_{E2}(x) \\
w_{2,1}e^{\frac{E_1 (t-t_0)}{i \hbar}}e^{i\gamma_{E1}}\sqrt{p_{E1}} \psi_{E1}(x)+w_{2,2}e^{\frac{E_2 (t-t_0)}{i \hbar}}e^{i\gamma_{E2}}\sqrt{p_{E2}} \psi_{E2}(x)
\end{pmatrix}= \nonumber \\
=w_{2,1}^{*}w_{1,1}e^{\frac{E_1 (t-t_0)}{i \hbar}}e^{i\gamma_{E1}}\sqrt{p_{E1}} +w_{2,2}^{*}w_{2,2}e^{\frac{E_2 (t-t_0)}{i \hbar}}e^{i\gamma_{E2}}\sqrt{p_{E2}}= \nonumber \\
=\Bigg[-\sqrt{1-\frac{(4+r^2)\pm 2r\sqrt{2-\frac{3}{4}r^2}}{2(r^4+4)}}\Bigg]^{\dag}\sqrt{\frac{(4+r^2)\pm 2r\sqrt{2-\frac{3}{4}r^2}}{2(r^4+4)}}e^{\frac{E_1 (t-t_0)}{i \hbar}}e^{i\gamma_{E1}}\sqrt{p_{E1}}+ \nonumber \\
+\Bigg[ +\sqrt{\frac{(4+r^2)\pm 2r\sqrt{2-\frac{3}{4}r^2}}{2(r^4+4)}} \Bigg]^{\dag}\Bigg[ +\sqrt{\frac{(4+r^2)\pm 2r\sqrt{2-\frac{3}{4}r^2}}{2(r^4+4)}} \Bigg]e^{\frac{E_2 (t-t_0)}{i \hbar}}e^{i\gamma_{E2}}\sqrt{p_{E2}}=\beta(t) \nonumber \\ .
\end{eqnarray}
\section{Schroedinger propagator formalism approach towards position-based qubits}
We have the equation of motion from Schroedinger formalism given as
\begin{equation}
(i\hbar \frac{d}{dt}+\frac{\hbar^2}{2m}\frac{d^2}{dx^2}+\frac{\hbar^2}{2m}\frac{d^2}{dy^2}-V(x,y,t)+c)\psi(x,y,t)=\hat{O}(x,y,t)\psi(x,y,t)=c \psi(x,y,t)
\end{equation}
and we introduce the following definition of the propagator $G(x_2,t_2,x_1,t_1)$ given as
\begin{eqnarray}
\psi(x_2,t_2)=\int_{-\infty}^{+\infty} dt_1 \int_{-\infty}^{+\infty}dx_1 (G(x_2,t_2,x_1,t_1)\psi(x_1,t_1))= \nonumber \\ =%
\int_{-\infty}^{\infty} dt_1 
\int_{-\infty}^{\infty}dx_1(\delta(t_2-t_1)\psi(x_2,t_1)\psi(x_1,t_1)^{*})[\psi(x_1,t_1)]=\int_{-\infty}^{\infty}dx_1(\psi(x_2,t_2)\psi^{*}(x_1,t_1))[\psi(x_1,t_1)].
\end{eqnarray}
We have consequently by postulating $\hat{O}(x_2,t_2)G(x_2,t_2,x_1,t_1)=c \delta(t_2-t_1)\delta(x_2-x_1)$ we obtain
\begin{eqnarray}
\hat{O}(x_2,t_2)\psi(x_2,t_2)=c \psi(x_2,t_2)=  \nonumber \\
=\int_{-\infty}^{\infty} dt_1 \int_{-\infty}^{+\infty}dx_1 [\hat{O}(x_2,t_2)G(x_2,t_2,x_1,t_1)]\psi(x_1,t_1)=\int_{-\infty}^{\infty} dt_1 \int_{-\infty}^{+\infty}dx_1 [c \delta(x_2-x_1)\delta(t_2-t_1)]\psi(x_1,t_1)= \nonumber \\
=c \psi(x_2,t_2).
\end{eqnarray}
Very last result confirms that indeed propagator equation of motion can be written by equation
\begin{equation}
\hat{O}(x_2,t_2)G(x_2,t_2,x_1,t_1)=c \delta(t_2-t_1)\delta(x_2-x_1).
\end{equation} what is equivalent to the following\textbf{ propagator in time and in space} in form as
\begin{eqnarray}
G(x_2,t_2,x_1,t_1)= \hat{O}(x_2,t_2)^{-1}[c \delta(t_2-t_1)\delta(x_2-x_1)]=\frac{c \delta(t_2-t_1)\delta(x_2-x_1)}{i\hbar \frac{d}{dt_2}+\frac{\hbar^2}{2m}\frac{d^2}{dx_2^2}-V(x_2,t_2)+c}=\frac{c \delta(t_2-t_1)\delta(x_2-x_1)}{i\hbar \frac{d}{dt_2}-\hat{H}(x_2,t_2)+c}= \nonumber \\
=\frac{\delta(t_2-t_1)\delta(x_2-x_1)}{i\hbar \frac{d}{dt_2}-\hat{H}(x_2,t_2)}.
\end{eqnarray}
and this propagator equation of motion remains valid for arbitrary small c but still c different from zero (c can be real or imaginary or complex value constant). Here $\hat{H}(x_2,t_2)$ is system Hamiltonian.
Defining only propagator in space we have $G_a(x_2,x_1,t_1)$ and it can be formally expressed as
 \begin{equation}
\int_{-\infty}^{+\infty}dx_1 (G_a(x_2,x_1,t_1)\psi(x_1,t_1))=\psi(x_2,t_1)=  \int_{-\infty}^{\infty}\psi(x_2,t_1)dx_1\psi(x_1,t_1)^{*}[\psi(x_1,t_1)].
\end{equation}
By postulating $\hat{O}(x_2,t_1)G_a(x_2,x_1,t_1)=c \delta(x_2-x_1)$ we obtain
\begin{eqnarray}
\hat{O}(x_2,t_1)\psi(x_2,t_1)=c \psi(x_2,t_1)=  \nonumber \\
= \int_{-\infty}^{+\infty}dx_1 [\hat{O}(x_2,t_2)G_a(x_2,x_1,t_1)]\psi(x_1,t_1)=\int_{-\infty}^{+\infty}dx_1 [c \delta(x_2-x_1)]\psi(x_1,t_1) 
=c \psi(x_2,t_1).
\end{eqnarray}
what is equivalent to the following \textbf{propagator in space (but not in the time!)} in form as
\begin{equation}
G_a(x_2,t_1,x_1,t_1)= \hat{O}(x_2,t_1)^{-1}[c \delta(x_2-x_1)]=\frac{c \delta(x_2-x_1)}{i\hbar \frac{d}{dt_1}+\frac{\hbar^2}{2m}\frac{d^2}{dx_2^2}-V(x_2,t_1)+c}=\frac{c \delta(x_2-x_1)}{i\hbar \frac{d}{dt_1}-\hat{H}(x_2,t_1)+c}.
\end{equation}
Defining only \textbf{propagator in time (but not in space)} we have $G_b(x_1,t_2,t_1)$ and it can be formally expressed as
 \begin{equation}
\int_{-\infty}^{+\infty}dt_1 (G_b(x_1,t_2,t_1)\psi(x_1,t_1))=\psi(x_1,t_2)=  \int_{-\infty}^{\infty}\psi(x_1,t_2)dx_1\psi(x_1,t_1)^{*}[\psi(x_1,t_1)].
\end{equation}
By postulating $\hat{O}(x_1,t_2)G_b(x_1,t_2,t_1)=c \delta(t_2-t_1)$ we obtain
\begin{eqnarray}
\hat{O}(x_1,t_2)\psi(x_1,t_2)=c \psi(x_1,t_2)=  \nonumber \\
= \int_{-\infty}^{+\infty}dt_1 [\hat{O}(x_1,t_2)G_b(x_1,t_2,t_1)]\psi(x_1,t_1)=\int_{-\infty}^{+\infty}dt_1 [c \delta(t_2-t_1)]\psi(x_1,t_1) 
=c \psi(x_1,t_2).
\end{eqnarray}
Further details on propagators are give in \ref{App1}.
\section{From Schroedinger formalism to tight-binding model via shifts in time and in space}
It is possible to obtain radical simplification of equation \ref{eqnq}.
We consider translational operator in x and in t that is naturally described in Taylor expansion and we have.
It is of the form
\begin{eqnarray}
(1+(t-t_0)\frac{d}{dt_0}+\frac{1}{2!}(t-t_0)^2\frac{d^2}{dt_0^2}+..)\psi(x,t_0)=e^{(t-t_0)\frac{d}{dt_0}}\psi(x,t_0)=e^{\Delta t\frac{1}{i\hbar}\hat{H}(t_0)}\psi(x,t_0)=\psi(x,t_0+\Delta t), \nonumber \\
(1+(x-x_0)\frac{d}{dx_0}+\frac{1}{2!}(x-x_0)^2\frac{d^2}{dx_0^2}+..)\psi(x_0,t_0)=e^{(x-x_0)\frac{d}{dx_0}}\psi(x_0,t_0)=e^{\Delta x\frac{i}{\hbar}\hat{p}(t_0)}\psi(x_0,t_0)=\psi(x_0+\Delta x,t_0), \nonumber \\
e^{(t-t_0)\frac{d}{dt_0}+(x-x_0)\frac{d}{dx_0}}\psi(x_0,t_0)=e^{\Delta t\frac{d}{dt_0}+\Delta x \frac{d}{dx_0}}\psi(x_0,t_0)=\psi(x_0+\Delta x,t_0+\Delta t)
\end{eqnarray}
and it implies
\begin{equation}
(e^{-\Delta t \frac{i}{\hbar}\hat{H}(x_0,t_0)+\Delta x \frac{i}{\hbar}\hat{p}(t_0)}\psi(x_0,t_0)))=e^{\hat{A}(t,t_0)+\hat{B}(x,x_0)}\psi(x_0,t_0)=\psi(x_0+\Delta x,t_0+\Delta t).
\end{equation}
We use formula Zassenhaus formula \cite{HB},\cite{HB1}
\begin{eqnarray}
e^{s[\hat{A}(t,t_0)+\frac{s_1}{s}\hat{B}(x,x_0)]}=e^{s\hat{A}(t,t_0)}e^{\frac{s_1}{s}\hat{B}(x,x_0)}e^{-\frac{s^2}{2}[\hat{A}(t,t_0),\frac{s_1}{s}\hat{B}(t,t_0)]}e^{+\frac{s^3}{6}(2[\hat{A}(t,t_0),[\frac{s_1}{s}\hat{B}(x,x_0),\hat{A}(t,t_0)]]+[\hat{A}(t,t_0),[\hat{A}(t,t_0),\frac{s_1}{s}\hat{B}(t,t_0)]])}\times .. \nonumber \\
\end{eqnarray}
We have the following commutator
\begin{eqnarray}
[\hat{A}(t,t_0),\hat{B}(t,t_0)]=\nonumber \\
=[-\Delta t \frac{i}{\hbar}\hat{H}(x_0,t_0),\Delta x \frac{i}{\hbar}\hat{p}(x_0,t_0)]=\frac{\Delta t \Delta x}{\hbar^2}[V(x_0,t_0),\frac{\hbar}{i}\frac{d}{dx_0}]=\frac{\Delta t \Delta x}{i\hbar}[V(x_0,t_0),\frac{d}{dx_0}]=\frac{\Delta t \Delta x}{i\hbar}(\frac{d}{dx_0}V(x_0,t_0)).
\end{eqnarray}
and
\begin{eqnarray}
[\hat{A}(t,t_0),[\hat{A}(t,t_0),\hat{B}(t,t_0)]]=\nonumber \\
=[-\Delta t \frac{i}{\hbar}\hat{H}(x_0,t_0),\frac{\Delta t \Delta x}{i\hbar}(\frac{d}{dx_0}V(x_0,t_0))]=[+\Delta t \frac{i}{\hbar}\frac{\hbar^2}{2m}\frac{d^2}{dx_0^2},\frac{\Delta t \Delta x}{i\hbar}(\frac{d}{dx_0}V(x_0,t_0))]= \nonumber \\
=\frac{(\Delta t)^2 \Delta x}{2m}[\frac{d^2}{dx_0^2},(\frac{d}{dx_0}V(x_0,t_0))]=\frac{(\Delta t)^2 \Delta x}{2m}((\frac{d^3}{dx_0^3}V(x_0,t_0))+(\frac{d^2}{dx_0^2}V(x_0,t_0))\frac{d}{dx_0}).
\end{eqnarray}
and
\begin{eqnarray}
[\hat{A}(t,t_0),[\hat{B}(t,t_0),\hat{A}(t,t_0)]] 
=-[\hat{A}(t,t_0),[\hat{A}(t,t_0),\hat{B}(t,t_0)]]=[\Delta t \frac{i}{\hbar}\hat{H}(x_0,t_0),\frac{\Delta t \Delta x}{i\hbar}(\frac{d}{dx_0}V(x_0,t_0))]= \nonumber \\
=[-\Delta t \frac{i}{2m \hbar} \hbar^2 \frac{d^2}{dx_0^2},\frac{\Delta t \Delta x}{i\hbar}(\frac{d}{dx_0}V(x_0,t_0))]
=-\frac{(\Delta t)^2 \Delta x}{2m}((\frac{d^3}{dx_0^3}V(x_0,t_0))+(\frac{d^2}{dx_0^2}V(x_0,t_0))\frac{d}{dx_0}).
\end{eqnarray}
Now we make approximation
\begin{eqnarray}
(e^{-\Delta t \frac{i}{\hbar}\hat{H}(x_0,t_0)+\Delta x \frac{i}{\hbar}\hat{p}(t_0)}\psi(x_0,t_0)))= \nonumber \\
(e^{-\Delta t \frac{i}{\hbar}\hat{H}(x_0,t_0)}(e^{+\Delta x \frac{i}{\hbar}\hat{p}(t_0)}\psi(x_0,t_0))= \nonumber \\
(e^{-\Delta t \frac{i}{\hbar}\hat{H}(x_0,t_0)}(\psi(x_0+\Delta x,t_0))=\psi(x_0+\Delta x,t_0+\Delta t).
\end{eqnarray}
We have two energies participating in hopping between two quantum dots or finite size quantum wells. The first one $t_{s12}$ is given as
\small
\begin{eqnarray}
t_{s12}=t_{R_1 \rightarrow R_2}(t_1,t_2)=t_{R_1 \rightarrow R_2}(t_1,t_1+\Delta t) =\int_{x_a}^{\frac{x_a+x_b}{2}} dx_1 \int_{\frac{x_a+x_b}{2}}^{x_b} dx_2 \psi^{*}(x_1,t_1)\hat{H}(x_2,t_2)\psi(x_2,t_2)= \nonumber \\
t_{R_1 \rightarrow R_2}(t_1,t_2)=t_{R_1 \rightarrow R_2}(t_1,t_1+\Delta t) =\int_{x_a}^{\frac{x_a+x_b}{2}} dx_1 \int_{\frac{x_a+x_b}{2}}^{x_b} dx_2 \psi^{*}(x_1,t_1)\hat{H}(x_2,t_2)\int_{-\infty}^{+\infty}\int_{-\infty}^{+\infty}dx_3dt_3G(x_2,t_2,x_3,t_3)\psi(x_3,t_3)= \nonumber \\
=\int_{x_a}^{\frac{x_a+x_b}{2}} dx_1 \int_{\frac{x_a+x_b}{2}}^{x_b} dx_2 \psi^{*}(x_1,t_1)\hat{H}\psi(x_1+(x_2-x_1),t_1+(t_2-t_1))=\nonumber \\
=\int_{x_a}^{\frac{x_a+x_b}{2}} dx_1 \int_{\frac{x_a+x_b}{2}}^{x_b} dx_2 \psi^{*}(x_1,t_1)E(t_2)\psi(x_1+(x_2-x_1),t_1+(t_2-t_1))=\nonumber \\
=E(t_2)\int_{x_a}^{\frac{x_a+x_b}{2}} dx_1 \int_{\frac{x_a+x_b}{2}}^{x_b} dx_2 \psi^{*}(x_1,t_1)e^{\frac{i}{\hbar}(x_2-x_1)\hat{p}(x_1,t_1)}e^{-\frac{i}{\hbar}(t_2-t_1)\hat{H}(t_1)} 
\psi(x_1,t_1)
=\nonumber \\
=\sum_{n=0}^{+\infty} \sum_{k=0}^{+\infty} c_{E_n(t_2)}c_{E_k(t_1)}^{*} E_n(t_2)e^{-(t_2-t_1) \frac{i}{\hbar}E_n(t_1)} \times \nonumber \\
\times\int_{x_a}^{\frac{x_a+x_b}{2}} dx_1 [\psi_k^{*}(x_1,t_1)(\frac{\psi_n(x_1,t_1)}{\frac{d}{dx_1}\psi_n(x_1,t_1)}(e^{(x_b-x_a)\frac{d}{dx_1}}-e^{(\frac{x_b+x_a}{2}-x_1)\frac{d}{dx_1}})\psi_n(x_1,t_1))].
\end{eqnarray}

It is important to underline that hopping energy from tight-binding model can be expressed in terms of average energies from Schroedinger equations as
\begin{eqnarray}
t_{s12}=t_{R_1 \rightarrow R_2}(t_1,t_2)=t_{R_1 \rightarrow R_2}(t_1,t_1+\Delta t) =\int_{x_a}^{\frac{x_a+x_b}{2}} dx_1 \int_{\frac{x_a+x_b}{2}}^{x_b} dx_2 \psi^{*}(x_1,t_1)\hat{H}(x_2,t_2)\psi(x_2,t_2)= \nonumber \\
=\frac{\hbar}{i}E(t_2)e^{-\frac{i}{\hbar}(t_2-t_1)E(t_1)}(\langle2\frac{Sin(\frac{i}{\hbar}(\frac{3}{4}x_b-\frac{1}{4}x_a-\frac{1}{2}x) \hat{p}(x,t_1))
Sin(\frac{\hbar}{i}(\frac{3}{4}x_a-\frac{1}{4}x_b-x)p(x,t_1))}{\hat{p}(x,t_1)}>\rangle_{x\in R_1} \nonumber \\
-i2\langle\frac{Sin(\frac{i}{\hbar}(\frac{x_b}{4}-\frac{3}{4}x_a+\frac{x}{2}) \hat{p}(x,t_1))Cos(\frac{\hbar}{i}(\frac{3}{4}x_b-\frac{1}{4}x_a-\frac{x}{2})p(x,t_1))}{\hat{p}(x,t_1)}\rangle_{x\in R_1})= \nonumber \\
=-2\hbar E(t_2)e^{\frac{i}{\hbar}(t_2-t_1)E(t_1)}(i\langle\frac{Sin(\frac{i}{\hbar}(\frac{3}{4}x_b-\frac{1}{4}x_a-\frac{1}{2}x) \hat{p}(x,t_1))
Sin(\frac{\hbar}{i}(\frac{3}{4}x_a-\frac{1}{4}x_b-x)p(x,t_1))}{\hat{p}(x,t_1)}\rangle_{x\in R_1} \nonumber \\
+\langle\frac{Sin(\frac{i}{\hbar}(\frac{x_b}{4}-\frac{3}{4}x_a+\frac{x}{2}) \hat{p}(x,t_1))Cos(\frac{\hbar}{i}(\frac{3}{4}x_b-\frac{1}{4}x_a-\frac{x}{2})p(x,t_1))}{\hat{p}(x,t_1)}\rangle_{x\in R_1})= \nonumber \\
-2\hbar E(t_1+\Delta t)e^{\frac{i}{\hbar}\Delta tE(t_1)}\Bigg[e^{i ArcTan \Bigg[ \frac{ \big \langle \frac{Sin(\frac{i}{\hbar}(\frac{3}{4}x_b-\frac{1}{4}x_a-\frac{1}{2}x) \hat{p}(x,t_1))
Sin(\frac{\hbar}{i}(\frac{3}{4}x_a-\frac{1}{4}x_b-x)p(x,t_1))}{\hat{p}(x,t_1)}\big \rangle_{x\in R_1}}{\big \langle \frac{Sin(\frac{i}{\hbar}(\frac{x_b}{4}-\frac{3}{4}x_a+\frac{x}{2}) \hat{p}(x,t_1))Cos(\frac{\hbar}{i}(\frac{3}{4}x_b-\frac{1}{4}x_a-\frac{x}{2})p(x,t_1))}{\hat{p}(x,t_1)}\big \rangle_{x\in R_1}} \Bigg] }\Bigg] \times  \nonumber \\
\times
\Bigg[
| \langle \frac{Sin(\frac{i}{\hbar}(\frac{3}{4}x_b-\frac{1}{4}x_a-\frac{1}{2}x) \hat{p}(x,t_1))Sin(\frac{\hbar}{i}(\frac{3}{4}x_a-\frac{1}{4}x_b-x)p(x,t_1))}{\hat{p}(x,t_1)} \rangle_{x\in R_1}|^2+ \nonumber \\
|\langle\frac{Sin(\frac{i}{\hbar}(\frac{x_b}{4}-\frac{3}{4}x_a+\frac{x}{2}) \hat{p}(x,t_1))Cos(\frac{\hbar}{i}(\frac{3}{4}x_b-\frac{1}{4}x_a-\frac{x}{2})p(x,t_1))}{\hat{p}(x,t_1)}\rangle_{x \in R_1}|^2 \Bigg]^{\frac{1}{2}}
=t_{s12}=t_{R_1 \rightarrow R_2}(t_1,t_1+\Delta t).
\label{eqn:fundamental}
\end{eqnarray}
In such way hopping constant in tight-binding model was obtained from time-dependent Schroedinger equation for single-electron. The electron presence was assumed to be in two regions $R_1 \in (x_a,\frac{x_a+x_b}{2})$ and in region $R_2 \in (\frac{x_a+x_b}{2},x_b)$.
We can also express hopping constant by Green functions in the way as
\begin{eqnarray}
t_{s12}=t_{R_1 \rightarrow R_2}(t_1,t_2)=t_{R_1 \rightarrow R_2}(t_1,t_1+\Delta t) =\int_{x_a}^{\frac{x_a+x_b}{2}} dx_1 \int_{\frac{x_a+x_b}{2}}^{x_b} dx_2 \psi^{*}(x_1,t_1)\hat{H}(x_2,t_2)\psi(x_2,t_2)= \nonumber \\
\int_{x_a}^{\frac{x_a+x_b}{2}} dx_1 \int_{\frac{x_a+x_b}{2}}^{x_b} dx_2 \psi^{*}(x_1,t_1)
\hat{H}(x_2,t_2)\times
\nonumber \\
\int_{x_1}^{x_2} dx_{1a} \int_{t_1}^{t_2} dt_{1a} \int_{-\infty}^{+\infty} \int_{-\infty}^{+\infty} \int_{-\infty}^{+\infty} \int_{-\infty}^{+\infty} \frac{dk_1dk_2 d\omega_1 d\omega_2 e^{i(k_{1}x_{1a}+k_2x_2+\omega_1t_{1a}+\omega_2t_2)}\psi(x_{1a},t_{1a})\frac{1}{(2\pi)^2}}{[
  (\hbar \omega_2+\frac{\hbar^2k_2^2}{2m})-\frac{1}{2\pi}\int_{-\infty}^{+\infty} \int_{-\infty}^{+\infty}d\omega_adk_aV(k_a-k_2,\omega_a-\omega_2)
  e^{-ix_2(k_a-k_2)}e^{-it_2(\omega_a-\omega_2)}]} \nonumber \\=
  \int_{x_a}^{\frac{x_a+x_b}{2}} dx_1 \int_{\frac{x_a+x_b}{2}}^{x_b}dx_2 \psi^{*}(x_1,t_1)
(-\frac{\hbar^2}{2m}\frac{d^2}{dx_2^2}+V(x_2,t_2)) \times
\nonumber \\
\times
\int_{x_1}^{x_2} dx_{1a} \int_{t_1}^{t_2} dt_{1a} \int_{-\infty}^{+\infty} \int_{-\infty}^{+\infty} \int_{-\infty}^{+\infty} \int_{-\infty}^{+\infty} \frac{dk_1dk_2 d\omega_1 d\omega_2 e^{i(k_1x_{1a}+k_2x_2+\omega_1t_{1a}+\omega_2t_2)}\psi(x_{1a},t_{1a})\frac{1}{(2\pi)^2}}{[
  (\hbar \omega_2+\frac{\hbar^2k_2^2}{2m})-V(x_2,t_2)]}, 
\end{eqnarray}
\normalsize
where the knowledge of electron wavefunction at time $t_1$ is preassumed and given as $\psi(x,t_1)$ and effective potential dependence with time and with position is given as $V(x,t)$.
Also second hopping energy $t_{s21}$ is given as
\begin{eqnarray}
t_{s21}=t_{R_2 \rightarrow R_1}(t_1,t_2)=t_{R_2 \rightarrow R_1}(t_1,t_1+\Delta t) =\int_{\frac{x_a+x_b}{2}}^{x_b} dx_2 \int_{x_a}^{\frac{x_a+x_b}{2}} dx_1 \psi^{*}(x_2,t_1)\hat{H}\psi(x_1,t_2)= \nonumber \\
=\int_{\frac{x_a+x_b}{2}}^{x_b} dx_2 \int_{x_a}^{\frac{x_a+x_b}{2}} dx_1 \psi^{*}(x_1+(x_2-x_1),t_1)\hat{H}\psi(x_1,t_2)=\nonumber \\
=E(t_2)\int_{\frac{x_a+x_b}{2}}^{x_b} dx_2 \int_{x_a}^{\frac{x_a+x_b}{2}} dx_1 \psi^{*}(x_1+(x_2-x_1),t_1)\psi(x_1,t_2)=\nonumber \\
=E(t_2)\int_{\frac{x_a+x_b}{2}}^{x_b} dx_2 \int_{x_a}^{\frac{x_a+x_b}{2}} dx_1 \psi^{*}(x_1+(x_2-x_1),t_1)e^{-\frac{i}{\hbar}(t_2-t_1)\hat{H}}\psi(x_1,t_1)=\nonumber \\
=E(t_2)e^{-\frac{i}{\hbar}(t_2-t_1)E(t_1)}\int_{\frac{x_a+x_b}{2}}^{x_b} dx_2 \int_{x_a}^{\frac{x_a+x_b}{2}} dx_1 \psi^{*}(x_1+(x_2-x_1),t_1)\psi(x_1,t_1)=\nonumber \\
=E(t_2)e^{-\frac{i}{\hbar}(t_2-t_1)E(t_1)}\int_{x_a}^{\frac{x_a+x_b}{2}} dx_1 \int_{\frac{x_a+x_b}{2}}^{x_b} dx_2  ((e^{(x_2-x_1) \frac{-i}{\hbar}\hat{p}(x_1,t_1)}\psi^{*}(x_1,t_1))) \psi(x_1,t_1)=\nonumber \\
=E(t_2)e^{-\frac{i}{\hbar}(t_2-t_1)E(t_1)} \int_{x_a}^{\frac{x_a+x_b}{2}} dx_1 \int_{\frac{x_a+x_b}{2}}^{x_b} dx_2  (e^{(x_2-x_1) \frac{-i}{\hbar}\hat{p}(x_1,t_1)}\psi^{*}(x_1,t_1))\psi(x_1,t_2)= \nonumber \\
E(t_2)e^{-\frac{i}{\hbar}(t_2-t_1)E(t_1)} \int_{x_a}^{\frac{x_a+x_b}{2}} dx_1 \int_{-(x_b-x_a)}^{-\frac{x_a+x_b}{2}+x_1} d\Delta x  (e^{\Delta x \frac{-i}{\hbar}\hat{p}(x_1,t_1)}\psi^{*}(x_1,t_1))\psi(x_1,t_2)=
\nonumber \\
=\sum_n \sum_k \sqrt{c_{E_n}(t_2)}\sqrt{c_{E_k}(t_1)}e^{i(\phi_{E_n(t_2)}-\phi_{E_n(t_1)})} e^{i(\phi_{E_n}(t_2)-\phi_{E_n}(t_1))} E_n(t_2)e^{+\frac{i}{\hbar}(t_2-t_1)E_n(t_1)}\times \nonumber \\ \times \int_{x_a}^{\frac{x_a+x_b}{2}} dx_1 \psi_n(x_1,t_1)(\frac{\psi_k^{*}(x_1,t)}{\frac{d}{dx_1}(\psi_k^{*}(x_1,t))}(e^{(-\frac{x_b-x_a}{2}+x_1)\frac{d}{dx_1} }-e^{(x_b-x_a)\frac{d}{dx_1} }))\psi_k(x_1,t_1)^{*}
\end{eqnarray}
\normalsize
where we have assumed that given wavefunction $\psi(x,t)$ can be decomposed into its eigenenergies so we have
\begin{eqnarray}
\psi(x,t)=\sum_n c_n(t) \psi_n(x,t)=\sum_n e^{i\phi_n(t)}\sqrt{p_{E_n}(t)} \psi_n(x,t),
\hat{H} \psi_n(x,t)=E_n(t) \psi_n(x,t), \nonumber \\
\hat{p}\psi_n(x,t)=\frac{\hbar}{i}\frac{d}{dx}\psi_n(x,t)=p_n(x,t)\psi_n(x,t),\hat{p}\psi_n^{*}(x,t)=-\frac{\hbar}{i}\frac{d}{dx}(\psi_n(x,t))^{*}=p_n(x,t)^{*}\psi_n(x,t)^{*}, \nonumber \\ \frac{\frac{\hbar}{i}\frac{d}{dx}\psi_n(x,t)}{\psi_n(x,t)}=p_n(x,t), \frac{-\frac{\hbar}{i}\frac{d}{dx}(\psi_n(x,t))^{*}}{(\psi_n(x,t))^{*}}=p_n(x,t)^{*}, \frac{1}{p_n(x,t)}=\frac{\psi_n(x,t)}{\frac{\hbar}{i}\frac{d}{dx}\psi_n(x,t)}, \nonumber \\
\int_{-\infty}^{+\infty} dx |\psi_n(x,t)|^2=1, c_{n}(t)=\int_{-\infty}^{+\infty} dx \psi_n(x,t)^{*}\psi(x,t)=\sqrt{p_{E_n(t)}}e^{i\phi_{E_n(t)}},
\end{eqnarray}
so $E_n(t)$ is n-th eigenenergy of the system at time t and $\psi_n(t)$ is the eigenenergy wavefunction.
Doing in the same fashion we can obtain the time dependent value of operator $E_{p1}$ in the form as
\begin{eqnarray}
E_{p1}=t_{R_1 \rightarrow R_1}(t_1,t_2)=t_{R_1 \rightarrow R_1}(t_1,t_1+\Delta t) =\int_{x_a}^{\frac{x_a+x_b}{2}} dx_1 \int_{x_a}^{\frac{x_a+x_b}{2}} dx_2 \psi^{*}(x_1,t_1)\hat{H}(x_2,t_2)\psi(x_2,t_2)= \nonumber \\
=\int_{x_a}^{\frac{x_a+x_b}{2}} dx_1 \int_{x_a}^{\frac{x_a+x_b}{2}} dx_2 \psi^{*}(x_1,t_1)\hat{H}(x_2,t_2)\int_{-\infty}^{+\infty}\int_{-\infty}^{+\infty}dx_3dt_3G(x_2,t_2,x_3,t_3)\psi(x_3,t_3)= \nonumber \\
=\int_{x_a}^{\frac{x_a+x_b}{2}} dx_1 \int_{x_a}^{\frac{x_a+x_b}{2}} dx_2 \psi^{*}(x_1,t_1)\hat{H}(x_2,t_2)\int_{t_1}^{t_2}\int_{x_1}^{x_2}dx_3dt_3G_q(x_2,t_2,x_3,t_3)\psi(x_3,t_3)= \nonumber \\
\int_{x_a}^{\frac{x_a+x_b}{2}} dx_1 \int_{x_a}^{\frac{x_a+x_b}{2}} dx_2 \psi^{*}(x_1,t_1)\hat{H}(x_2,t_2)\psi(x_1+(x_2-x_1),t_1+(t_2-t_1))=\nonumber \\
=2\hbar E(t_2)e^{-\Delta t \frac{i}{\hbar}E(t_1)}
( \langle
 \frac{Sin((\frac{x_b+3x_a}{4}-x)\hat{p}(t_1))Sin((x+\frac{x_a+x_b}{4})\hat{p}(t_1))}{i\hat{p}(t_1)}
\rangle 
+i \langle
 \frac{Cos((\frac{x_b+3x_a}{4}-x)\hat{p}(t_1))Sin((+\frac{x_b-x_a}{4}-x)\hat{p}(t_1))}{i\hat{p}(t_1)}
\rangle ) \nonumber \\
=\hbar E(t_2)e^{-\Delta t \frac{i}{\hbar}E(t_1)} \Bigg[ e^{iArcTan(\frac{\frac{Cos((\frac{x_b+3x_a}{4}-x)\hat{p}(t_1))Sin((+\frac{x_b-x_a}{4}-x)\hat{p}(t_1))}{i\hat{p}(t_1)}}{\langle
 \frac{Sin((\frac{x_b+3x_a}{4}-x)\hat{p}(t_1))Sin((x+\frac{x_a=x_b}{4})\hat{p}(t_1))}{i\hat{p}(t_1)}|\langle [\frac{\hbar}{i\hat{p}(t_1)}(e^{(\frac{x_a+x_b}{2}-x) \frac{i}{\hbar}\hat{p}(t_1)}-e^{(-x+x_a) \frac{i}{\hbar}\hat{p}(t_1)})]
\rangle
\rangle})} \Bigg] \times \nonumber \\ \times |\langle [\frac{\hbar}{i\hat{p}(t_1)}(e^{(\frac{x_a+x_b}{2}-x) \frac{i}{\hbar}\hat{p}(t_1)}-e^{(-x+x_a) \frac{i}{\hbar}\hat{p}(t_1)})]
\rangle| \nonumber \\
\label{given}
\end{eqnarray}
It can be shown that the change of occupancy of energies can bring the complex value of $E_{p1}$ and $E_{p2}$ constants of tight-binding model.
In quite straightforward way such complex values of $E_{p1}$ or $E_{p2}$ can be interpreted as describing the decoherence effects taking place in Wannier qubit for the case of isolated qubit but being described by time-dependent Hamiltonian. It is quite easy to generalize the conducted considerations and computations for k coupled and electrostatically biased quantum dots with one electron distributed between them. In such case Wannier qudit (position-based qudit) would correspond to k=3 coupled quantum dots. In conducted considerations the spin degree of freedom present in electron description was omitted. Yet it can be easily incorporated by introduction of spin index to wavefunction.
\section{Two Wannier qubits interacting electrostatically in Schroedinger formalism and correlation functions}
Let us consider the Hamiltonian of the electrostatic interaction of 2 Wannier qubits given in the form given as
\begin{eqnarray}
\hat{H}=\hat{H}_{p1}+\hat{H}_{p2}+\hat{V}_{C,p1-p2}=\hat{T}_{p1}+\hat{V}_{p1}+\hat{T}_{p2}+\hat{V}_{p2}+\hat{V}_{C,p1-p2}=\nonumber \\
-\frac{\hbar^2}{2m_{p1}}\frac{d^2}{dx_1^2}+V_{p1}(x_1)-\frac{\hbar^2}{2m_{p2}}\frac{d^2}{dx_2^2}+V_{p1}(x_2)+\frac{q^2}{4\pi\epsilon_0\sqrt{|x_1-x_2|^2+d^2}}.
\end{eqnarray}
Consequently we are going to use

\begin{eqnarray}
\psi(x_{1,p1}+\Delta x_{1,p1},x_{1,p2}+\Delta x_{1,p2},t)=e^{(\Delta x_{1,p1}\frac{d}{dx_1}+\Delta x_{2,p2}\frac{d}{dx_2})}\psi(x_{1,p1},x_{1,p2},t) 
=e^{\frac{i}{\hbar}(\Delta x_{1,p1}\hat{p}_{1}+\Delta x_{2,p2}\hat{p}_2)}\psi(x_{1,p1},x_{1,p2},t), \nonumber \\
\psi(x_{1,p1}+\Delta x_{1,p1},x_{1,p2}+\Delta x_{1,p2},t+\Delta t)=e^{(\Delta x_{1,p1}\frac{d}{dx_1}+\Delta x_{2,p2}\frac{d}{dx_2}+\Delta t \frac{d}{dt})}\psi(x_{1,p1},x_{1,p2},t)=\nonumber \\
=e^{\frac{i}{\hbar}[(\Delta x_{1,p1}\hat{p}_{1}+\Delta x_{2,p2}\hat{p}_2)-\Delta t \hat{H}]}\psi(x_{1,p1},x_{1,p2},t), \nonumber \\
\end{eqnarray}
\small
\begin{eqnarray}
t_{s12,p1-p2,AntiCorr1}=t_{R_{1,p1} \rightarrow R_{2,p1},R_{2,p2} \rightarrow R_{1,p2}}(t_1,t_2) 
=t_{R_{1,p1} \rightarrow R_{2,p1},R_{2,p2} \rightarrow R_{1,p1}}(t_1,t_1+\Delta t)= \nonumber \\ =\int_{x_a}^{\frac{x_a+x_b}{2}} dx_{1,p1} \int_{\frac{x_a+x_b}{2}}^{x_b} dx_{2,p1}
\int_{\frac{x_a+x_b}{2}}^{x_b} dx_{1,p2} \int_{x_a}^{\frac{x_a+x_b}{2}} dx_{2,p2}\psi^{*}(x_{1,p1},x_{1,p2},t_1)\hat{H}(x_{2,p1},x_{2,p2},t_2)\psi(x_{2,p1},x_{2,p2},t_2)=  \nonumber \\
=\int_{x_a}^{\frac{x_a+x_b}{2}} dx_{1,p1} \int_{\frac{-x_a+x_b}{2}}^{\frac{x_a-x_b}{2}} d\Delta x_{2,p1}
\int_{\frac{x_a+x_b}{2}}^{x_b} dx_{1,p2} \int_{-\frac{x_a-x_b}{2}}^{\frac{x_a-x_b}{2}} d\Delta x_{2,p2}\psi^{*}(x_{1,p1},x_{1,p2},t_1)* \nonumber \\
*[\hat{H}_c(x_{2,p1},x_{2,p2})+\hat{H}_{p1}(x_{2,p1})+\hat{H}_{p2}(x_{2,p2})]e^{\frac{i}{\hbar}[(\Delta x_{1,p1}\hat{p}_{1}+\Delta x_{2,p2}\hat{p}_2)-\Delta t \hat{H}]}\psi(x_{1,p1},x_{1,p2},t_1)= \nonumber \\
=\int_{x_a}^{\frac{x_a+x_b}{2}} dx_{1,p1}
\int_{\frac{x_a+x_b}{2}}^{x_b} dx_{1,p2} \int_{\frac{-x_a+x_b}{2}}^{\frac{x_a-x_b}{2}} d\Delta x_{2,p1} \int_{-\frac{x_a-x_b}{2}}^{\frac{x_a-x_b}{2}} d\Delta x_{2,p2}\psi^{*}(x_{1,p1},x_{1,p2},t_1)* \nonumber \\
*[\hat{H}_c(x_{2,p1},x_{2,p2})+\hat{H}_{p1}(x_{1,p1})+\hat{H}_{p2}(x_{1,p2})]e^{\frac{i}{\hbar}[(\Delta x_{1,p1}\hat{p}_{1}+\Delta x_{2,p2}\hat{p}_2)-\Delta t \hat{H}]}\psi(x_{1,p1},x_{1,p2},t_1)=\nonumber \\
=\int_{x_a}^{\frac{x_a+x_b}{2}} dx_{1,p1}
\int_{\frac{x_a+x_b}{2}}^{x_b} dx_{1,p2} 
\psi(x_{1,p1},x_{1,p2},t_1)^{*}
[\hat{H}_c(x_{2,p1},x_{2,p2})+\hat{H}_{p1}(x_{1,p1})+\hat{H}_{p2}(x_{1,p2})] * \nonumber \\ \frac{1}{\frac{i}{\hbar}[(\Delta x_{1,p1}\hat{p}_{p1,x1})]}\frac{1}{\frac{i}{\hbar}[(\Delta x_{1,p2}\hat{p}_{p2,x1})]}* 
*e^{\frac{i}{\hbar}[(\Delta x_{1,p1}\hat{p}_{1}+\Delta x_{2,p2}\hat{p}_2)-\Delta t \hat{H}]}\psi(x_{1,p1},x_{1,p2},t_1)]|_{\Delta x_{2,p2}=\frac{-x_a+x_b}{2}}^{\Delta x_{2,p2}=\frac{x_a-x_b}{2}}|_{\Delta x_{2,p1}=-\frac{x_a-x_b}{2}}^{\Delta x_{2,p1}=\frac{x_a-x_b}{2}}.
\end{eqnarray}

Alternative measure is given by

\begin{eqnarray}
t_{Q,p1:R2,p2:R2 \rightarrow R1}(t_1,t_2)=t_{Q,R_{2,p1} \rightarrow R_{2,p1},R_{2,p2} \rightarrow R_{1,p2}}(t_1,t_2) 
=t_{Q,R_{2,p1} \rightarrow R_{2,p1},R_{2,p2} \rightarrow R_{1,p2}}(t_1,t_1+\Delta t)= \nonumber \\ =\int_{\frac{x_a+x_b}{2}}^{x_b} dx_{1,p1} \int_{\frac{x_a+x_b}{2}}^{x_b} dx_{2,p1}
\int_{\frac{x_a+x_b}{2}}^{x_b} dx_{1,p2} \int_{x_a}^{\frac{x_a+x_b}{2}} dx_{2,p2}|\psi(x_{1,p1},x_{1,p2},t_1)|^2 \psi(x_{2,p1},x_{2,p2},t_2)^{*}\hat{H}(x_{2,p1},x_{2,p2},t_2)\psi(x_{2,p1},x_{2,p2},t_2)  \nonumber \\
\end{eqnarray}
or by
\begin{eqnarray}
CE1_{Q,p1:R2,p2:R2 \rightarrow R1}(t_1,t_2)=t_{Q,R_{2,p1} \rightarrow R_{2,p1},R_{2,p2} \rightarrow R_{1,p2}}(t_1,t_2) 
=t_{Q,R_{2,p1} \rightarrow R_{2,p1},R_{2,p2} \rightarrow R_{1,p2}}(t_1,t_1+\Delta t)= \nonumber \\ =\int_{\frac{x_a+x_b}{2}}^{x_b} dx_{1,p1} \int_{\frac{x_a+x_b}{2}}^{x_b} dx_{2,p1}
\int_{\frac{x_a+x_b}{2}}^{x_b} dx_{1,p2} \int_{x_a}^{\frac{x_a+x_b}{2}} dx_{2,p2}|\psi(x_{1,p1},x_{1,p2},t_1)|^2 \psi(x_{2,p1},x_{2,p2},t_2)^{*}\hat{H}(x_{2,p1},x_{2,p2},t_2)\psi(x_{2,p1},x_{2,p2},t_2)  \nonumber \\
\end{eqnarray}
and by
\begin{eqnarray}
CE2_{Q,p1:R2,p2:R1 \rightarrow R1}(t_1,t_2)=t_{Q,R_{2,p1} \rightarrow R_{2,p1},R_{1,p2} \rightarrow R_{1,p2}}(t_1,t_2) 
=t_{Q,R_{2,p1} \rightarrow R_{2,p1},R_{1,p2} \rightarrow R_{1,p2}}(t_1,t_1+\Delta t)= \nonumber \\ =\int_{\frac{x_a+x_b}{2}}^{x_b} dx_{1,p1} \int_{\frac{x_a+x_b}{2}}^{x_b} dx_{2,p1}
\int_{\frac{x_a+x_b}{2}}^{x_b} dx_{1,p2} \int_{x_a}^{\frac{x_a+x_b}{2}} dx_{2,p2}|\psi(x_{1,p1},x_{1,p2},t_1)|^2 \psi(x_{2,p1},x_{2,p2},t_2)^{*}\hat{H}(x_{2,p1},x_{2,p2},t_2)\psi(x_{2,p1},x_{2,p2},t_2)  \nonumber \\ .
\end{eqnarray}
Other correlation functions are given by \ref{TwoPCorrelation}.
\section{Shift in time and space for single particle in non-zero vector potential in Schroedinger formalism}
\setcounter{equation}{0}
The single-electron in non-zero vector potential field can be described by equation
\begin{eqnarray}
\hat{H}_t=[\frac{1}{2m}(\frac{\hbar}{i}\frac{d}{dx}-\frac{e}{c}A_x(x,t))^2+\frac{e^2}{2cm}(A_y(x,t)^2+A_z(x,t)^2)+V(x,t)]=\nonumber \\
=-\frac{\hbar^2}{2m}\frac{d^2}{dx^2}-\frac{2\hbar}{i m}A_x(x,t)\frac{d}{dx}+(-\frac{\hbar}{i m}(\frac{d}{dx}A_x(x,t))+\frac{e^2}{2cm}(A_x(x,t)^2+A_y(x,t)^2+A_z(x,t)^2)+V(x,t))
\end{eqnarray}
In such way the we can write the definition of canonical momentum $\hat{p}_c$ (with presence of non-zero vector potential) and obtain the relation between $\psi(x,t)$ and $\psi(x_0,t)$ given as
\begin{eqnarray*}
\hat{p}_c= (\frac{\hbar}{i}\frac{d}{dx}-\frac{e}{c}A_x)=\hat{p}-\frac{e}{c}A_x, \frac{i}{\hbar}(\frac{e}{c}A_x+\hat{p}_c)=\frac{d}{dx}, \nonumber \\
(1+\frac{1}{1!}(x-x_0)^1\frac{d^1}{dx^1}+\frac{1}{2!}(x-x_0)^2\frac{d^2}{dx^2}+..)\psi(x_0,t)=e^{(x-x_0)\frac{i}{\hbar}(\frac{e}{c}A_x+\hat{p}_c)}\psi(x_0,t)=\psi(x,t).
\end{eqnarray*}
We can also obtain the relation between $\psi(x,t_0)$ and $\psi(x,t)$ in the form as
\begin{eqnarray}
e^{\frac{1}{i\hbar}(t-t_0)\hat{H}_{t_0}}\psi(x,t_0)=\nonumber \\
=e^{\frac{1}{i\hbar}(t-t_0)[-\frac{\hbar^2}{2m}\frac{d^2}{dx^2}-\frac{2\hbar}{i m}A_x(x,t_0)\frac{d}{dx}+(-\frac{\hbar}{i m}(\frac{d}{dx}A_x(x,t_0))+\frac{e^2}{2cm}(A_x(x,t_0)^2+A_y(x,t_0)^2+A_z(x,t_0)^2)+V(x,t_0))]}\psi(x,t_0).
\end{eqnarray}
\subsection{Schroedinger spinor and tight-binding model}
\setcounter{equation}{0}
We can account for existence of non-zero magnetic field.
In such case can define the Hamiltonian for Schroedinger particle with spin 1/2 in magnetic field in the form as
\tiny
\begin{eqnarray*}
\hat{H}_t= \nonumber \\
\begin{pmatrix}
[\frac{1}{2m}(\frac{\hbar}{i}\frac{d}{dx}-\frac{e}{c}A_x(x,t))^2+\frac{e^2}{2cm}(A_y(x,t)^2+A_z(x,t)^2)+V(x,t)]+\mu_0 B_z(t) & \mu_0(B_x-iB_y) \\
\mu_0(B_x+iB_y) & [\frac{1}{2m}(\frac{\hbar}{i}\frac{d}{dx}-\frac{e}{c}A_x(x,t))^2+\frac{e^2}{2cm}(A_y(x,t)^2+A_z(x,t)^2)+V(x,t)]-\mu_0 B_z(t)
\end{pmatrix}=\nonumber \\
\begin{pmatrix}
[\frac{1}{2m}(\frac{\hbar}{i}\frac{d}{dx}-\frac{e}{c}A_x(x,t))^2+V_{eff}]+\mu_0(\frac{d}{dx}A_y-\frac{d}{dy}A_x) & \mu_0((\frac{d}{dy}A_x-\frac{d}{dz}A_y)-i(\frac{d}{dx}A_z-\frac{d}{dz}A_x)) \\
\mu_0((\frac{d}{dy}A_x-\frac{d}{dz}A_y)+i(\frac{d}{dx}A_z-\frac{d}{dz}A_x)) & [\frac{1}{2m}(\frac{\hbar}{i}\frac{d}{dx}-\frac{e}{c}A_x(x,t))^2+V_{eff}]-\mu_0(\frac{d}{dx}A_y-\frac{d}{dy}A_x)
\end{pmatrix}, \nonumber \\
V_{eff}(x,t)=+\frac{e^2}{2cm}(A_y(x,t)^2+A_z(x,t)^2)+V(x,t).
\end{eqnarray*}
\normalsize
We can write Taylor expansion in time for spinor field as
\begin{eqnarray}
\begin{pmatrix}
\psi_{\uparrow}(x,t) \\
\psi_{\downarrow}(x,t) \\
\end{pmatrix}=
\Bigg[ \sum_{k=0}^{+\infty}\frac{1}{k!}
\begin{pmatrix}
(t-t_0)^k\frac{d^k}{dt_0^k} & 0 \\
0 & (t-t_0)^k\frac{d^k}{dt_0^k} \\
\end{pmatrix}
\Bigg]
\begin{pmatrix}
\psi_{\uparrow}(x,t_0) \\
\psi_{\downarrow}(x,t_0) \\
\end{pmatrix}= \nonumber \\
=
e^{\frac{1}{i\hbar}(t-t_0)\hat{H}_{t0}}
\begin{pmatrix}
\psi_{\uparrow}(x,t_0) \\
\psi_{\downarrow}(x,t_0) \\
\end{pmatrix}=(\sum_{k}\frac{(-i)^k}{(\hbar)^{k} k!}(t-t_0)^k\hat{H}_{t_0}^k)
\begin{pmatrix}
\psi_{\uparrow}(x,t_0) \\
\psi_{\downarrow}(x,t_0) \\
\end{pmatrix}.
\end{eqnarray}
Rewriting the Hamiltonian in detailed way we obtain
\tiny
\begin{eqnarray}
e^{\frac{1}{\hbar i}(t-t_0)\begin{pmatrix}
[\frac{1}{2m}(\frac{\hbar}{i}\frac{d}{dx}-\frac{e}{c}A_x(x,t))^2+V_{eff}]+\mu_0(\frac{d}{dx}A_y-\frac{d}{dy}A_x) & \mu_0((\frac{d}{dy}A_x-\frac{d}{dz}A_y)-i(\frac{d}{dx}A_z-\frac{d}{dz}A_x)) \\
\mu_0((\frac{d}{dy}A_x-\frac{d}{dz}A_y)+i(\frac{d}{dx}A_z-\frac{d}{dz}A_x)) & [\frac{1}{2m}(\frac{\hbar}{i}\frac{d}{dx}-\frac{e}{c}A_x(x,t))^2+V_{eff}]-\mu_0(\frac{d}{dx}A_y-\frac{d}{dy}A_x)
\end{pmatrix}_{t=t_0}}
\begin{pmatrix}
\psi_{\uparrow}(x,t_0) \\
\psi_{\downarrow}(x,t_0) \\
\end{pmatrix}=
\begin{pmatrix}
\psi_{\uparrow}(x,t) \\
\psi_{\downarrow}(x,t) \\
\end{pmatrix}
\end{eqnarray}
\normalsize
We can write Taylor expansion in position for spinor field as
\begin{eqnarray}
\begin{pmatrix}
\psi_{\uparrow}(x,t) \\
\psi_{\downarrow}(x,t) \\
\end{pmatrix}=
\Bigg[ \sum_{k=0}^{+\infty}\frac{1}{k!}
\begin{pmatrix}
(x-x_0)^k\frac{d^k}{dx_0^k} & 0 \\
0 & (x-x_0)^k\frac{d^k}{dx_0^k} \\
\end{pmatrix}
\Bigg]
\begin{pmatrix}
\psi_{\uparrow}(x_0,t) \\
\psi_{\downarrow}(x_0,t) \\
\end{pmatrix}=
\begin{pmatrix}
e^{(x-x_0)\frac{d}{dx_0}} & 0 \\
0 & e^{(x-x_0)\frac{d}{dx_0}} \\
\end{pmatrix}
\begin{pmatrix}
\psi_{\uparrow}(x_0,t) \\
\psi_{\downarrow}(x_0,t) \\
\end{pmatrix}
\nonumber \\
=e^{(x-x_0)\frac{d}{dx_0}}
\begin{pmatrix}
\psi_{\uparrow}(x_0,t) \\
\psi_{\downarrow}(x_0,t) \\
\end{pmatrix}
=
\begin{pmatrix}
e^{\frac{i}{\hbar}(x-x_0)(\hat{p}_{c,x_0}+\frac{e}{c}A_x)} & 0 \\
0 & e^{\frac{i}{\hbar}(x-x_0)(\hat{p}_{c,x_0}+\frac{e}{c}A_x)} \\
\end{pmatrix}
\begin{pmatrix}
\psi_{\uparrow}(x_0,t) \\
\psi_{\downarrow}(x_0,t) \\
\end{pmatrix}
\end{eqnarray}
Finally we obtain $\psi(x,t)$ in relation to $\psi(x_0,t_0)$ in the form as
  \tiny
\begin{eqnarray*}
e^{\frac{(t-t_0)}{\hbar i}\begin{pmatrix}
[\frac{1}{2m}(\frac{\hbar}{i}\frac{d}{dx}-\frac{e}{c}A_x(x,t_0))^2+V_{eff}]+\mu_0(\frac{d}{dx}A_y-\frac{d}{dy}A_x) & \mu_0((\frac{d}{dy}A_x-\frac{d}{dz}A_y)-i(\frac{d}{dx}A_z-\frac{d}{dz}A_x)) \\
\mu_0((\frac{d}{dy}A_x-\frac{d}{dz}A_y)+i(\frac{d}{dx}A_z-\frac{d}{dz}A_x)) & [\frac{1}{2m}(\frac{\hbar}{i}\frac{d}{dx}-\frac{e}{c}A_x(x,t_0))^2+V_{eff}]-\mu_0(\frac{d}{dx}A_y-\frac{d}{dy}A_x)
\end{pmatrix}+
(x-x_0)
\begin{pmatrix}
\frac{d}{dx_0} & 0 \\
0 & \frac{d}{dx_0} \\
\end{pmatrix}}
\begin{pmatrix}
\psi_{\uparrow}(x_0,t_0) \\
\psi_{\downarrow}(x_0,t_0) \\
\end{pmatrix}\nonumber \\
=
\begin{pmatrix}
\psi_{\uparrow}(x,t) \\
\psi_{\downarrow}(x,t) \\
\end{pmatrix}\nonumber \\
\end{eqnarray*}
\normalsize
\section{System of 2 electrostatically coupled charged Wannier qubits in tight-binding model} 
Later it will be shown that particularly interesting case is Berry phase that can be acquired by single-electron charge qubits.
This will be based on the analogies of tight-binding Hamiltonian of single qubit to spin 1/2 particle. Such analogies was firstly expressed by work \cite{SEL}.
We refer to the physical situation two symmetric qubits as by \cite{SEL} and utilize the correlation function C to capture as to what extent the two electrons are in a correlated state being both either on the left or on the right side that is corresponding to terms N(-,-), N(+,+), or in an anticorrelated state (expressed by terms N(+,-) and N(-,+). Such function is commonly used in spin systems and is a measure of non-classical correlations. Using a tight-binding model describing two electrostatically coupled
SELs (Single Electron Lines as it is the case of 2 parallel placed double quantum dots) and using the same correlation function applicable in the test of Bell theory of entangled spins, we obtain the correlation function C given by formula
\begin{eqnarray}
  C =\frac{N(+,+)+N(-,-)-N(-,+)-N(+,-)}{N(+,+)+N(-,-)+N(-,+)+N(+,-)}=\nonumber \\ =4\Big[ \frac{\sqrt{p_{E1}}\sqrt{p_{E2}}(t_{s1}-t_{s2})cos(-t\sqrt{(E_{c1}-E_{c2})^2+4(t_{s1}-t_{s2})^2}+\phi_{E10}-\phi_{E20})}{\sqrt{(E_{c1}-E_{c2})^2+4(t_{s1}-t_{s2})^2}}   \nonumber \\
  +\frac{\sqrt{p_{E3}}\sqrt{p_{E4}}(t_{s1}+t_{s2})cos(-t\sqrt{(E_{c1}-E_{c2})^2+4(t_{s1}+t_{s2})^2}+\phi_{E30}-\phi_{E40})}{\sqrt{(E_{c1}-E_{c2})^2+4(t_{s1}+t_{s2})^2}} \Big]
   \nonumber \\
  -(E_{c1}-E_{c2})\Big[ \frac{p_{E1}-p_{E2}}{\sqrt{(E_{c1}-E_{c2})^2+4(t_{s1}-t_{s2})^2}}+\frac{p_{E3}-p_{E4}}{\sqrt{(E_{c1}-E_{c2})^2+4(t_{s1}+t_{s2})^2}} \Big],
\end{eqnarray}
where
\begin{eqnarray}
|\psi>=\sqrt{p_{E1}}e^{i \phi_{E_{10}}}e^{\frac{E_1t}{\hbar i}}|E_1>+\sqrt{p_{E2}}e^{i \phi_{E_{20}}}e^{\frac{E_2t}{\hbar i}}|E_2>+\sqrt{p_{E3}}e^{i \phi_{E_{30}}}e^{\frac{E_3t}{\hbar i}}|E_3>+\sqrt{p_{E4}}e^{i \phi_{E_{40}}}e^{\frac{E_4t}{\hbar i}}|E_4>.
\end{eqnarray}
 and the Hamiltonian of the system is given as

\begin{eqnarray}
\hat{H}=
 \begin{pmatrix}
  2E_p+E_{c1} & t_{s2}e^{i\beta} & t_{s1}e^{i\alpha} & 0 \\
  t_{s2}e^{-i\beta} & 2E_p+E_{c2} & 0 & t_{s1}e^{i\alpha} \\
  t_{s1}e^{-i\alpha} & 0 & 2E_p+E_{c2} & t_{s2}e^{+i\beta}  \\
  0 & t_{s1}e^{-i\alpha} & t_{s2}e^{-i\beta} & 2E_p+E_{c1} \\
 \end{pmatrix}
\end{eqnarray}

 The Hamiltonian obtained for 2 coupling 1/2 spin particle is analogical to the Hamiltonian for 2 electrostatically interacting Wannier qubits (each having 2 coupled quantum dots) that is expressed as
\begin{eqnarray}
\hat{H}=
 \begin{pmatrix}
E_{pA1}+E_{pB1}+E_{c{(1A-1B)}}& t_{1A \rightarrow 1A,2B \rightarrow 1B}   &  t_{2A \rightarrow 1A}         &  t_{2A \rightarrow 1A,2B \rightarrow 1B} \\
t_{1A \rightarrow 1A,1B \rightarrow 2B}   & E_{pA1}+E_{pB2}+E_{c(1A-2B)}  &  t_{2A \rightarrow 1A, 1B \rightarrow 2B} &  t_{2A \rightarrow 1A}        \\
 t_{1A \rightarrow 2A}          &  t_{1A \rightarrow 2A, 2B \rightarrow 1B}  &  E_{pA2}+E_{pB1}+E_{c(2A-1B)} & t_{2B \rightarrow 1B} \\
t_{1A \rightarrow 2A,1B \rightarrow 2B}   &  t_{1A \rightarrow 2A}          & t_{1B \rightarrow 2B} &  E_{pA2}+E_{pB2}+E_{c(2A-2B)}  \\
\end{pmatrix}. \nonumber \\
\end{eqnarray}

It shall be underlined that process when 2 particles A and B are moving in synchronous way is less probable than probability when 2 particles are moving in opposite directions since first process is generating higher magnetic field in outside environment. Those processes can be properly expressed by tight-binding model validation procedure.

Rigorous treatment of correlation function shall be expressed by 2-body wavefunction $\psi(x_1,x_2,t)$
\begin{eqnarray}
  C(t) =\frac{N(+,+)+N(-,-)-N(-,+)-N(+,-)}{N(+,+)+N(-,-)+N(-,+)+N(+,-)}=\nonumber \\ =
 [  \int_{x_{p1} \in R_2}dx_{p1}\int_{x_{p2} \in R_2}dx_{p2}|\psi(x_{p1},x_{p2},t)|^2+ 
  \int_{x_{p1} \in R_1}dx_{p1}\int_{x_{p2} \in R_1}dx_{p2}|\psi(x_{p1},x_{p2},t)|^2 \nonumber \\
  -\int_{x_{p1} \in R_1}dx_{p1}\int_{x_{p2} \in R_2}dx_{p2}|\psi(x_{p1},x_{p2},t)|^2 
  -\int_{x_{p1} \in R_2}dx_{p1}\int_{x_{p2} \in R_1}dx_{p2}|\psi(x_{p1},x_{p2},t)|^2 ]. 
\end{eqnarray}

	\section{Two electrostatically interacting straight single-electron nanowires (Wannier qubits) in Schroedinger description}
We consider 2 body Schroedinger equation with omission spin degrees of freedom. Two bodies are placed at different semiconductor nanowires and in each nanowire the single
electron is injected. In such case we are dealing with the $\hat{T}$ kinetic energy terms, with potential terms corresponding to each nanowire denoted by $V_1$ and $V_1$ as well
as Coulomb interacting term $V_C$. We divide each of nanowires into $N_1$ and $N_2$ slices and set $N=N_1=N_2$, so $\Delta x_1=\frac{max(x_1)-min(x_1)}{N_1}$ and $\Delta x_2=\frac{max(x_2)-min(x_2)}{N_2}$. We obtain 	
\begin{eqnarray}
\hat{T}=
-\frac{\hbar^2}{2m}\frac{1}{(\Delta x_1)^2}
\begin{pmatrix}
-2 & 1 & 0  & 0  & 0  & 0   \\
1 & -2 & 1  & 0  & 0  & 0    \\
0 & 1 & -2 & 1  & 0  & 0    \\
0 & 0 &  1 & -2 & 1  & 0    \\
0 & 0 &  0 &  1 & -2 & 1    \\
0 & 0 &  0 &  0 & 1  & -2   \\
\end{pmatrix} \times
\begin{pmatrix}
1 & 0 & 0  &  0  & 0  & 0   \\
0 & 1 & 0  &  0  & 0  & 0    \\
0 & 0 & 1  &  0  & 0  & 0    \\
0 & 0 & 0  &  1  & 0  & 0    \\
0 & 0 & 0  &  0  & 1  & 0    \\
0 & 0 & 0  &  0  & 0  & 1   \\
\end{pmatrix}+ \nonumber \\
-\frac{\hbar^2}{2m}\frac{1}{(\Delta x_2)^2}
\begin{pmatrix}
1 & 0 & 0  &  0  & 0  & 0   \\
0 & 1 & 0  &  0  & 0  & 0    \\
0 & 0 & 1  &  0  & 0  & 0    \\
0 & 0 & 0  &  1  & 0  & 0    \\
0 & 0 & 0  &  0  & 1  & 0    \\
0 & 0 & 0  &  0  & 0  & 1   \\
\end{pmatrix} \times
\begin{pmatrix}
-2 & 1 & 0  & 0  & 0  & 0   \\
1 & -2 & 1  & 0  & 0  & 0    \\
0 & 1 & -2 & 1  & 0  & 0    \\
0 & 0 &  1 & -2 & 1  & 0    \\
0 & 0 &  0 &  1 & -2 & 1    \\
0 & 0 &  0 &  0 & 1  & -2   \\
\end{pmatrix}
\end{eqnarray}

\begin{eqnarray}
\hat{V}=
\begin{pmatrix}
V_1(1) & 0 & 0  & 0  & 0  & 0   \\
0 & V_1(2) & 0  & 0  & 0  & 0    \\
0 & 0 & V_1(3) & 0  & 0  & 0    \\
0 & 0 &  0 & V_1(4) & 0  & 0    \\
0 & 0 &  0 &  0 & V_1(5) & 1    \\
0 & 0 &  0 &  0 & 0  & V_1(6)   \\
\end{pmatrix} \times
\begin{pmatrix}
1 & 0 & 0  &  0  & 0  & 0   \\
0 & 1 & 0  &  0  & 0  & 0    \\
0 & 0 & 1  &  0  & 0  & 0    \\
0 & 0 & 0  &  1  & 0  & 0    \\
0 & 0 & 0  &  0  & 1  & 0    \\
0 & 0 & 0  &  0  & 0  & 1   \\
\end{pmatrix}+ \nonumber \\
+
\begin{pmatrix}
1 & 0 & 0  &  0  & 0  & 0   \\
0 & 1 & 0  &  0  & 0  & 0    \\
0 & 0 & 1  &  0  & 0  & 0    \\
0 & 0 & 0  &  1  & 0  & 0    \\
0 & 0 & 0  &  0  & 1  & 0    \\
0 & 0 & 0  &  0  & 0  & 1   \\
\end{pmatrix} \times
\begin{pmatrix}
V_2(1) & 0 & 0  & 0  & 0  & 0   \\
0 & V_2(2) & 0  & 0  & 0  & 0    \\
0 & 0 & V_2(3) & 0  & 0  & 0    \\
0 & 0 &  0 & V_2(4) & 0  & 0    \\
0 & 0 &  0 &  0 & V_2(5) & 0    \\
0 & 0 &  0 &  0 & 0  & V_2(6)   \\
\end{pmatrix}
\end{eqnarray}

\begin{eqnarray}
\hat{V}_{1-2}=
\begin{pmatrix}
V_C(1,1) & 0  & ..  & 0 & 0 & .. & 0 & .. & 0 \\ 
0 & V_C(1,2)  & ..  & 0 & 0 & .. & 0 & .. & 0 \\ 
0 & 0  & ..  & 0 & 0 & .. & 0 & .. & 0\\
0 & 0  & ..  & V_C(1,N)& 0 & .. & 0 & .. & 0\\
0 & 0  & ..  & 0 & V_C(2,1) & .. & 0 & .. & 0\\
0 & 0  & ..  & 0 & 0 & .. & 0 & .. & 0 \\
0 & 0  & ..  & 0 & 0 & .. & V_C(2,N) & .. & 0 \\
0 & 0  & ..  & 0 & 0 & .. & 0 & .. & 0 \\
0 & 0  & ..  & 0 & 0 & .. & 0 & .. & V_C(N,N) \\
\end{pmatrix}
\end{eqnarray}
Here we have
\begin{eqnarray}
V(i,j)=\frac{q^2}{ d(i,j) }.
\end{eqnarray}
For 2 interacting bodies and sampling the potential and wavefunction of each object to N pieces we have Hamiltonian matrix of dimensions $N^2$ by $N^2$.
Therefore we obtain the Hamiltonian that is N by N matrix given in the form as
$\hat{H}=\hat{T}+\hat{V}+\hat{V}_{1-2}$.
We have density matrix of the system defined as
\begin{landscape}
\tiny
\begin{eqnarray}
\hat{\rho}_{1-2, N^2 \times N^2}=
\begin{pmatrix}
\psi_1(1)^{*}\psi_1(1) & \psi_1(1)^{*}\psi_1(2) & \psi_1(1)^{*}\psi_1(3) \nonumber \\
\psi_1(2)^{*}\psi_1(1) & \psi_1(2)^{*}\psi_1(2) & \psi_1(2)^{*}\psi_1(3) \nonumber \\
\psi_1(3)^{*}\psi_1(1) & \psi_1(3)^{*}\psi_1(2) & \psi_1(3)^{*}\psi_1(3) \nonumber \\
\end{pmatrix}_{N \times N} \times
\begin{pmatrix}
\psi_2(1)^{*}\psi_2(1) & \psi_2(1)^{*}\psi_2(2) & \psi_2(1)^{*}\psi_2(3) \nonumber \\
\psi_2(2)^{*}\psi_2(1) & \psi_2(2)^{*}\psi_2(2) & \psi_2(2)^{*}\psi_2(3) \nonumber \\
\psi_2(3)^{*}\psi_2(1) & \psi_2(3)^{*}\psi_2(2) & \psi_2(3)^{*}\psi_2(3) \nonumber \\
\end{pmatrix}_{N \times N} = \nonumber \\
=
\begin{pmatrix}
\psi_1(1)^{*}\psi_1(1)\begin{pmatrix}
\psi_2(1)^{*}\psi_2(1) & \psi_2(1)^{*}\psi_2(2) & \psi_2(1)^{*}\psi_2(3) \nonumber \\
\psi_2(2)^{*}\psi_2(1) & \psi_2(2)^{*}\psi_2(2) & \psi_2(2)^{*}\psi_2(3) \nonumber \\
\psi_2(3)^{*}\psi_2(1) & \psi_2(3)^{*}\psi_2(2) & \psi_2(3)^{*}\psi_2(3) \nonumber \\
\end{pmatrix}_{N \times N}  & \psi_1(1)^{*}\psi_1(2)\begin{pmatrix}
\psi_2(1)^{*}\psi_2(1) & \psi_2(1)^{*}\psi_2(2) & \psi_2(1)^{*}\psi_2(3) \nonumber \\
\psi_2(2)^{*}\psi_2(1) & \psi_2(2)^{*}\psi_2(2) & \psi_2(2)^{*}\psi_2(3) \nonumber \\
\psi_2(3)^{*}\psi_2(1) & \psi_2(3)^{*}\psi_2(2) & \psi_2(3)^{*}\psi_2(3) \nonumber \\
\end{pmatrix}_{N \times N} & \psi_1(1)^{*}\psi_1(3)\begin{pmatrix}
\psi_2(1)^{*}\psi_2(1) & \psi_2(1)^{*}\psi_2(2) & \psi_2(1)^{*}\psi_2(3) \nonumber \\
\psi_2(2)^{*}\psi_2(1) & \psi_2(2)^{*}\psi_2(2) & \psi_2(2)^{*}\psi_2(3) \nonumber \\
\psi_2(3)^{*}\psi_2(1) & \psi_2(3)^{*}\psi_2(2) & \psi_2(3)^{*}\psi_2(3) \nonumber \\
\end{pmatrix}_{N \times N} \nonumber \\
\psi_1(2)^{*}\psi_1(1)\begin{pmatrix}
\psi_2(1)^{*}\psi_2(1) & \psi_2(1)^{*}\psi_2(2) & \psi_2(1)^{*}\psi_2(3) \nonumber \\
\psi_2(2)^{*}\psi_2(1) & \psi_2(2)^{*}\psi_2(2) & \psi_2(2)^{*}\psi_2(3) \nonumber \\
\psi_2(3)^{*}\psi_2(1) & \psi_2(3)^{*}\psi_2(2) & \psi_2(3)^{*}\psi_2(3) \nonumber \\
\end{pmatrix}_{N \times N}  & \psi_1(2)^{*}\psi_1(2)\begin{pmatrix}
\psi_2(1)^{*}\psi_2(1) & \psi_2(1)^{*}\psi_2(2) & \psi_2(1)^{*}\psi_2(3) \nonumber \\
\psi_2(2)^{*}\psi_2(1) & \psi_2(2)^{*}\psi_2(2) & \psi_2(2)^{*}\psi_2(3) \nonumber \\
\psi_2(3)^{*}\psi_2(1) & \psi_2(3)^{*}\psi_2(2) & \psi_2(3)^{*}\psi_2(3) \nonumber \\
\end{pmatrix}_{N \times N} & \psi_1(2)^{*}\psi_1(3)\begin{pmatrix}
\psi_2(1)^{*}\psi_2(1) & \psi_2(1)^{*}\psi_2(2) & \psi_2(1)^{*}\psi_2(3) \nonumber \\
\psi_2(2)^{*}\psi_2(1) & \psi_2(2)^{*}\psi_2(2) & \psi_2(2)^{*}\psi_2(3) \nonumber \\
\psi_2(3)^{*}\psi_2(1) & \psi_2(3)^{*}\psi_2(2) & \psi_2(3)^{*}\psi_2(3) \nonumber \\
\end{pmatrix}_{N \times N} \nonumber \\
\psi_1(3)^{*}\psi_1(1)\begin{pmatrix}
\psi_2(1)^{*}\psi_2(1) & \psi_2(1)^{*}\psi_2(2) & \psi_2(1)^{*}\psi_2(3) \nonumber \\
\psi_2(2)^{*}\psi_2(1) & \psi_2(2)^{*}\psi_2(2) & \psi_2(2)^{*}\psi_2(3) \nonumber \\
\psi_2(3)^{*}\psi_2(1) & \psi_2(3)^{*}\psi_2(2) & \psi_2(3)^{*}\psi_2(3) \nonumber \\
\end{pmatrix}_{N \times N}  & \psi_1(3)^{*}\psi_1(2)\begin{pmatrix}
\psi_2(1)^{*}\psi_2(1) & \psi_2(1)^{*}\psi_2(2) & \psi_2(1)^{*}\psi_2(3) \nonumber \\
\psi_2(2)^{*}\psi_2(1) & \psi_2(2)^{*}\psi_2(2) & \psi_2(2)^{*}\psi_2(3) \nonumber \\
\psi_2(3)^{*}\psi_2(1) & \psi_2(3)^{*}\psi_2(2) & \psi_2(3)^{*}\psi_2(3) \nonumber \\
\end{pmatrix}_{N \times N}  & \psi_1(3)^{*}\psi_1(3)\begin{pmatrix}
\psi_2(1)^{*}\psi_2(1) & \psi_2(1)^{*}\psi_2(2) & \psi_2(1)^{*}\psi_2(3) \nonumber \\
\psi_2(2)^{*}\psi_2(1) & \psi_2(2)^{*}\psi_2(2) & \psi_2(2)^{*}\psi_2(3) \nonumber \\
\psi_2(3)^{*}\psi_2(1) & \psi_2(3)^{*}\psi_2(2) & \psi_2(3)^{*}\psi_2(3) \nonumber \\
\end{pmatrix}_{N \times N} \nonumber \\
\end{pmatrix}_{N^2 \times N^2}
\end{eqnarray}
\normalsize
We recognize that
\begin{eqnarray}
\hat{\rho}_{1, N \times N}=
\hat{\rho}_{1-2, N^2 \times N^2}[(1,1),(N,N)]+\hat{\rho}_{1-2, N^2 \times N^2}[(N+1,N+1),(2N,2N)]+..+\hat{\rho}_{1-2, N^2 \times N^2}[(N^2-N+1,N^2-N+1),(N^2,N^2)].
\end{eqnarray}
and that
\small
\begin{eqnarray}
\hat{\rho}_{1, N \times N}=
\begin{pmatrix}
Tr(\hat{\rho}_{1-2, N^2 \times N^2}[(1,1),(N,N)]) & Tr(\hat{\rho}_{1-2, N^2 \times N^2}[(N+1,1),(2N,N)]) & .. & Tr(\hat{\rho}_{1-2, N^2 \times N^2}[(N^2-N+1,1),(N^2,N)]) \\ 
Tr(\hat{\rho}_{1-2, N^2 \times N^2}[(1,N+1),(N,2N)]) & Tr(\hat{\rho}_{1-2, N^2 \times N^2}[(N+1,N+1),(2N,2N)]) & .. & .. \\
.. & .. & .. & .. \\
Tr(\hat{\rho}_{1-2, N^2 \times N^2}[(1,N^2-N+1),(N,N^2)]) & Tr(\hat{\rho}_{1-2, N^2 \times N^2}[(N+1,N^2-N+1),(2N,N^2)]) & .. & Tr(\hat{\rho}_{1-2, N^2 \times N^2}[(N^2-N+1,N^2-N+1),(N^2,N^2)]) \\
\end{pmatrix}_{N \times N}.
\end{eqnarray}
\end{landscape}

\normalsize

\section{Open curvy loops confining single electron in Cartesian coordinates in Schroedinger formalism}

\begin{figure}
    \centering
    \includegraphics[scale=0.8]{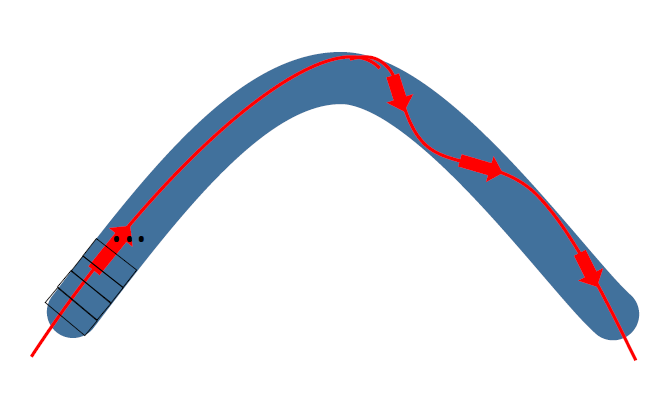}
    \caption{Schematic movement of wave-packet across nanowire that can be simplified as quasi-one dimensional object after proper transformation from 3D or 2D to 1D (Dimension) in visualization by Marcin Piontek. }
    \label{fiber}
    \end{figure}

We consider the set of open curvy quasi-one dimensional loops (that can be straight or curved smooth semiconductor nanowires with single electron) described by $x(s)$,$y(s)$ and $z(s)$, where s is the distance from beginning to the end of loop. We have
\begin{eqnarray} \label{eqnondiss1}
[-\frac{\hbar^2}{2m}(\frac{d^2}{dx^2}+\frac{d^2}{dy^2}+\frac{d^2}{dz^2})+V(x,y,z)] \psi(x,y,z)=E \psi(x,y,z)
\end{eqnarray}
In such case the wavepacket moves as in depicted Fig.\ref{fiber}.
We notice that $\frac{d}{dx}=\frac{ds}{dx}\frac{d}{ds}$ and similarly with y and z.
We have $\frac{d^2}{dx^2}=\frac{ds}{dx}\frac{d}{ds}(\frac{ds}{dx}\frac{d}{ds})=(\frac{1}{x'(s)})^2\frac{d^2}{ds^2}-[\frac{x''(s)}{(x'(s))^3}]\frac{d}{ds}$.
We thus end up in equation of structure as
\begin{eqnarray}
\Bigg[-\frac{\hbar^2}{2m}\Bigg[([(\frac{1}{x'(s)})^2+(\frac{1}{y'(s)})^2+(\frac{1}{z'(s)})^2]\frac{d^2}{ds^2}-[\frac{x''(s)}{(x'(s))^3}+\frac{y''(s)}{(y'(s))^3}+\frac{z''(s)}{(z'(s))^3}]\frac{d}{ds})\Bigg]+V(s)\Bigg] \psi(s)=E \psi(s)
\end{eqnarray}
and it can be summarized as
\begin{eqnarray}
\Bigg[-\frac{\hbar^2}{2m}\Bigg[([(\frac{1}{x'(s)})^2+(\frac{1}{y'(s)})^2+(\frac{1}{z'(s)})^2]\frac{d^2}{ds^2}-[\frac{x''(s)}{(x'(s))^3}+\frac{y''(s)}{(y'(s))^3}+\frac{z''(s)}{(z'(s))^3}]\frac{d}{ds})\Bigg]+V(s)\Bigg] \psi(s)=E \psi(s)
\end{eqnarray}
so we have
\begin{eqnarray} \label{eqndiss1}
\Bigg[-\frac{\hbar^2}{2m}\Bigg[(f(s)\frac{d^2}{ds^2}-g(s)\frac{d}{ds})\Bigg]+V(s)\Bigg] \psi(s)=E \psi(s),
\end{eqnarray}
where $f(s)=[(\frac{1}{x'(s)})^2+(\frac{1}{y'(s)})^2+(\frac{1}{z'(s)})^2]$, $g(s)=[\frac{x''(s)}{(x'(s))^3}+\frac{y''(s)}{(y'(s))^3}+\frac{z''(s)}{(z'(s))^3}]$.
The most prominent feature that can be observed from equation transformation from \ref{eqnondiss1} to \ref{eqndiss1} is the occurrence of dissipation term that is proportional to operator $\frac{d}{ds}$ as analogical to friction force (that is usually proportional to particle momentum). Indeed wavepacket travelling in semiconductor nanowire is being bent that corresponds to occurrence of force changing the
direction of wavepacket momentum. However from another perspective we can say that bending straight trajectory of particle (flat space) we generate dissipation. Thinking reversely we can say that fact of having certain type of dissipation in given system in given coordinates we can change by moving to space with another curvature so dissipation
is reduced or cancelled. The consequence of having dissipation in quantum system (or more precisely dissipation-like) will imply the fact of nonhermicity of Hamiltonian matrix that shall imply the existence of complex value of eigenergies. What is interesting we can observe dissipation like term in the system of classical description of electron moving in nanowire what is given by equation \ref{cleqnodiss1} and \ref{cleqdiss1}.

Using Cartesian coordinates we can formulate the following Schroedinger equation of motion
\begin{eqnarray}
-\frac{\hbar^2}{2m}[(1+\frac{1}{(\frac{d}{dx}y(x)^2)})\frac{d^2}{dx^2}-\frac{\frac{d^2}{dx^2}y(x)}{(\frac{d}{dx}y(x))^2}\frac{d}{dx}]\psi(x,y(x))+V(x,y(x))\psi(x,y(x))=E \psi(x,y(x)).
\end{eqnarray}
Local confining potential is given by $V(x,y(x))$ and can simply take into account the existence of 1, 2, 3 and more quantum dots across semiconductor nanowire or can
omit the existence of quantum dots and external polarizing electric and magnetic fields by being constant.
Effectively we have obtained modified quasi-one dimensional Schroedinger equation of the form
\begin{eqnarray}
-\frac{\hbar^2}{2m}[(1+\frac{1}{(\frac{d}{dx}y(x)^2)})\frac{d^2}{dx^2}-\frac{\frac{d^2}{dx^2}y(x)}{(\frac{d}{dx}y(x))^2}\frac{d}{dx}]\psi(x)+V(x)\psi(x)=E \psi(x).
\end{eqnarray}
Here the shape of open-loop nanowire is encoded in $y(x)$ function dependence (what is reflected in functions measuring cable curvature as by $(\frac{d}{dx}y(x)$ and by $(\frac{d^2}{dx^2}y(x)$). The last \textbf{CM-Schroedinger} equation (\textbf{C}urvature \textbf{M}odified Schroedinger equation) can easily be generalized to open-loop nanowire in 3 dimensions that brings quasi-one dimensional CM-Schroedinger equation as well. We have obtained the following results for Tanh square nanocable as given by Fig.17.
Furthermore two parallel lines in $><$ configuration with single-electron distributed at each line are expressed by 2 body Schroedinger modified equations in the form as
\begin{eqnarray}
-\frac{\hbar^2}{2m_A}[(1+\frac{1}{(\frac{d}{dx_A}y_A(x_A)^2)})\frac{d^2}{dx_A^2}-\frac{\frac{d^2}{dx_A^2}y_A(x_A)}{(\frac{d}{dx_A}y(x_A))^2}\frac{d}{dx_A}]\psi(x_A,y_A) \nonumber \\
-\frac{\hbar^2}{2m_B}[(1+\frac{1}{(\frac{d}{dx_B}y_B(x_B)^2)})\frac{d^2}{dx_B^2}-\frac{\frac{d^2}{dx_B^2}y_B(x_B)}{(\frac{d}{dx_B}y(x_B))^2}\frac{d}{dx_B}]\psi(x_A,y_A)+ \nonumber \\
+[V_A(x_A)+V_B(x_B)+V_{A-B}(x_A,y_A,x_B,y_B)]\psi(x_A,y_A)=E \psi(x_A,y_A), \nonumber \\
\end{eqnarray}
where $V_A$ and $V_B$ are local confining potentials for electron A and B, while Coulomb interaction between electrons $V_{A-B}(x_A,y_A,x_B,y_B)=\frac{q^2}{d((x_A,y_A),(x_B,y_B))}$.
This is the case of 2 body interaction that is considered with omission spin degrees of freedom. If each of nanowires (bent or straight) is divided into m pieces we deal with Hamiltonian matrix of size $M^{p}$ by $M^{p}$ that has $M^{p=2}$ energy eigenfunctions and eigenvalues. Here we set number of particles p to 2 (for q-swap gate p=2 and p=3 for CNOT gate). Those particles are interacting and represented by number of electrons placed at different open loop semiconductor nanowires. We set M=7 for two symmetric around OX axes V lines and assumed $\alpha_A=\alpha_B$.
We have obtained the following results as depicted in Fig.\ref{supsup}. We can trace electron anticorrelation under different circumstances. Yet anticorrelation and correlation is occurring partly for the case of electrostatically interacting electrons placed at different nanocables as it was indicated already in \cite{SEL}. For the purpose of computations in this work all eigenenergies were set with equal probability of occupancy. Quite clearly the presented results
go beyond tight-binding model expressed by \cite{SEL} and \cite{Cryogenics}.
\begin{figure}
\centering
\includegraphics[scale=0.3]{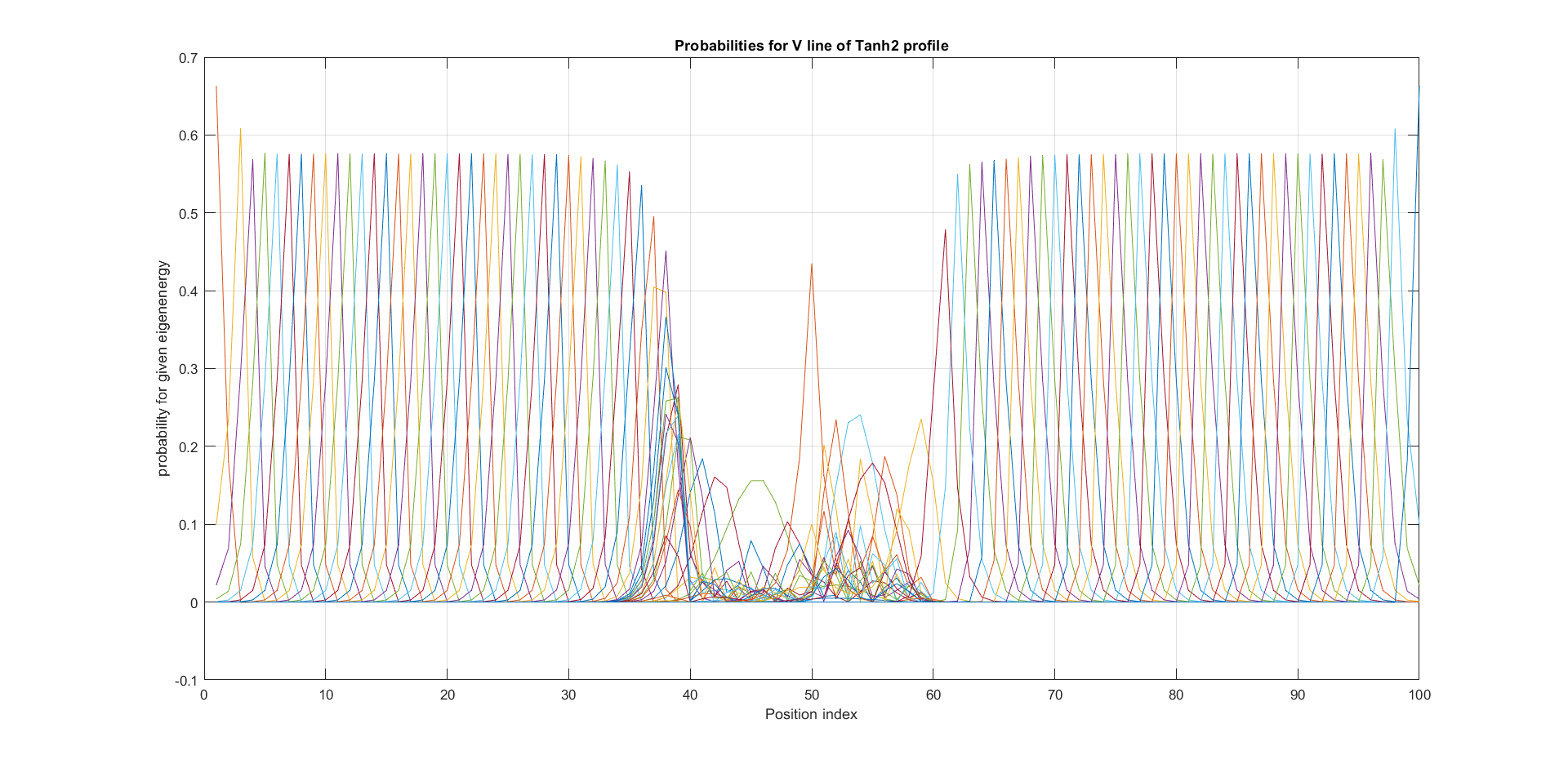} 
\includegraphics[scale=0.3]{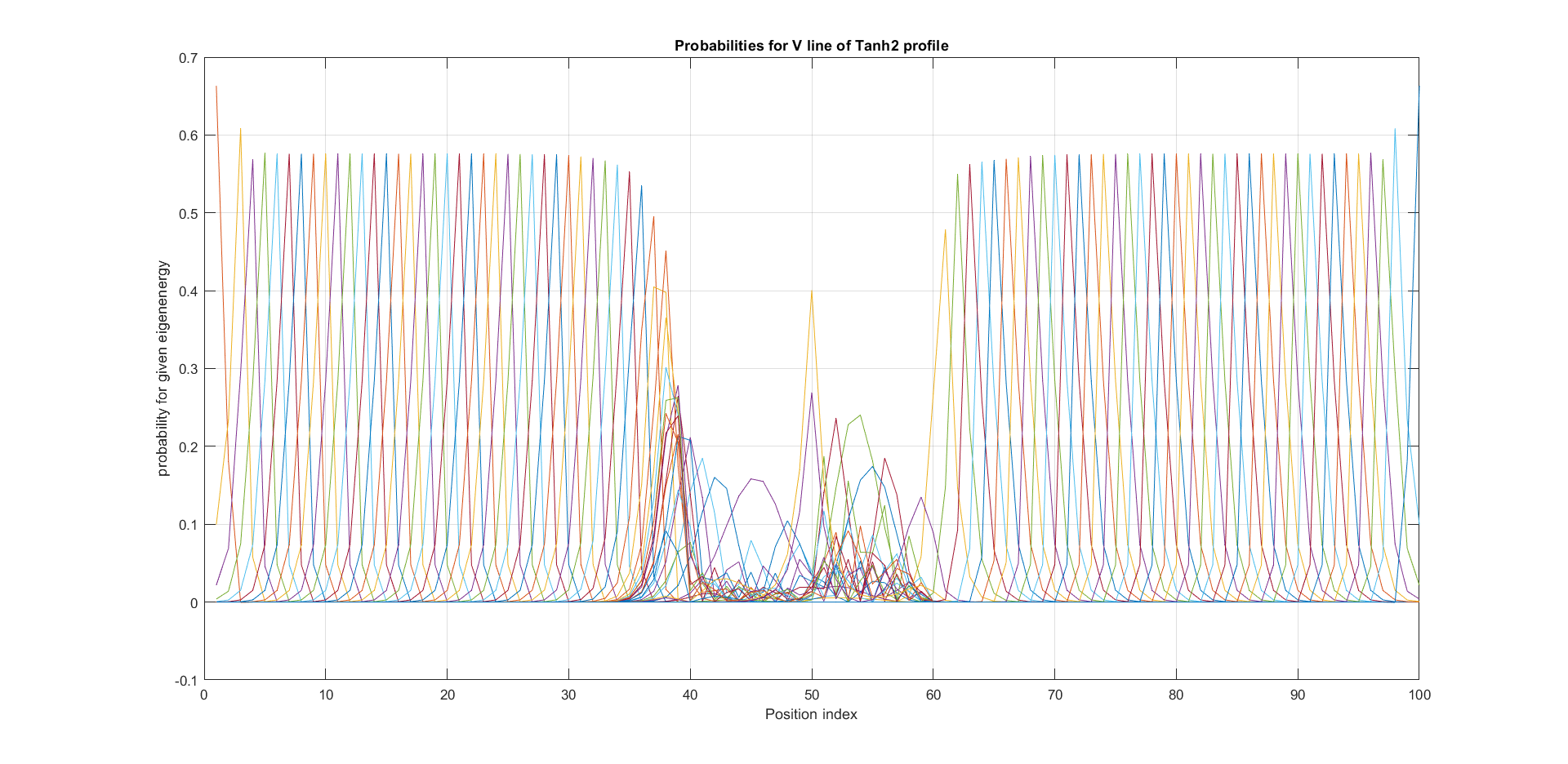} 
\includegraphics[scale=0.3]{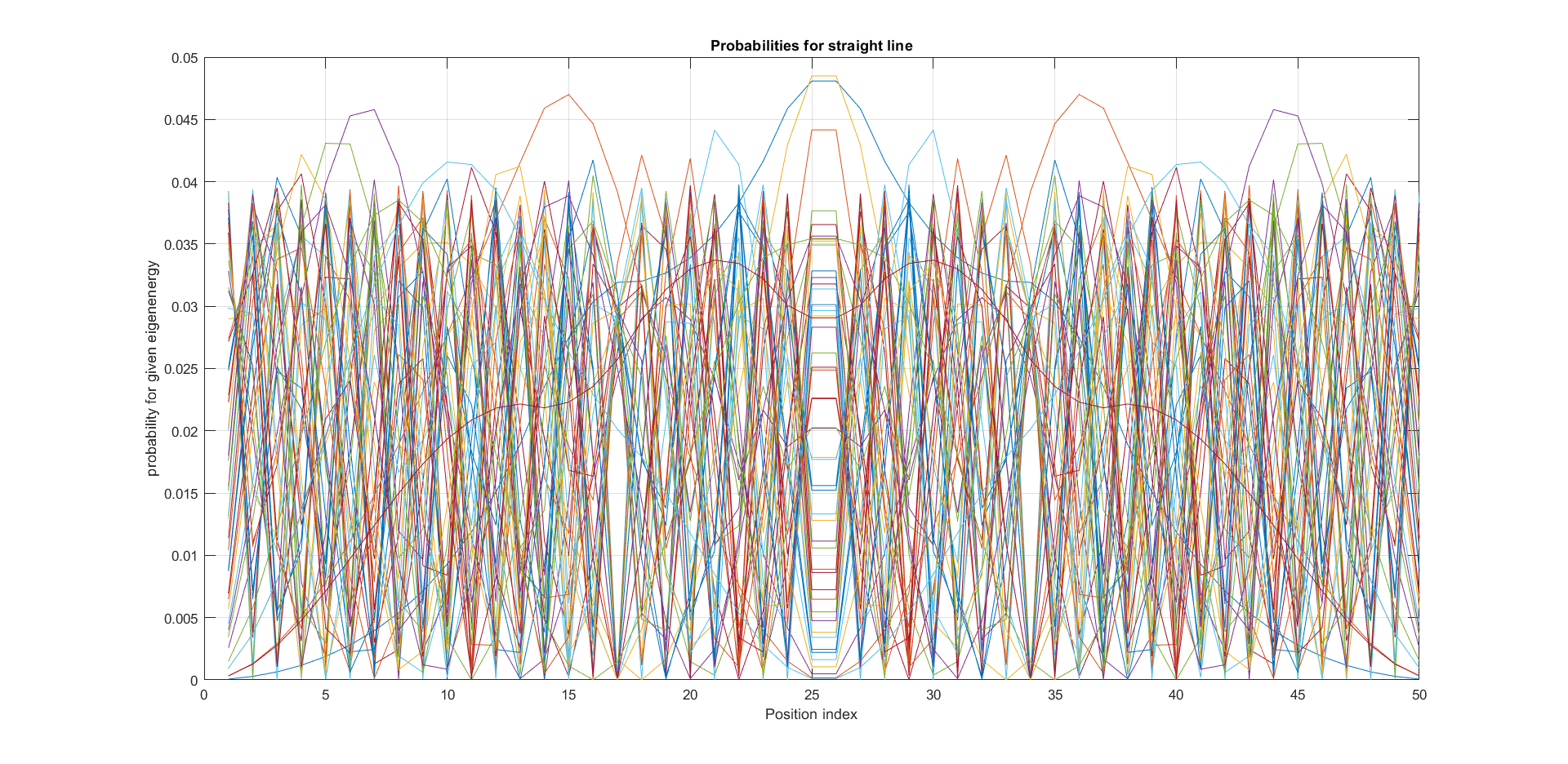}
\caption{Probability distributions corresponding to eigenenergy wavefunctions for tanh square nanowire ($\alpha=10$) with 3 built-in q-wells (UPPER), no q-wells (MIDDLE) and straight nanowire with 2 built-in q-wells (LOWER). }
\end{figure}
\begin{figure}
\includegraphics[scale=0.35]{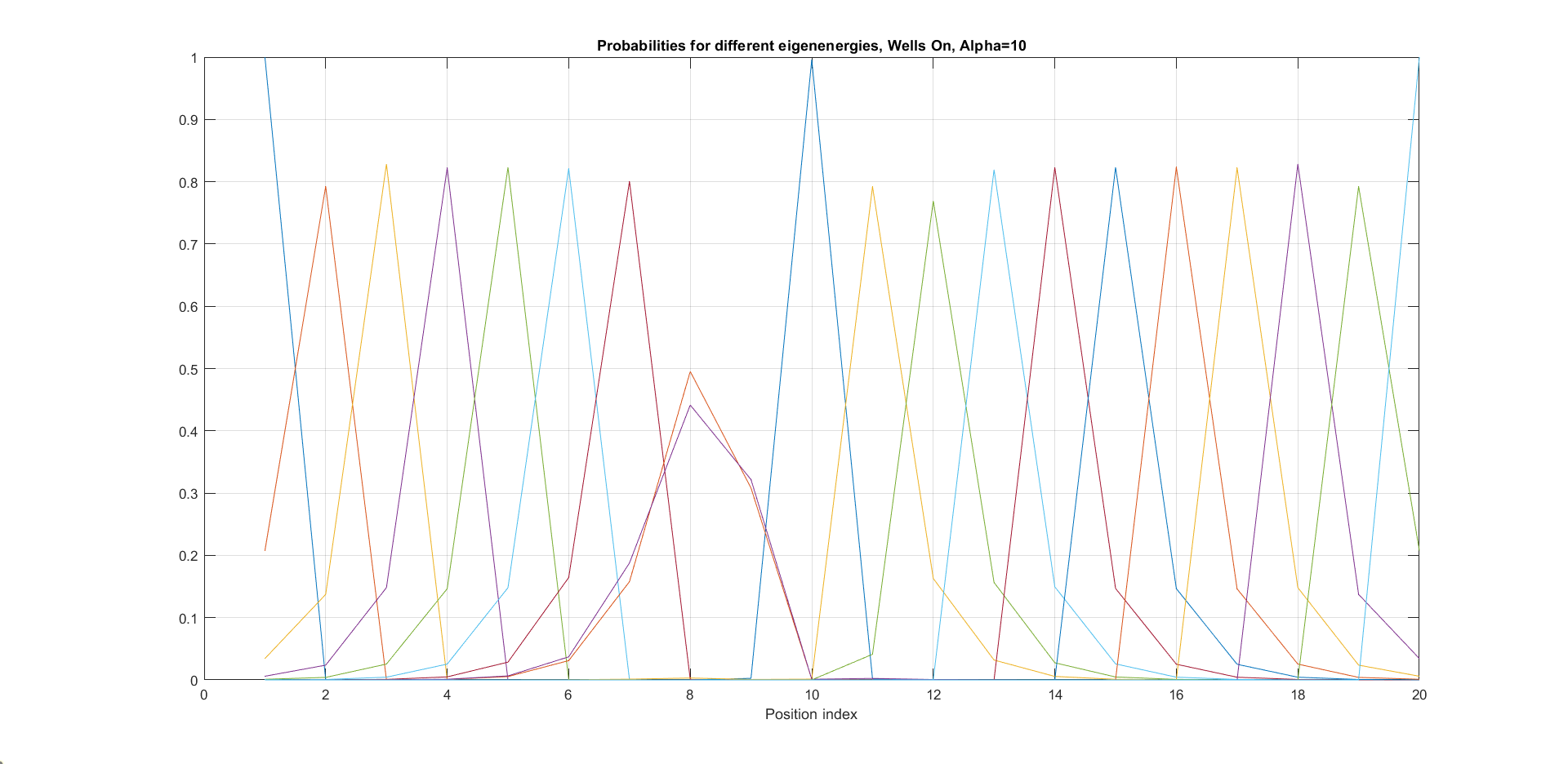}
\includegraphics[scale=0.35]{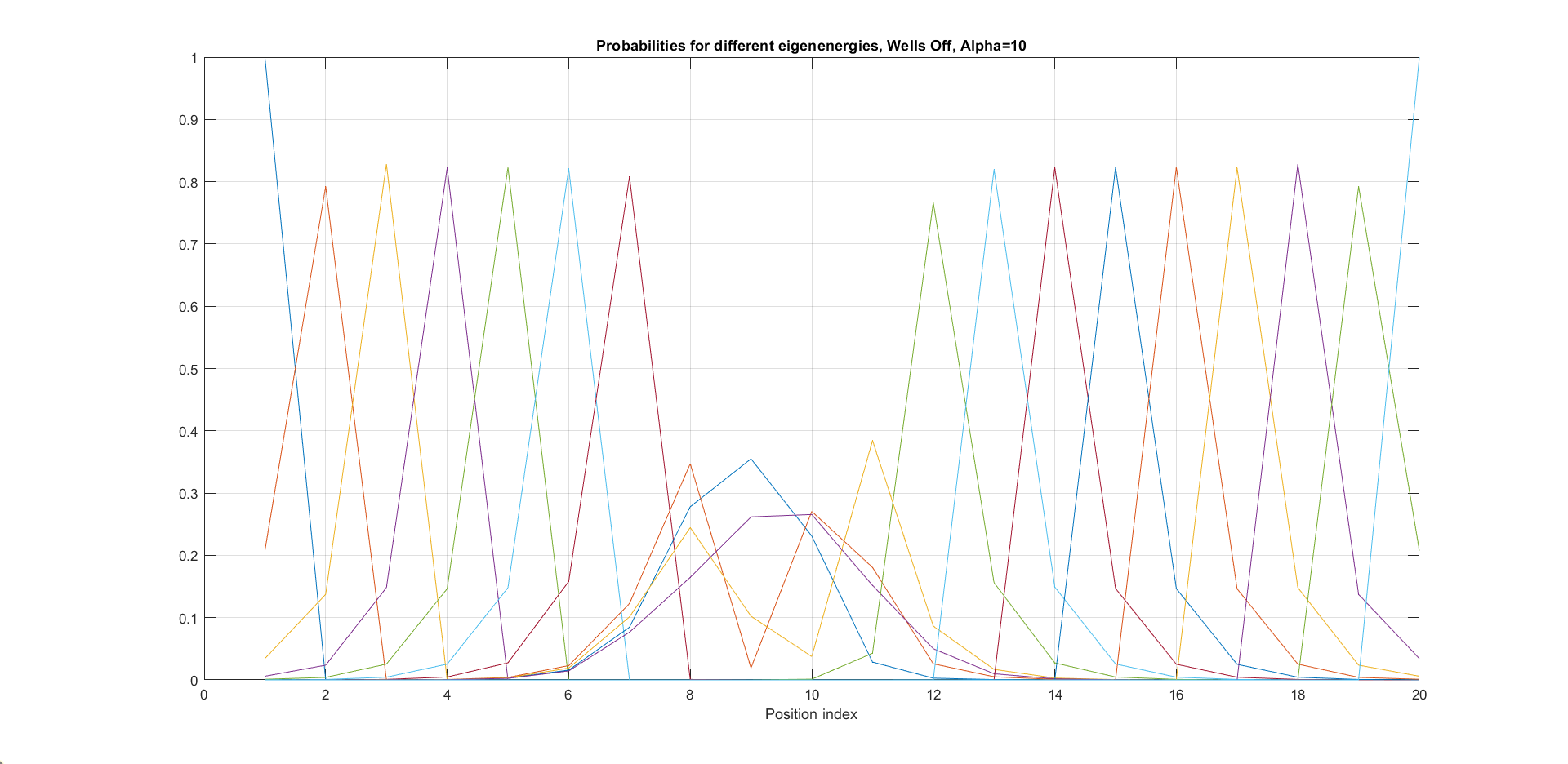}
\caption{Detailed analysis of probability distributions (for 20 cases) for Tanh square nanowires shows that eigenenergy wavefunctions are strongly localized due to the fact that nanowire has non-zero curvature (equivalent to condition that $\frac{\frac{d^2}{dx^2}}y(x){(\frac{d}{dx}y(x))^2} \neq 0$). }
\end{figure}
\begin{figure}
\includegraphics[scale=0.35]{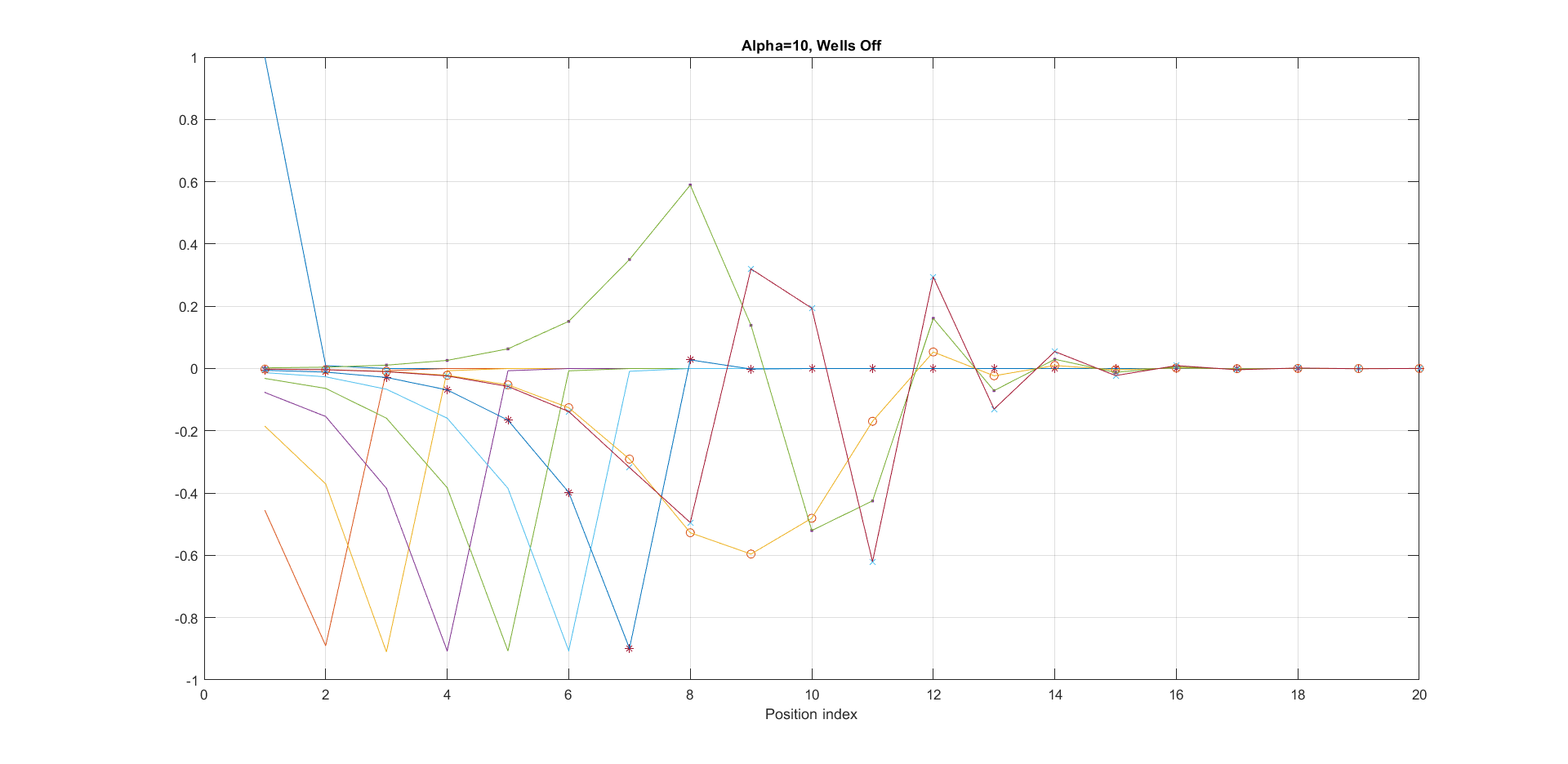}
\includegraphics[scale=0.35]{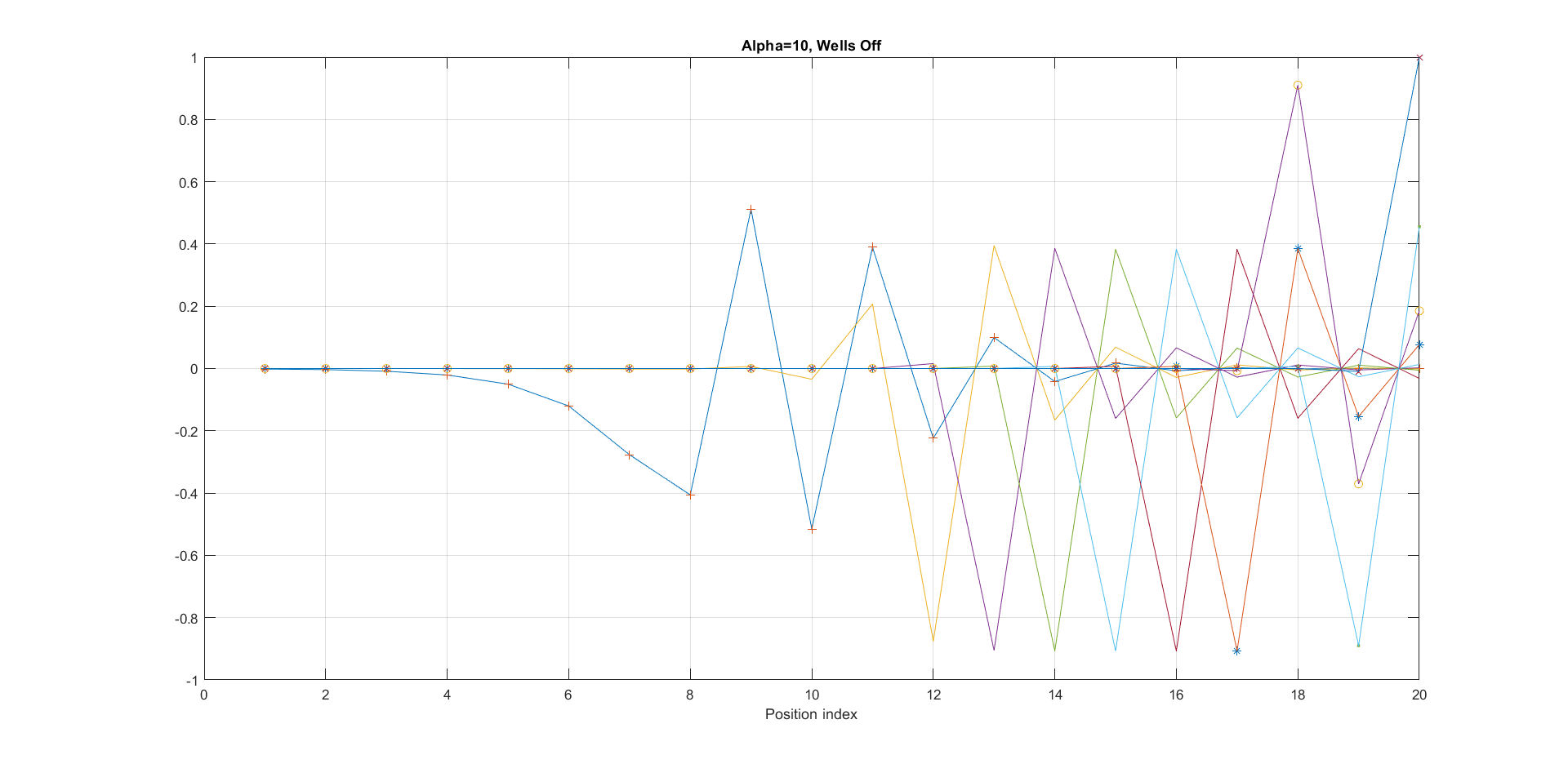}
\caption{Detailed analysis of eigenergy wavefunctions (for 20 cases) for Tanh square nanowires shows that wavefunctions are strongly localized due to the fact that nanowire has non-zero curvature (equivalent to condition that $\frac{\frac{d^2}{dx^2}}y(x){(\frac{d}{dx}y(x))^2} \neq 0$). }
\end{figure}

\begin{figure}
\includegraphics[scale=0.35]{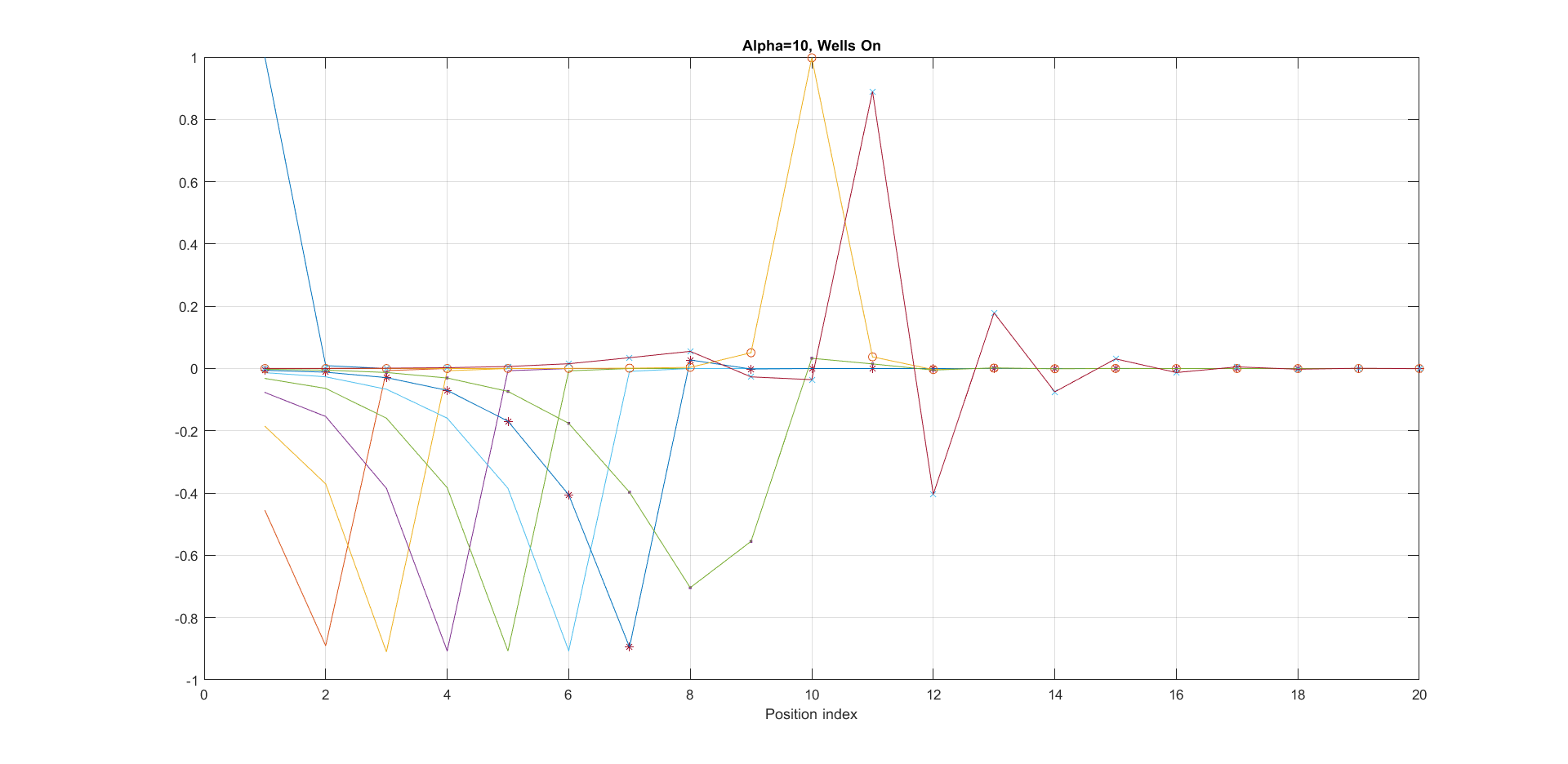}
\includegraphics[scale=0.35]{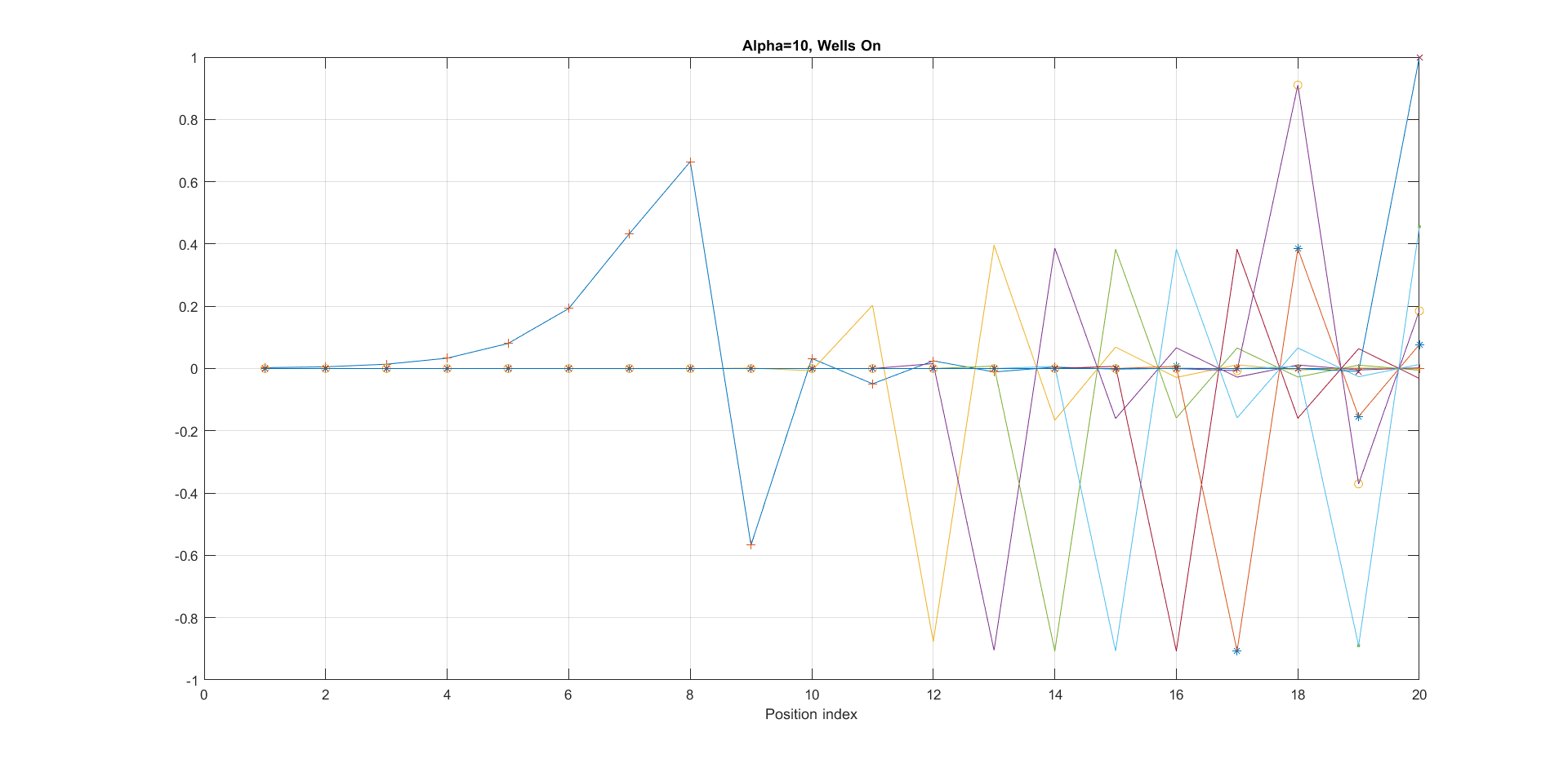}
\caption{Detailed analysis of eigenergy wavefunctions (for 20 cases) for Tanh square nanowires shows strong localization due to cable bending expressed by non-zero curvature (equivalent to condition that $\frac{\frac{d^2}{dx^2}}y(x){(\frac{d}{dx}y(x))^2} \neq 0$). }
\end{figure}

\begin{figure}
\centering
\includegraphics[scale=0.35]{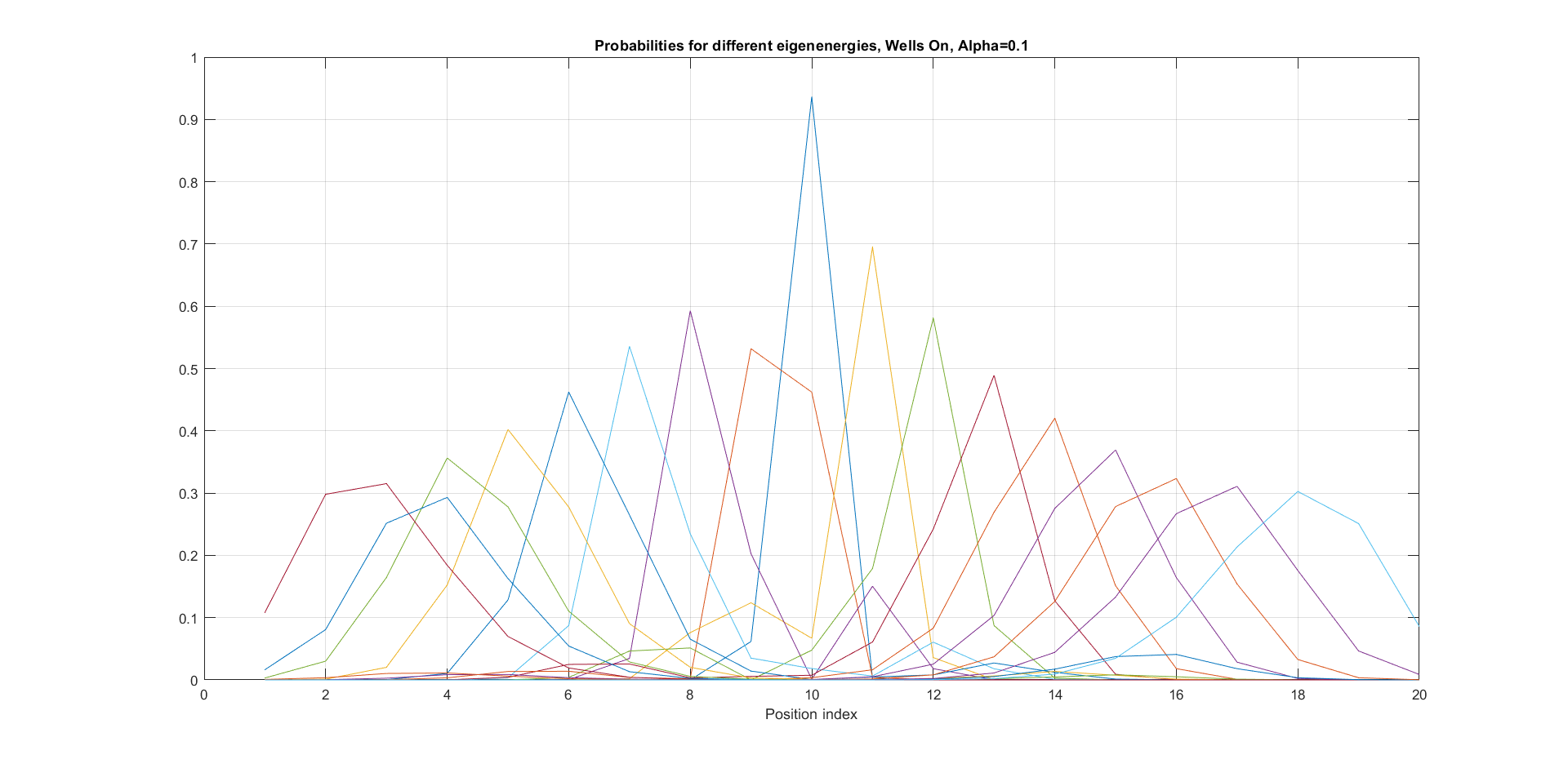}
\includegraphics[scale=0.35]{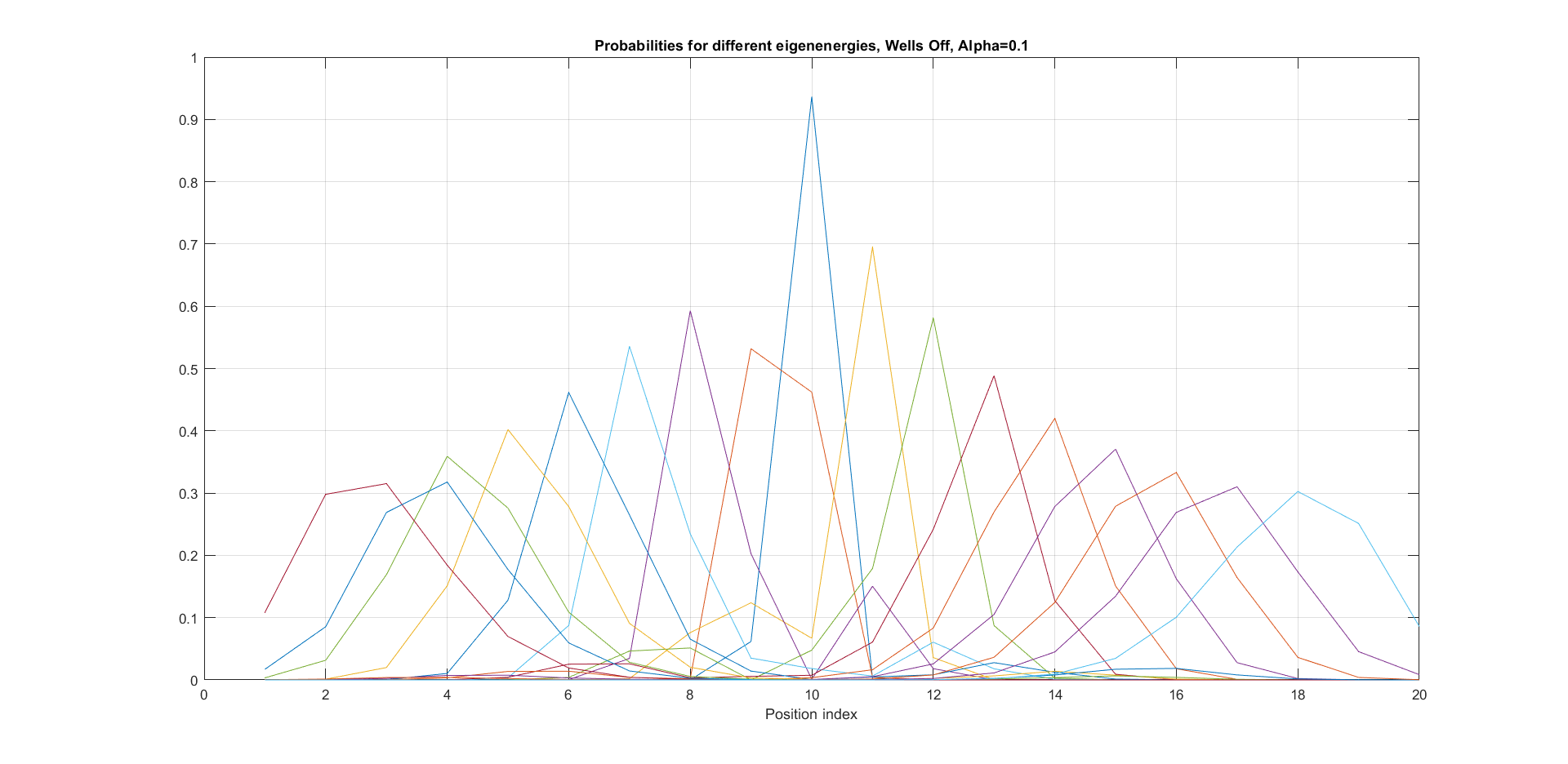}
\caption{Case of coefficient $\alpha=0.1$ reveals very similar probability distributions for cable with 3 built-in q-wells as given above and lack of built-in q-wells. }
\end{figure}

\begin{figure}
\centering
\includegraphics[scale=0.35]{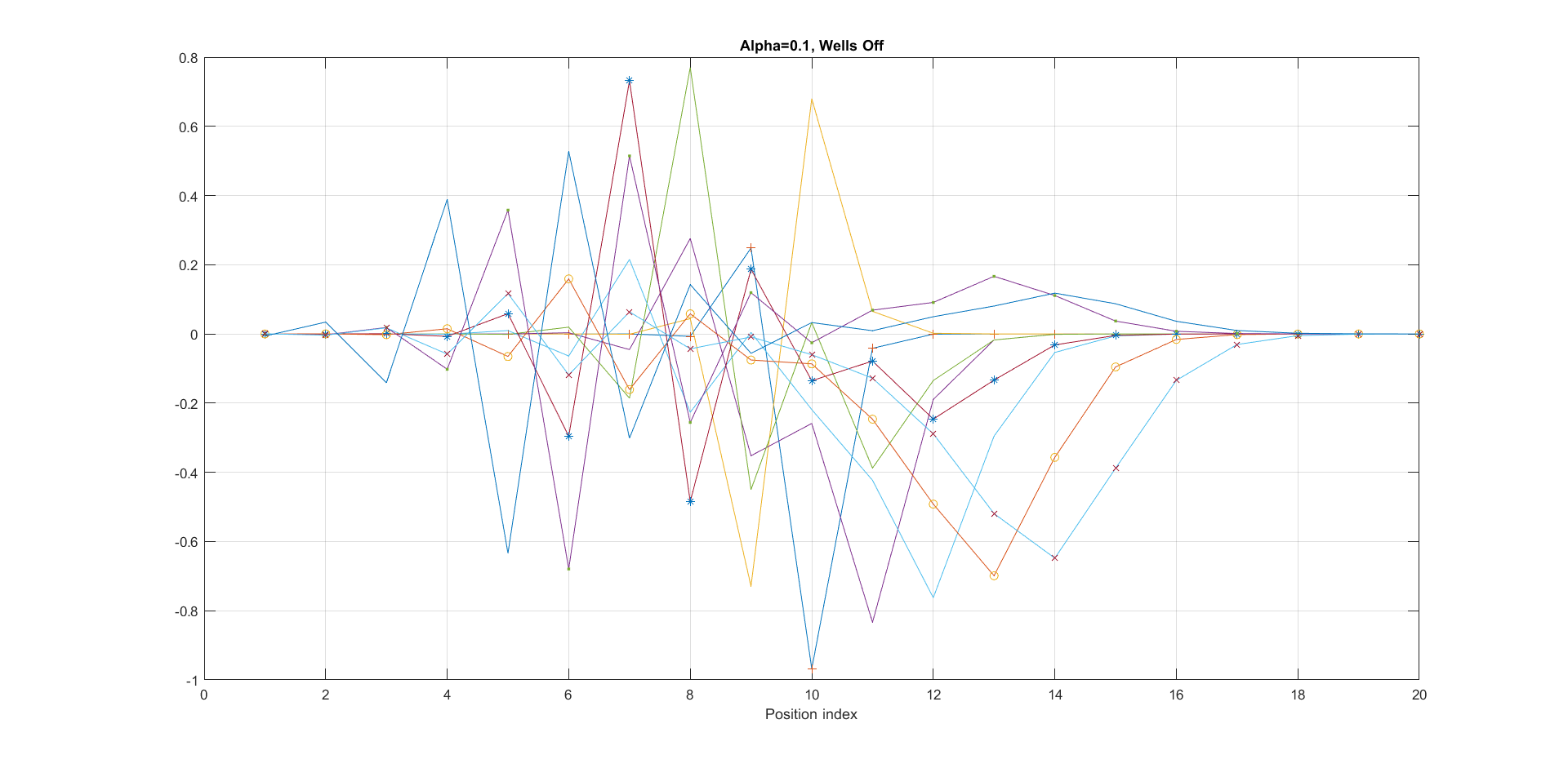}
\includegraphics[scale=0.35]{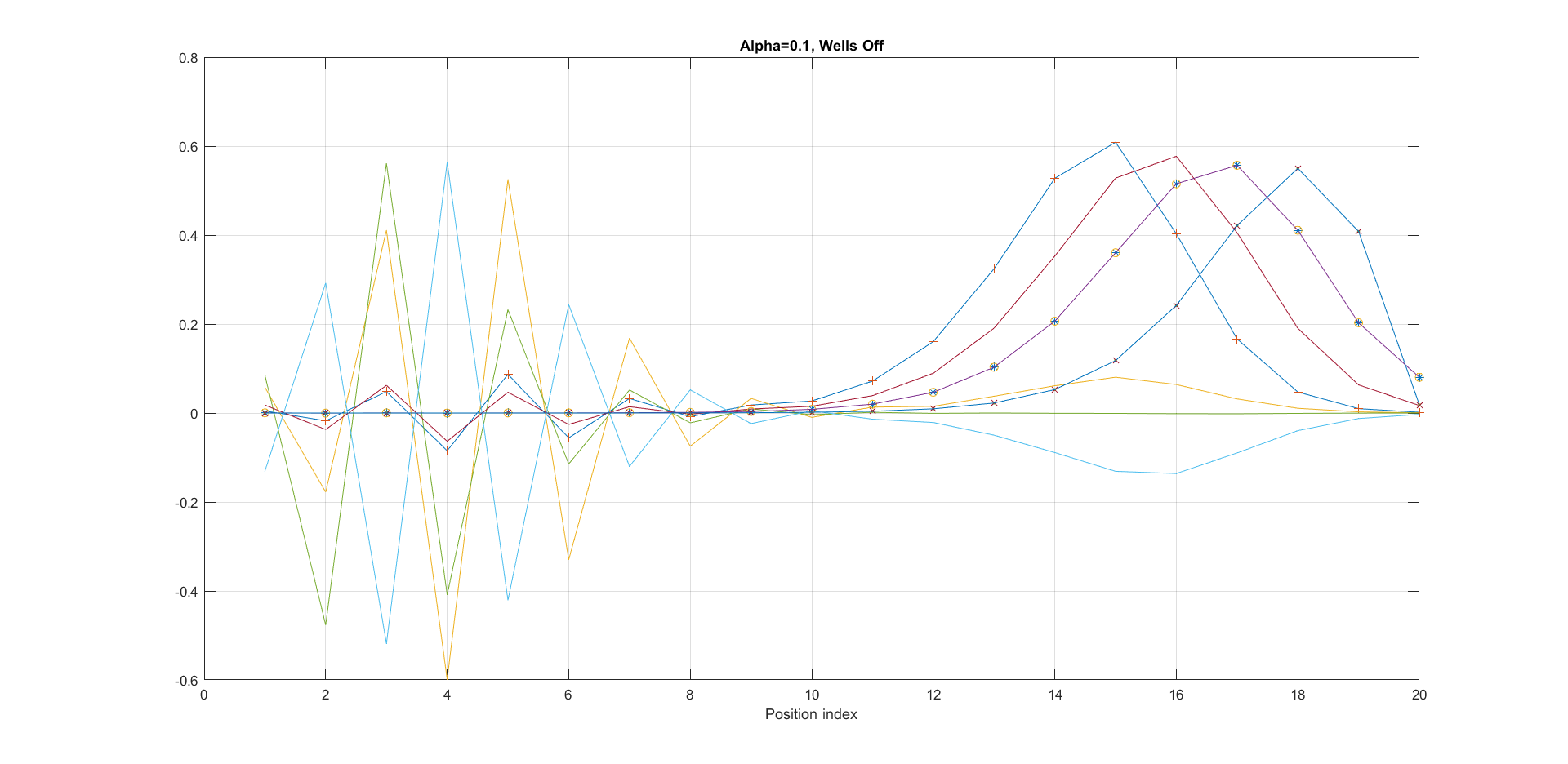}
\caption{Case of coefficient $\alpha=0.1$ reveals interesting eigenenergy wavefunction distitributions for cable no built-in q-well. }
\end{figure}

\begin{figure}
\centering
\includegraphics[scale=0.35]{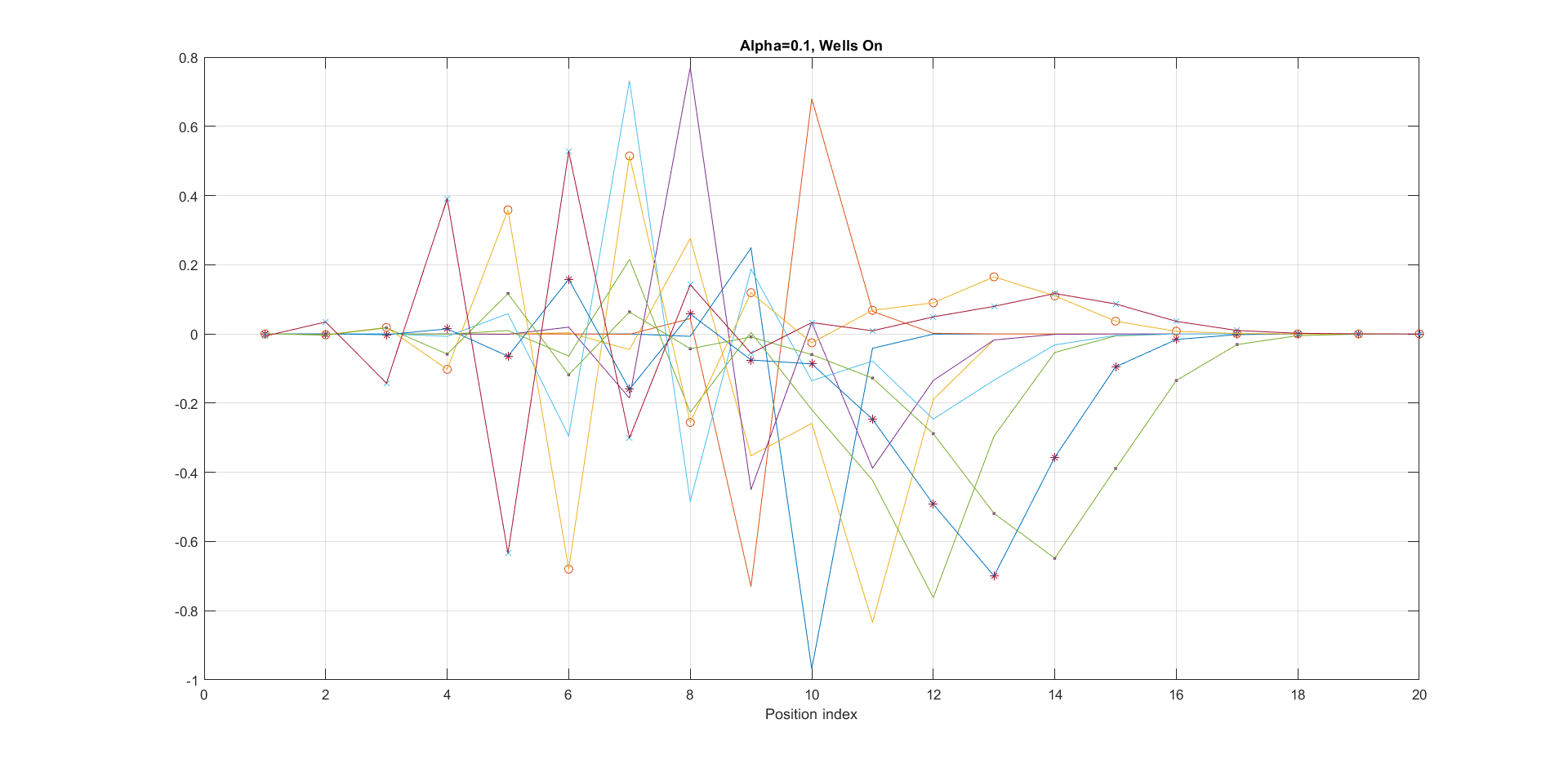} 
\includegraphics[scale=0.35]{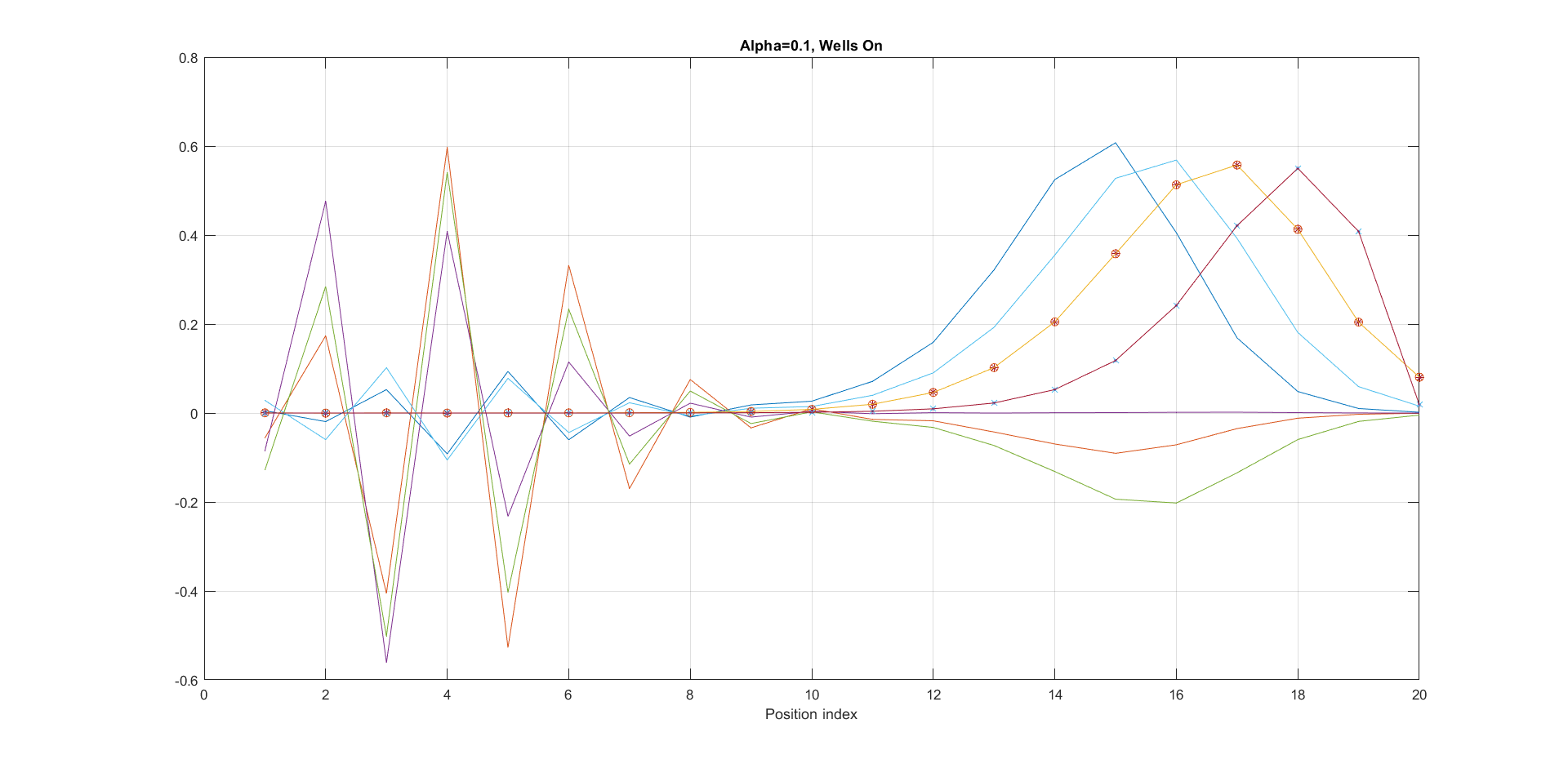} 
\caption{Case of coefficient $\alpha=0.1$ reveals interesting eigenenergy wavefunction distitributions for nanowire cable with 3 built-in q-wells. }
\end{figure}

\begin{figure}
\centering
\includegraphics[scale=0.15]{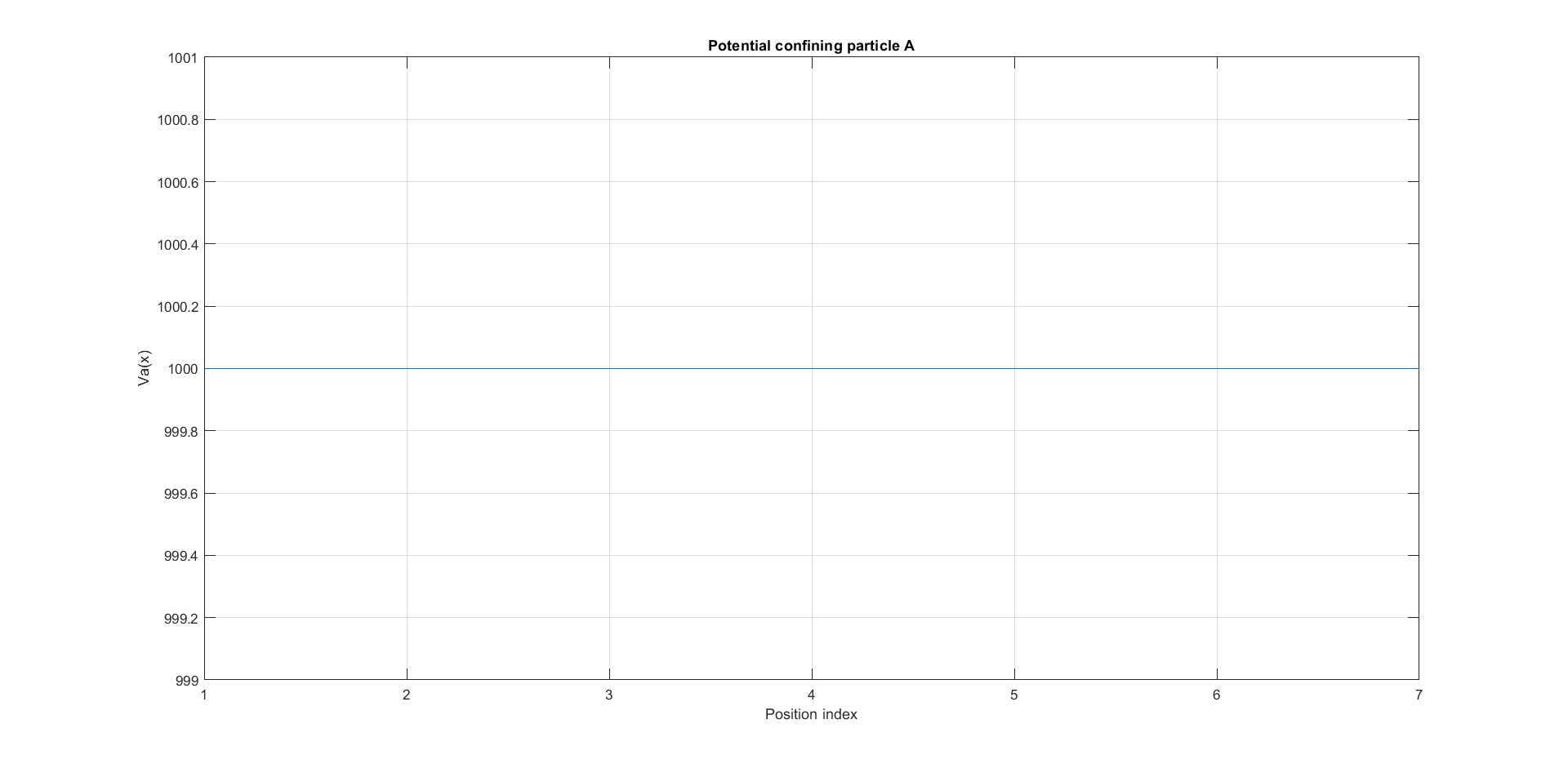} 
\includegraphics[scale=0.15]{PotAWellsOff.png} 
\includegraphics[scale=0.15]{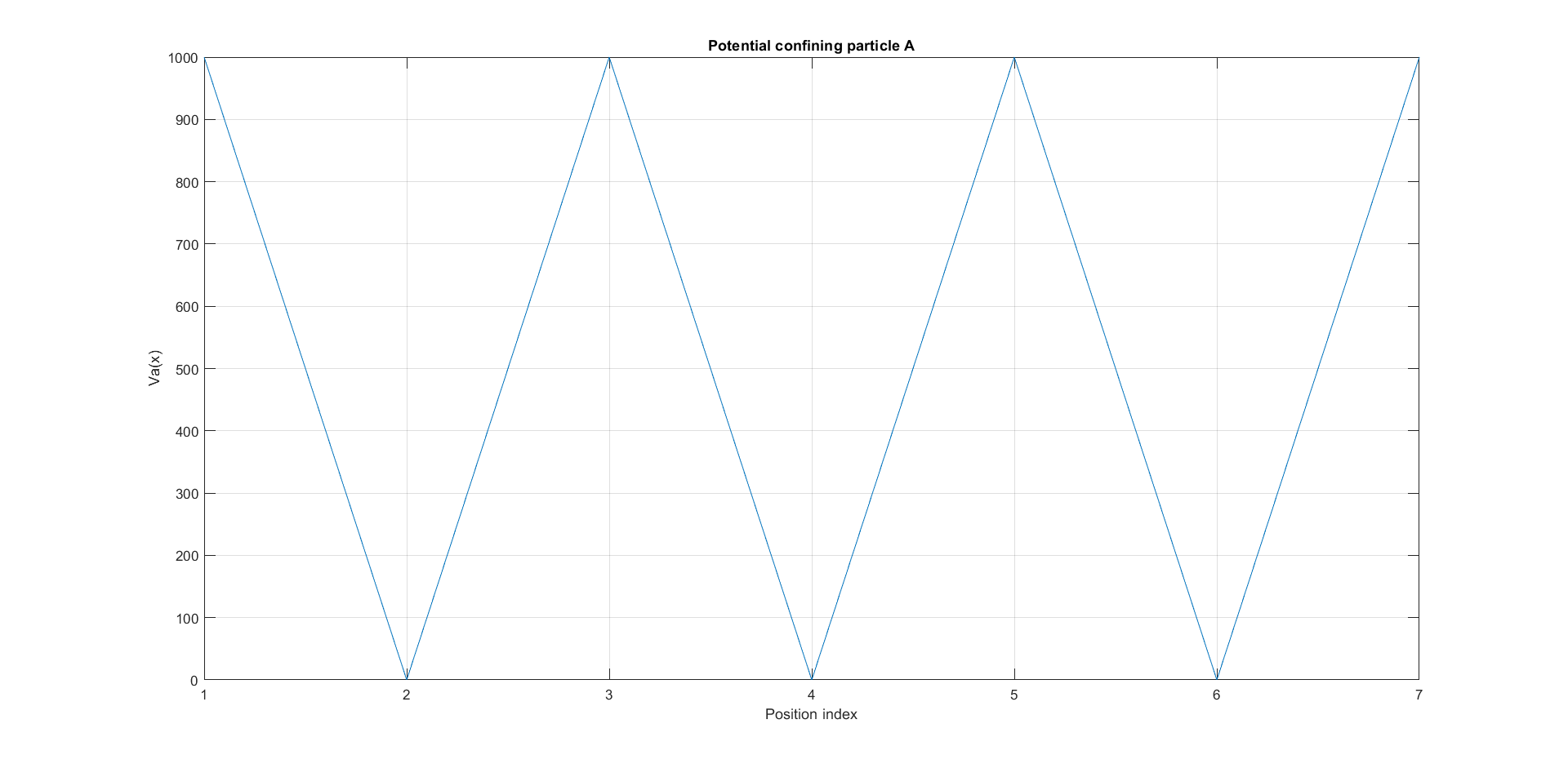} 
\includegraphics[scale=0.15]{PotAWellsOn.png} 
\caption{Case of local confining potential for Tanh Square interacting nanowire cables with no (UPPER) and built-in (LOWER) q-wells. }
\end{figure}

\begin{figure}
\includegraphics[scale=0.35]{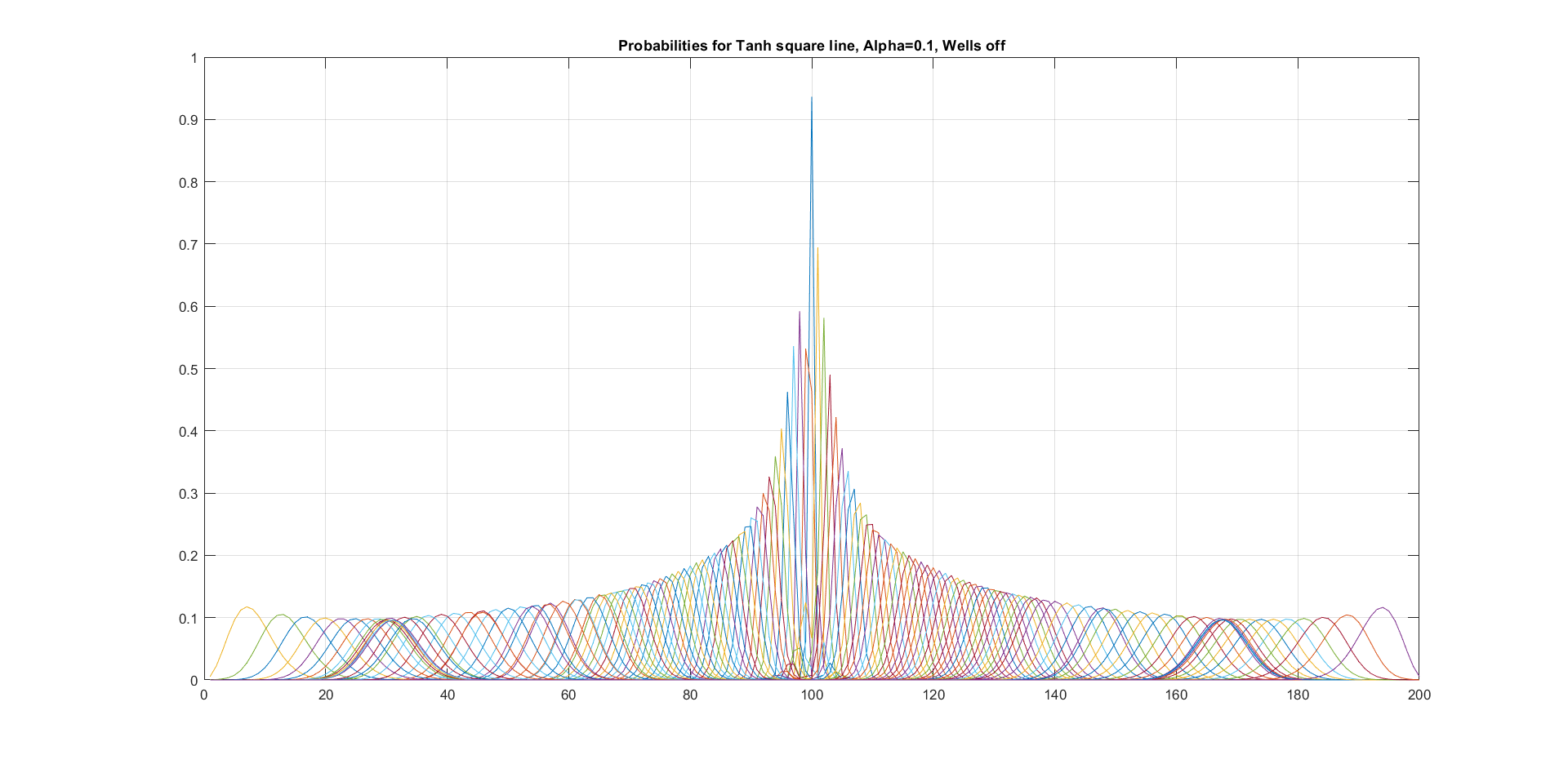} 
\includegraphics[scale=0.35]{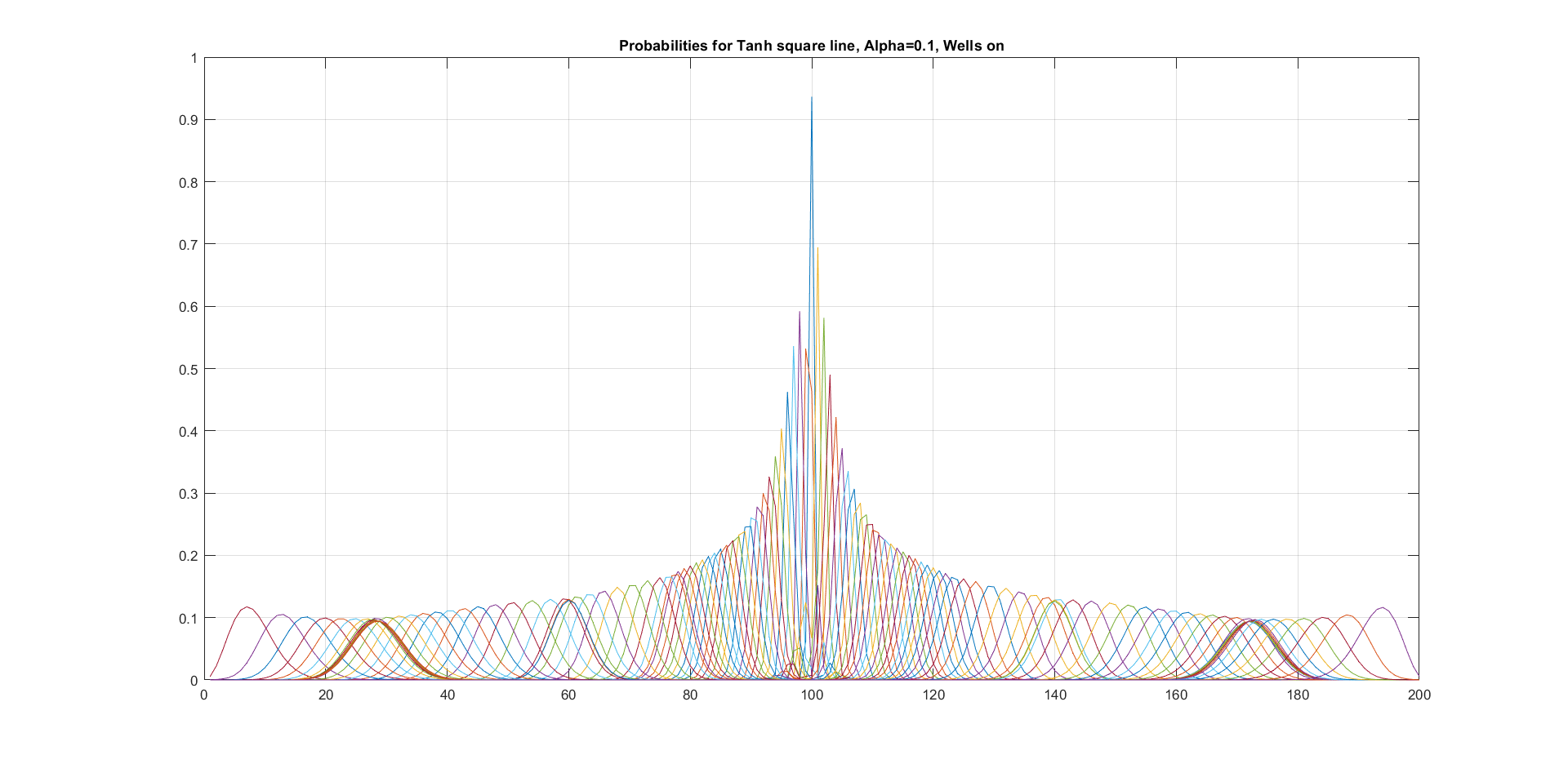}
\caption{Probability distributions for 200 different eigenenergies in function of position index for $\alpha=0.1$ with no built-in (upper) and built-in q-wells (lower). Difference is very minor but it takes place. }
\end{figure}

\begin{figure}
\centering
\includegraphics[scale=0.35]{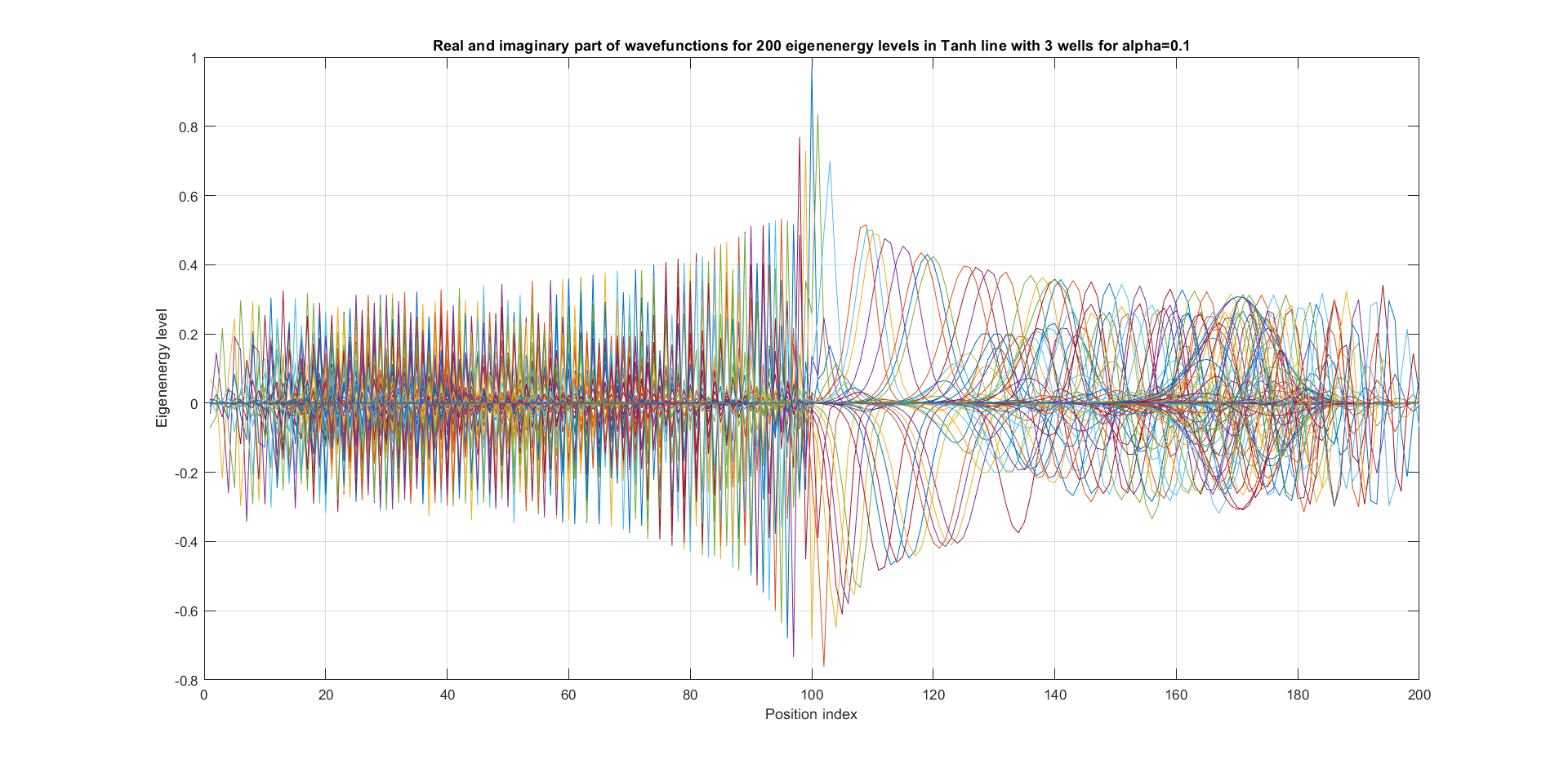}
\caption{First 200 eigen-energy wavefunctions for Tanh Square V shape nanowire with 3 built-in quantum wells. }
\includegraphics[scale=0.35]{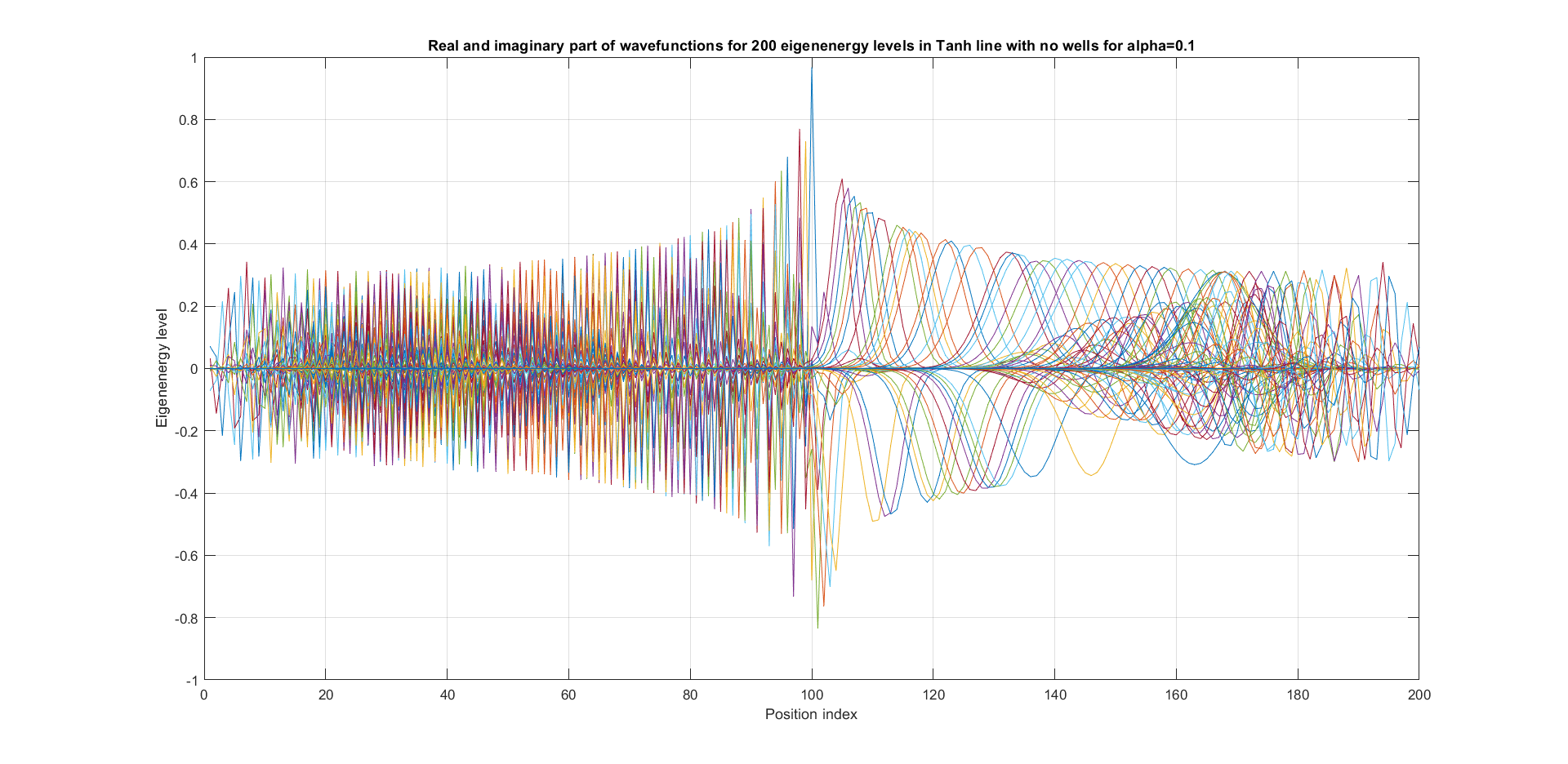}
\caption{First 200 eigen-energy wavefunctions for Tanh Square V shape nanowire with no built-in quantum wells. }
\end{figure}
\begin{figure} \label{supsup}
\centering
\includegraphics[scale=0.25]{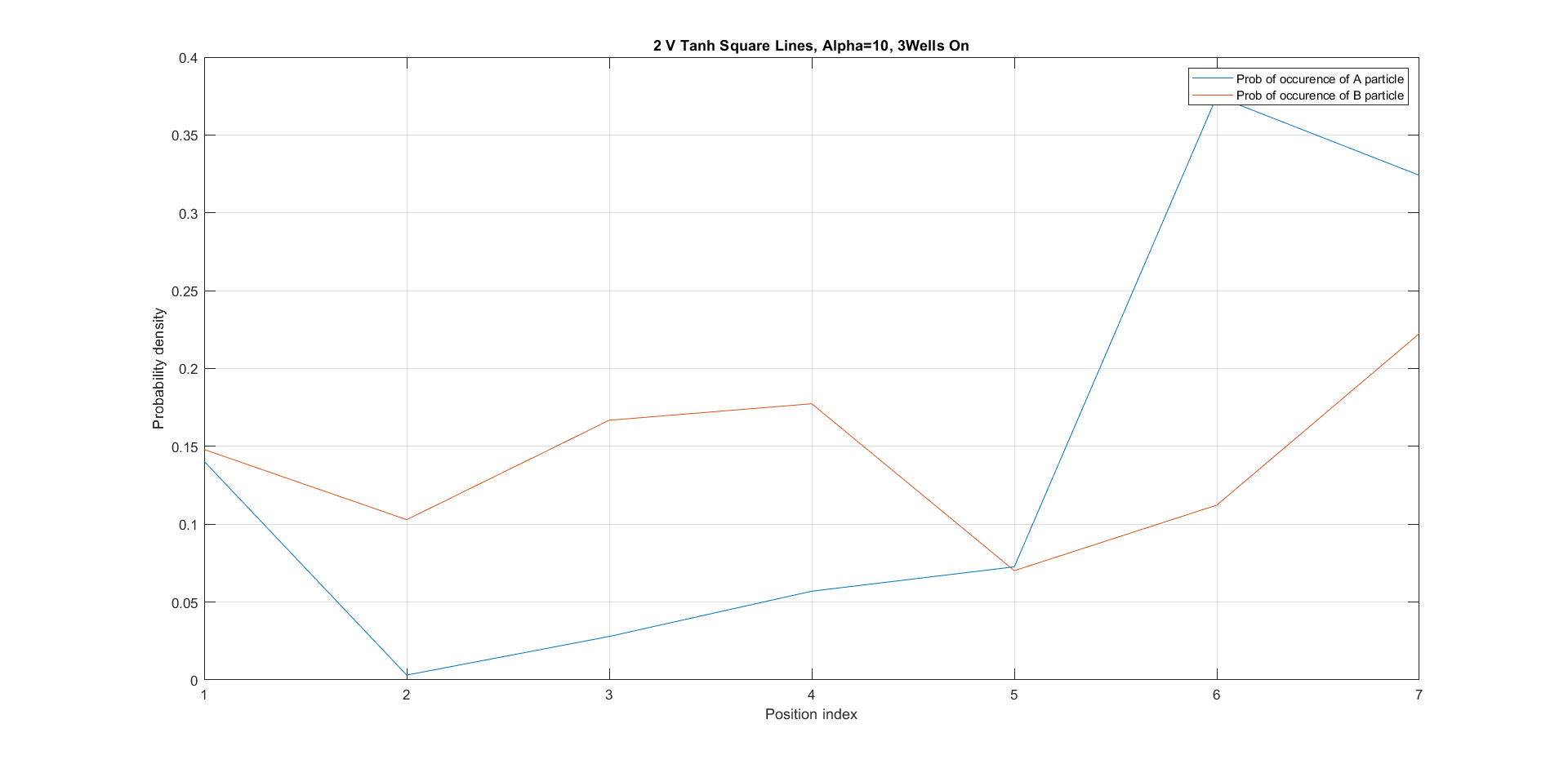}
\includegraphics[scale=0.25]{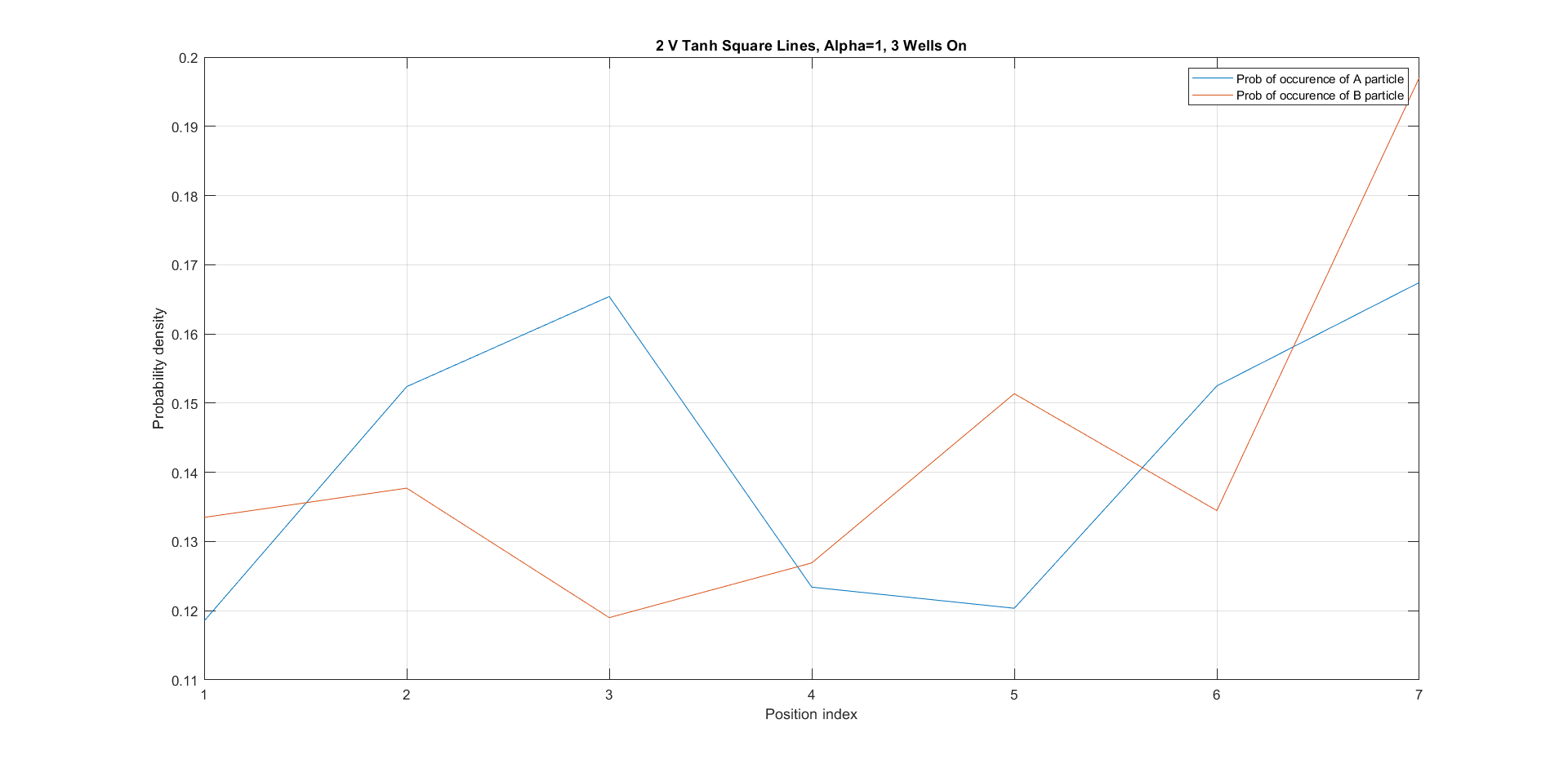}
\includegraphics[scale=0.25]{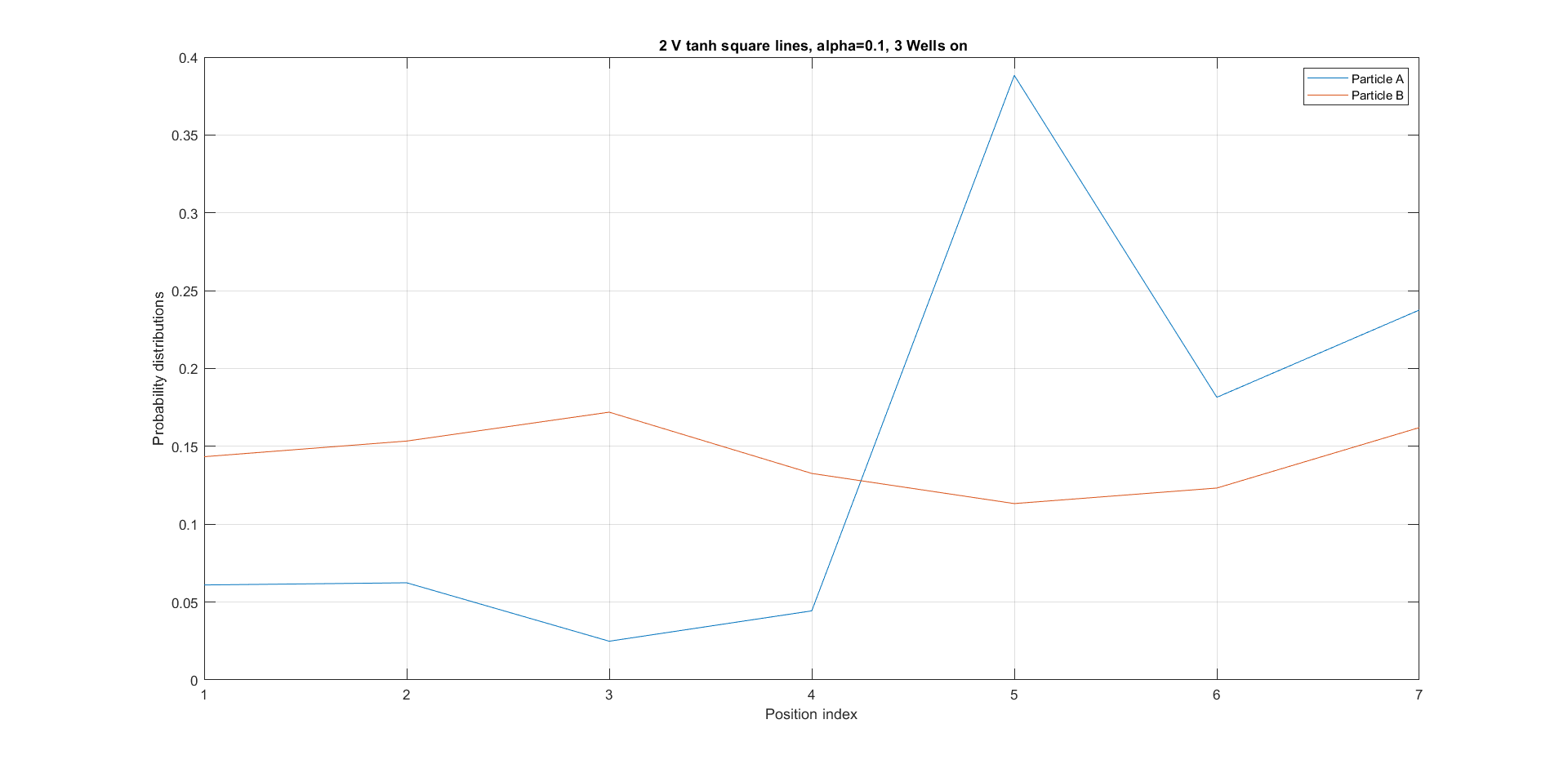}
\caption{Case of 2 V Tanh Squares Lines interacting and probability distributions around each line for electrons A and B with $\alpha=$ (10, 1, 0.1) for (UPPER, MIDDLE, LOWER) pictures with 3 quantum wells built-in.  }
\end{figure}
\begin{figure} \label{supsup1}
\centering
\includegraphics[scale=0.25]{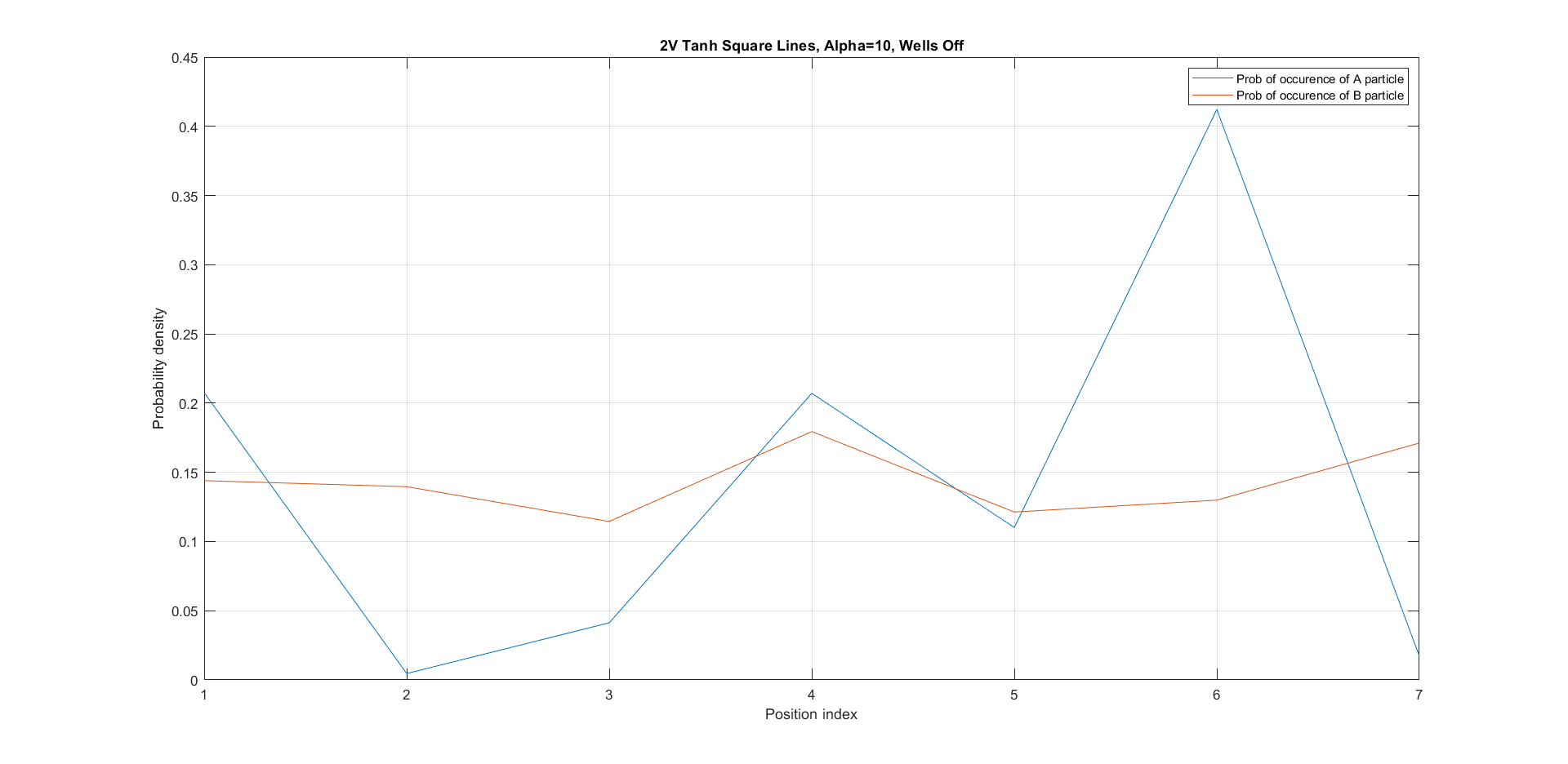}
\includegraphics[scale=0.25]{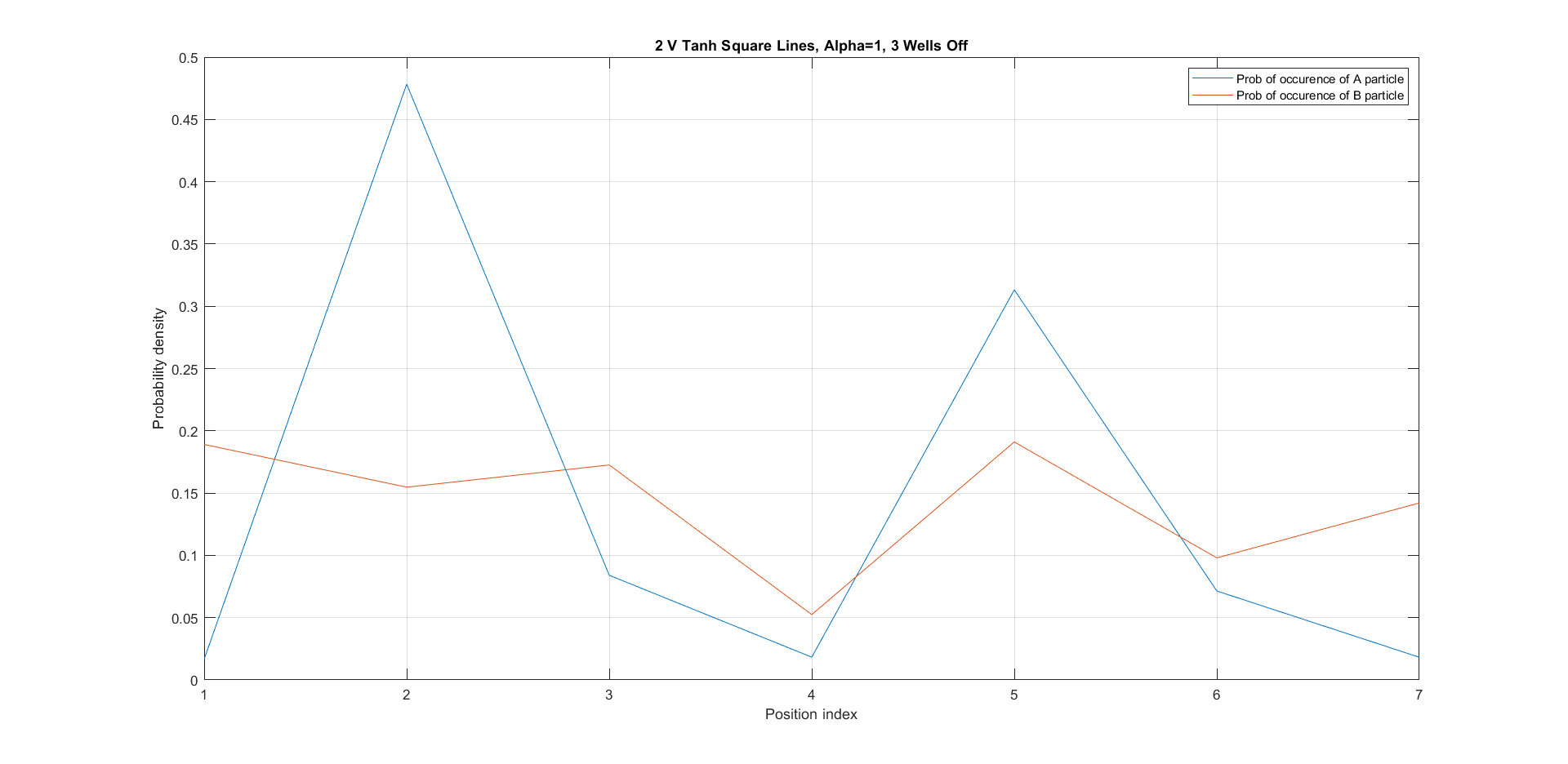}
\includegraphics[scale=0.25]{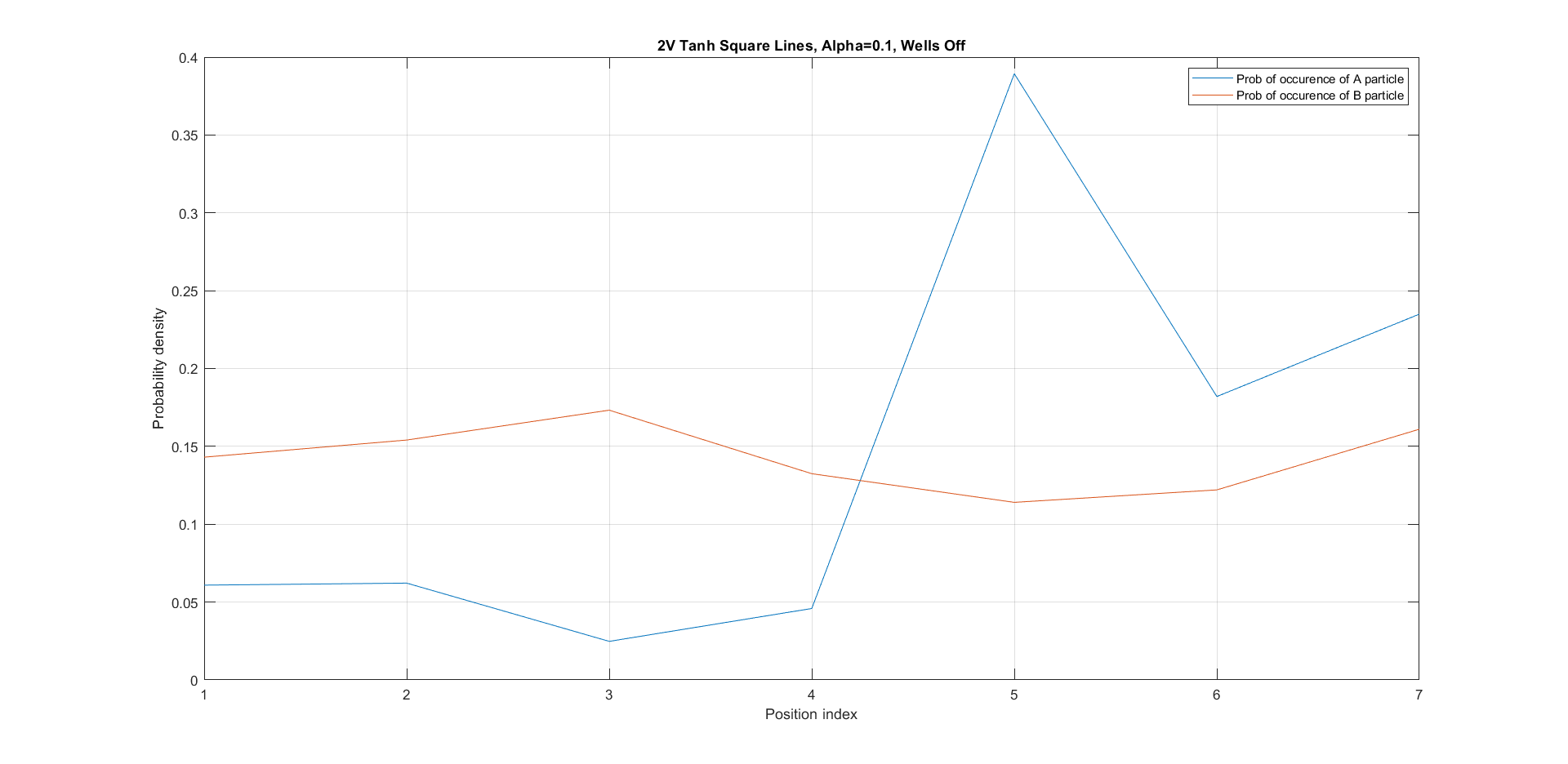}
\caption{Case of 2 V Tanh Squares Lines interacting and probability distributions around each line for electrons A and B with $\alpha=$ (10, 1, 0.1) for (UPPER, MIDDLE, LOWER) pictures with no q-wells built-in.  }
\end{figure}
\section{Transformation from curvy into non-curvy Schroedinger equation}
Let us consider the equation structure
\begin{eqnarray}
\Bigg[-\frac{\hbar^2}{2m}\Bigg[ a(s)\frac{d}{ds}-b(s)\Bigg]^2+V_1(s) \Bigg]\psi(s)=E \psi(s).
\end{eqnarray}
with introduced $a(s)$, $b(s)$,
$V_1(s)$ (in functions of f(s), g(s) and V(s)) and equivalently we obtain
\begin{eqnarray}
\Bigg[-\frac{\hbar^2}{2m}\Bigg[ a(s)^2\frac{d^2}{ds^2}+a(s)(\frac{da}{ds})\frac{d}{ds}-a(s)b(s)\frac{d}{ds}-(a(s)(\frac{db}{ds}))+b(s)^2\Bigg]+V_1(s)\Bigg] \psi(s)=E \psi(s).
\end{eqnarray}
so we can write
\begin{eqnarray}
\Bigg[-\frac{\hbar^2}{2m}\Bigg[ a(s)^2\frac{d^2}{ds^2}+a(s)[(\frac{da}{ds})-b(s)]\frac{d}{ds}+(-a(s)(\frac{db}{ds})+b(s)^2)\Bigg]+V_1(s)\Bigg] \psi(s)=E \psi(s).
\end{eqnarray}
and we have
\begin{eqnarray}
f(s)=\sqrt{a(s)}, g(s)=-a(s)[(\frac{da}{ds})-b(s)],
\end{eqnarray}
what implies
\begin{eqnarray}
\frac{g(s)}{a(s)}+\frac{da}{ds}=b(s)=\frac{g(s)}{(f(s))^2}+\frac{d(f(s)^2)}{ds}=\frac{g(s)}{(f(s))^2}+2f(s)\frac{df(s)}{ds}
\end{eqnarray}
and
\begin{eqnarray}
V(s)=V_1(s)-\frac{\hbar^2}{2m}(-a(s)(\frac{db}{ds})+b(s)^2),
\end{eqnarray}
implies that
\begin{eqnarray}
V_1(s)=V(s)+\frac{\hbar^2}{2m}(-a(s)(\frac{db}{ds})+b(s)^2).
\end{eqnarray}
Further considerations of curvy semiconductor nanowire is described by \ref{spherical} and \ref{cylindrical}.
\section{Analytical solutions of Schroedinger equation in curvlinear coordinates}
If we deal with equation
\begin{eqnarray}
\Bigg[-\frac{\hbar^2}{2m}\Bigg[ a(s)\frac{d}{ds}-b(s)\Bigg]^2+V_1(s) \Bigg]\psi(s)=E \psi(s).
\end{eqnarray}
we observe that
\begin{eqnarray}
\Bigg[ a(s)\frac{d}{ds}-b(s)\Bigg]\psi(s)=\Bigg[ \Bigg[ a(s)\frac{d}{ds}\Bigg](e^{-\int_0^{s}ds_1\frac{b(s_1)}{a(s_1)}}\psi(s))\Bigg]\frac{1}{e^{-\int_0^{s}ds_1\frac{b(s_1)}{a(s_1)}}}.
\end{eqnarray} and consequently we have
\begin{eqnarray}
\Bigg[ a(s)\frac{d}{ds}-b(s)\Bigg]^2\psi(s)=\Bigg[ \Bigg[ a(s)\frac{d}{ds}-b(s)\Bigg] \Bigg[ \Bigg[ a(s)\frac{d}{ds}\Bigg](e^{-\int_0^{s}ds_1\frac{b(s_1)}{a(s_1)}}\psi(s))\Bigg]\frac{1}{e^{-\int_0^{s}ds_1\frac{b(s_1)}{a(s_1)}}}= \nonumber \\
=  \Bigg[ \Bigg[ \Bigg[ \Bigg[ \Bigg[ \Bigg[ \Bigg[ \Bigg[ a(s)\frac{d}{ds}\Bigg]^2-b(s)\Bigg[ a(s)\frac{d}{ds}\Bigg]\Bigg]\Bigg] ( e^{-\int_0^{s}ds_1\frac{b(s_1)}{a(s_1)}} \psi(s))\Bigg] \Bigg] \frac{1}{e^{-\int_0^{s}ds_1\frac{b(s_1)}{a(s_1)}}} .
\end{eqnarray}
Now we characterize the operator
\begin{eqnarray}
\Bigg[ a(s)\frac{d}{ds} \Bigg]^2=[a(s)]^2 \frac{d^2}{ds^2}+[a(s)(\frac{da}{ds})]\frac{d}{ds}=[a(s)]^2 \frac{d^2}{ds^2}+\frac{1}{2}[(\frac{d(a(s)^2)}{ds})]\frac{d}{ds}.
\end{eqnarray}
In similar way we obtain higher orders of power of operator $\Bigg[ a(s)\frac{d}{ds} \Bigg]^N$ by induction principle.
Let us analyze the eigenstate of operator $a(s)\frac{d}{ds}$. We have
\begin{eqnarray}
[a(s)\frac{d}{ds}]\psi_o(s)=p(s)\psi_o(s).
\end{eqnarray}
Last equation implies
\begin{eqnarray}
\frac{d}{ds}\psi_o(s)=(\frac{p(s)}{a(s)})\psi_o(s).
\end{eqnarray}
The last equation has solution
\begin{eqnarray}
\psi_o(s)=e^{\int_{s1=0}^{s1=s}(ds_1\frac{p(s_1)}{a(s_1)})}u_0.
\end{eqnarray}

\section{Physical system implementing Wannier qubit swap gate in classical description}
\begin{figure} 
    \centering
    \includegraphics[scale=0.8]{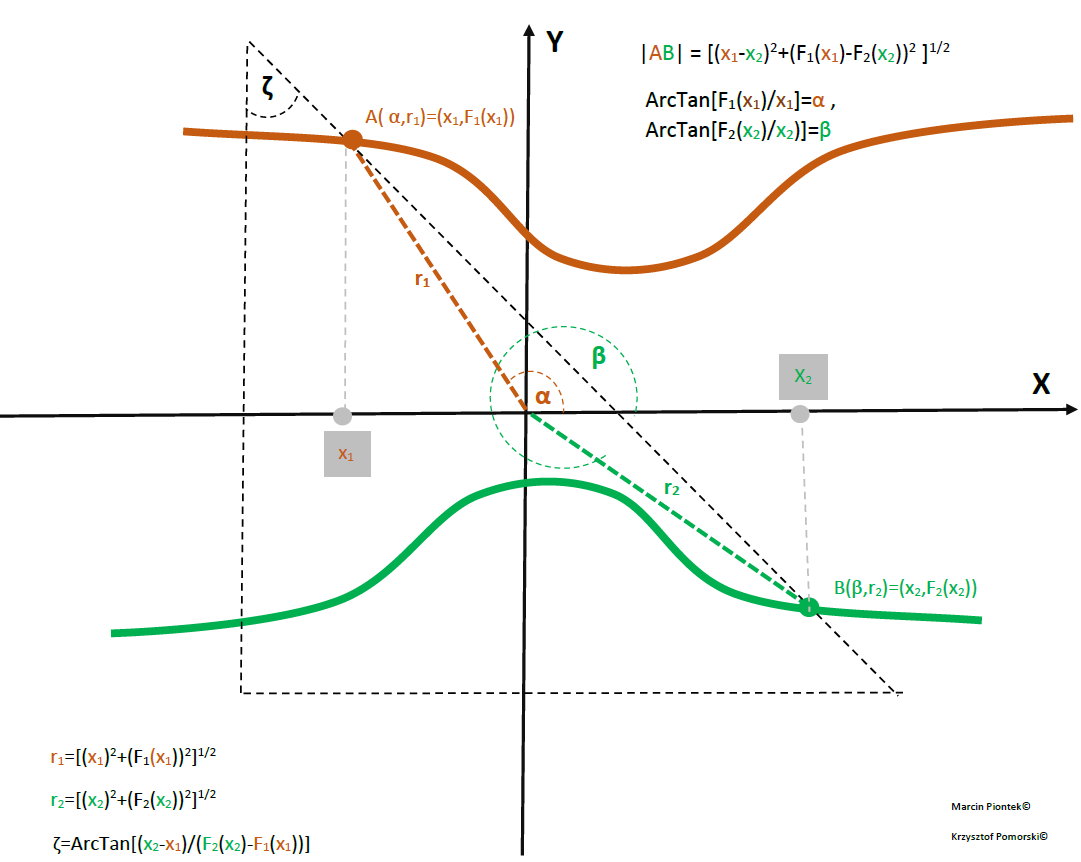}
    \includegraphics[scale=0.8]{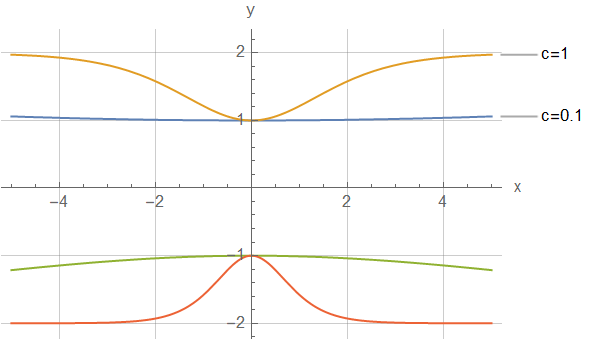}
    \includegraphics[scale=0.6]{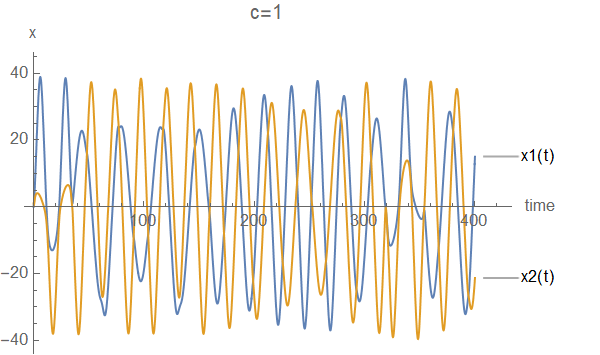}\includegraphics[scale=0.6]{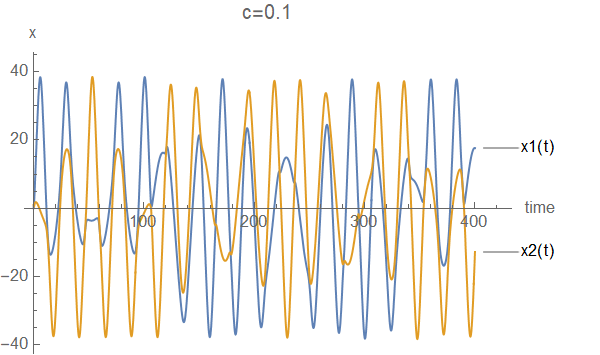}
    \caption{Two particles (electrons) placed at semiconductor nanowires interacting electrostatically and family of V shape lines parametrized by $F_{1(2)}(x)=a+b*(Tanh(c*x+d))^2$, so nanolines are given by $(x,F_{1}(x))$ and $(x,F_{2}(x))$ points. }
    \label{fig1q}
    \end{figure}

We assume movement of 2 particles at trajectories $(x_1(s_1),y_1(s_1))$ and $(x_2(s_2),y_2(s_2))$ as given by Fig.\ref{fig1q}.
    \begin{eqnarray}
      -\frac{d}{dx_1}H=\frac{d}{dt}p_{1,x}, -\frac{d}{dy_1}H=\frac{d}{dt}p_{1,y}, \frac{d}{dp_{1,x}}H=\frac{d}{dt}x_{1}, \frac{d}{dp_{1,y}}H=\frac{d}{dt}y_{1},   \nonumber \\
      -\frac{d}{dx_2}H=\frac{d}{dt}p_{2,x}, -\frac{d}{dy_2}H=\frac{d}{dt}p_{2,y}, \frac{d}{dp_{2,x}}H=\frac{d}{dt}x_{2}, \frac{d}{dp_{2,y}}H=\frac{d}{dt}y_{2}.
    \end{eqnarray}

    We have the Hamiltonian of the structure
    \begin{eqnarray}
      H =H_1+H_2+H_{1-2}= \nonumber \\
      =\frac{(p_{1,x})^2}{2m_1}+\frac{(p_{1,y})^2}{2m_1}+\frac{(p_{2,x})^2}{2m_2}+\frac{(p_{2,y})^2}{2m_2}+V(x_1(s_1),y_1(s_1))+V(x_2(s_2),y_2(s_2))+H_C(s_1,s_2), \nonumber
      \\
    \end{eqnarray}
    and we have
     \begin{eqnarray} \label{cleqnodiss1}
     \frac{d}{dx_1}=\frac{ds_1}{dx_1}\frac{d}{ds_1},  \frac{d}{dy_1}=\frac{ds_1}{dy_1}\frac{d}{ds_1}, \nonumber \\
     \frac{d}{dx_2}=\frac{ds_2}{dx_2}\frac{d}{ds_2},  \frac{d}{dy_2}=\frac{ds_2}{dy_2}\frac{d}{ds_2}, \nonumber \\
     \frac{d}{dt}x_1=\frac{dx_1}{ds_1}\frac{d}{dt}s_1,  \frac{d}{dt}y_1=\frac{dy_1}{ds_1}\frac{d}{dt}s_1, \nonumber \\
     \frac{d}{dt}x_2=\frac{dx_1}{dx_1}\frac{d}{dt}s_2,  \frac{d}{dt}y_2=\frac{dy_2}{ds_2}\frac{d}{dt}s_2, \nonumber \\
    \frac{1}{m_1}\frac{d}{dt}p_{1,x}=\frac{d^2}{dt^2}x_1=\frac{dx_1}{ds_1}\frac{d}{dt}s_1[\frac{dx_1}{ds_1}\frac{d}{dt}s_1]=[(\frac{dx_1}{ds_1})^2\frac{d^2}{dt^2}s_1+\frac{d^2x_1}{ds_1^2}\frac{dx_1}{ds_1}\frac{d}{dt}s_1], \nonumber \\
     \frac{1}{m_2}\frac{d}{dt}p_{2,x}=\frac{d^2}{dt^2}x_2=\frac{dx_2}{ds_2}\frac{d}{dt}s_2[\frac{dx_2}{ds_2}\frac{d}{dt}s_2]=[(\frac{dx_2}{ds_2})^2\frac{d^2}{dt^2}s_2+\frac{d^2x_2}{ds_2^2}\frac{dx_2}{ds_2}\frac{d}{dt}s_2], \nonumber \\
     \frac{1}{m_1}\frac{d}{dt}p_{1,y}=\frac{d^2}{dt^2}y_1=\frac{dy_1}{ds_1}\frac{d}{dt}s_1[\frac{dy_1}{ds_1}\frac{d}{dt}s_1]=[(\frac{dy_1}{ds_1})^2\frac{d^2}{dt^2}s_1+\frac{d^2y_1}{ds_1^2}\frac{dy_1}{ds_1}\frac{d}{dt}s_1], \nonumber \\
     \frac{1}{m_2}\frac{d}{dt}p_{2,y}=\frac{d^2}{dt^2}y_2=\frac{dy_2}{ds_2}\frac{d}{dt}s_2[\frac{dy_2}{ds_2}\frac{d}{dt}s_2]=[(\frac{dy_2}{ds_2})^2\frac{d^2}{dt^2}s_2+\frac{d^2y_2}{ds_2^2}\frac{dy_2}{ds_2}\frac{d}{dt}s_2], \nonumber \\
    \end{eqnarray}
    We set $V(s_1,s_2)=q^2/((x_1(s_1)-x_2(s_2))^2+(f_1(x_1(s_1))-f_2(x_2(s_2)))^2)^{1/2}$ and thus obtain $\frac{d}{dx1}H=\frac{ds_1}{dx_1}\frac{d}{ds_1}H$
    and $\frac{d}{dx2}H=\frac{ds_2}{dx_1}\frac{d}{ds_2}H$.
    Consequently have
     \begin{eqnarray}
     H= \frac{(p_{1,x})^2}{2m_1}+\frac{(p_{1,y})^2}{2m_1}+\frac{(p_{2,x})^2}{2m_2}+\frac{(p_{2,y})^2}{2m_2}+V_1(x_1(s_1),y_1(s_1))+V_2(x_2(s_2),y_2(s_2))+H_C(s_1,s_2), \nonumber \\
     =\frac{q^2}{((x_1(s_1)-x_2(s_2))^2+(f_1(x_1(s_1))-f_2(x_2(s_2)))^2){\frac{1}{2}}}+V_1(s_1)+V_2(s_2)+ \nonumber \\
     +\frac{1}{2}m_1[(\frac{dx_1}{ds_1})^2+(\frac{dy_1}{ds_1})^2](\frac{ds_1}{dt})^2+\frac{1}{2}m_2[(\frac{dx_2}{ds_2})^2+(\frac{dy_2}{ds_2})^2](\frac{ds_2}{dt})^2.
     \end{eqnarray}
     and

     \begin{eqnarray}
     \frac{d}{ds_1}H=  \frac{d}{ds_1}[ \frac{(p_{1,x})^2}{2m_1}+\frac{(p_{1,y})^2}{2m_1}+V_1(x_1(s_1),y_1(s_1))+H_C(s_1,s_2)]= \nonumber \\
     \frac{d}{ds_1}[ \frac{q^2}{((x_1(s_1)-x_2(s_2))^2+(f_1(x_1(s_1))-f_2(x_2(s_2)))^2)^{\frac{1}{2}}}] + \frac{d}{ds_1} [V_1(s_1)] 
     + \frac{d}{ds_1}[\frac{1}{2}m_1[(\frac{dx_1}{ds_1})^2+(\frac{dy_1}{ds_1})^2](\frac{ds_1}{dt})^2]= \nonumber \\
    -\frac{(  q^2[ ( ( x_1(s_1)-x_2(s_2) ) \frac{d}{ds_1} x_1(s_1) )+ (( f_1(x_1(s_1))-f_2(x_2(s_2)) ) \frac{d}{ds_1} f_1(s_1) ) ] )}{((x_1(s_1)-x_2(s_2))^2+(f_1(x_1(s_1))-f_2(x_2(s_2)))^2)^{\frac{3}{2}}} 
    + \nonumber \\
      \frac{d}{ds_1}[V_1(s_1)]+[m_1[(\frac{dx_1}{ds_1}\frac{d^2x_1}{ds_1^2})+(\frac{dy_1}{ds_1}\frac{d^2y_1}{ds_1^2})](\frac{ds_1}{dt})^2]= 
      -m_1(\frac{dx_1}{ds1})[(\frac{dx_1}{ds_1})^2\frac{d^2}{dt^2}s_1+\frac{d^2x_1}{ds_1^2}\frac{dx_1}{ds_1}\frac{d}{dt}s_1]. \nonumber \\
     \end{eqnarray}
     Dealing in similar fashion but for $s_2$ variable we obtain
     \begin{eqnarray}
    +\frac{(  q^2[ ( ( x_1(s_1)-x_2(s_2) ) \frac{d}{ds_1} x_2(s_2) )+ (( f_1(x_1(s_1))-f_2(x_2(s_2)) ) \frac{d}{ds_2} f_2(s_1) ) ] )}{((x_1(s_1)-x_2(s_2))^2+(f_1(x_1(s_1))-f_2(x_2(s_2)))^2)^{\frac{3}{2}}} 
    + \nonumber \\
      \frac{d}{ds_2}[V_2(s_2)]+[m_2[(\frac{dx_2}{ds_2}\frac{d^2x_2}{ds_2^2})+(\frac{dy_2}{ds_2}\frac{d^2y_2}{ds_2^2})](\frac{ds_2}{dt})^2]= 
      -m_2(\frac{dx_2}{ds_2})[(\frac{dx_2}{ds_2})^2\frac{d^2}{dt^2}s_2+\frac{d^2x_2}{ds_2^2}\frac{dx_2}{ds_2}\frac{d}{dt}s_2].
     \end{eqnarray}
     Last 2 coupled non-ODE differential equations can be written as
     \begin{eqnarray}
      +\frac{(  q^2[ ( ( x_1(s_1)-x_2(s_2) ) \frac{d}{ds_2} x_2(s_2) )+ (( f_1(x_1(s_1))-f_2(x_2(s_2)) ) \frac{dx_2}{ds_2}\frac{d}{dx_2} f_2(x_2(s_2)) ) ] )}{((x_1(s_1)-x_2(s_2))^2+(f_1(x_1(s_1))-f_2(x_2(s_2)))^2)^{\frac{3}{2}}} 
    + \nonumber \\
     + ( \frac{dx_2}{ds_2}\frac{d}{dx_2}V_2(x_2,y_2))= \nonumber \\
      =-[m_2[(\frac{dx_2}{ds_2}\frac{d^2x_2}{ds_2^2})+(\frac{dy_2}{ds_2}\frac{d^2y_2}{ds_2^2})](\frac{ds_2}{dt})^2]= 
      -m_2(\frac{dx_2}{ds_2})[(\frac{dx_2}{ds_2})^2\frac{d^2}{dt^2}s_2+\frac{d^2x_2}{ds_2^2}\frac{dx_2}{ds_2}\frac{d}{dt}s_2], \nonumber \\
      -\frac{(  q^2[ ( ( x_1(s_1)-x_2(s_2) ) \frac{d}{ds_1} x_1(s_1) )+ (( f_1(x_1(s_1))-f_2(x_2(s_2)) ) \frac{dx_1}{ds_1}\frac{d}{dx_1} f_1(x_1(s_1)) ) ] )}{((x_1(s_1)-x_2(s_2))^2+(f_1(x_1(s_1))-f_2(x_2(s_2)))^2)^{\frac{3}{2}}} 
    + \nonumber \\
      ((\frac{dx_1}{ds_1})\frac{d}{dx_1})[V_1(x_1,y_1)]+[m_1[(\frac{dx_1}{ds_1}\frac{d^2x_1}{ds_1^2})+(\frac{dy_1}{ds_1}\frac{d^2y_1}{ds_1^2})](\frac{ds_1}{dt})^2]= \nonumber \\
      -m_1(\frac{dx_1}{ds_1})[(\frac{dx_1}{ds_1})^2\frac{d^2}{dt^2}s_1+\frac{d^2x_1}{ds_1^2}\frac{dx_1}{ds_1}\frac{d}{dt}s_1]
     \end{eqnarray}
     We introduce $x_1'=\frac{dx_1}{ds_1}$, $x_2'=\frac{dx_2}{ds_2}$, $y_1'=\frac{dy_1}{ds_1}$ and $y_2'=\frac{dy_2}{ds_2}$.
     Last equations can be written as
     \begin{eqnarray} \label{cleqdiss1}
     \frac{1}{m_1}F_1(x_1(s_1),y_1(s_1),x_2(s_2),y_2(s_2))+
      \frac{1}{m_1}((\frac{dx_1}{ds_1})\frac{d}{dx_1})[V_1(x_1,y_1)]=F1Q(s_1,s_2)=\nonumber \\
      -(x_1'(s_1)x''(s_1)+y_1'(s_1)y_1''(s_1))(\frac{ds_1}{dt})^2
      -((x_1(s_1)')^3(\frac{d^2}{dt^2}s_1(t_1))-x_1''(s_1)(x_1'(s_1))^2(\frac{d}{dt}(s_1)) = \nonumber \\
      -g_{1,1}(s_1)(\frac{ds_1}{dt})^2-g_{1,2}(s_1)(\frac{d^2}{dt^2}s_1(t_1))-g_{1,3}(s_1)(\frac{d}{dt}(s_1))=F1Q(s_1,s_2), \nonumber \\
      -g_{2,1}(s_2)(\frac{ds_2}{dt})^2-g_{2,2}(s_2)(\frac{d^2}{dt^2}s_2(t_1))-g_{2,3}(s_2)(\frac{d}{dt}(s_2))=F2Q(s_1,s_2). \nonumber \\
     \end{eqnarray}
     Here we have
     \begin{eqnarray}
     g_{1,1}(s_1)=(x_1'(s_1)x''(s_1)+y_1'(s_1)y_1''(s_1)), \nonumber \\
     g_{1,2}(s_1)=((x_1(s_1)')^3, \nonumber \\
     g_{1,3}(s_1)=x_1''(s_1)(x_1'(s_1))^2. \nonumber \\
     \end{eqnarray}
     We can generalize the results for the case of 3 and N separated lines with charged balls in 2 and in 3 geometrical dimensions. In such case we have
     \begin{eqnarray}
      -g_{1,1}(s_1)(\frac{ds_1}{dt})^2-g_{1,2}(s_1)(\frac{d^2}{dt^2}s_1(t_1))-g_{1,3}(s_1)(\frac{d}{dt}(s_1))=F1QQ(s_1,s_2,s_3), \nonumber \\
      -g_{2,1}(s_2)(\frac{ds_2}{dt})^2-g_{2,2}(s_2)(\frac{d^2}{dt^2}s_2(t_1))-g_{2,3}(s_2)(\frac{d}{dt}(s_2))=F2QQ(s_1,s_2,s_3), \nonumber \\
      -g_{3,1}(s_2)(\frac{ds_3}{dt})^2-g_{3,2}(s_2)(\frac{d^2}{dt^2}s_3(t_1))-g_{3,3}(s_3)(\frac{d}{dt}(s_3))=F3QQ(s_1,s_2,s_3). \nonumber \\
     \end{eqnarray}
Let us solve the practical set of coupled non-linear ODE equations by setting $s_1=x_1$ and $s_2=x_2$, so we have
\begin{eqnarray}
  m_1\frac{d^2}{dt^2}x_1(t) &=& -m_1(\frac{d}{dx_1}f_1(x_1))(\frac{d^2}{dx_1^2}f_1(x_1))(\frac{d}{dt}x_1(t))^2-\frac{d}{dx_1}V_1(x_1,f_1(x_1))-\frac{d}{dx_1}\frac{q^2}{((x_1-x_2)^2+(f_1(x_1)-f_2(x_2)))^{\frac{1}{2}}}, \nonumber  \\
  m_2\frac{d^2}{dt^2}x_2(t) &=& -m_2(\frac{d}{dx_2}f_2(x_2))(\frac{d^2}{dx_2^2}f_2(x_2))(\frac{d}{dt}x_2(t))^2-\frac{d}{dx_2}V_2(x_2,f_2(x_2))-\frac{d}{dx_2}\frac{q^2}{((x_1-x_2)^2+(f_1(x_1)-f_2(x_2)))^{\frac{1}{2}}}, \nonumber  \\
  \end{eqnarray}
  where $V_1(x_1(t),y_1(t))$ and $V_2(x_2(t),y_2(t))$ are local confining potentials in nanowires. In particular we have set $V_1(x_1)= e^{0.1 \sqrt{x_1^2 +(f_1(x_1))^2 }}$ and
 $V_2(x_2)= e^{0.1 \sqrt{x_2^2 +(f_2(x_2))^2 }}$ with $f_1(x_1)=1 + Tanh(c 0.5 x_1)^2$ and $f_2(x_2)=1 + Tanh(c x_2)^2$. There is occurrence of additional "dissipative" terms in equations of motions as by expressions \newline $-m_1(\frac{d}{dx_1}f_1(x_1))(\frac{d^2}{dx_1^2}f_1(x_1))(\frac{d}{dt}x_1(t))^2$ and $ -m_2(\frac{d}{dx_2}f_2(x_2))(\frac{d^2}{dx_2^2}f_2(x_2))(\frac{d}{dt}x_2(t))^2$ due to non-zero nanocable's curvature.  One can observe the emergence of deterministic chaos as depicted in Fig. \ref{fig1q}.

\begin{eqnarray}\hat{CF}[p_1:R_k \rightarrow R_l,p_2:R_s \rightarrow R_w][t_1,t_2][\hat{O}_{p1}(t_1)][\hat{O}_{p2}(t_2)]=\hat{CF}[p_1:R_k \rightarrow R_l,p_2:R_s \rightarrow R_w][t_1,t_1+\Delta t]= \nonumber \\
\begin{pmatrix}
CF[R_1R_1,R_1R_1][t_1,t_2] & CF[R_1R_1,R_2R_1][t_1,t_2] & CF[R_2R_1,R_1R_1][t_1,t_2] & CF[R_2R_1,R_2R_1][t_1,t_2] \\
CF[R_1R_1,R_1R_2][t_1,t_2] & CF[R_1R_1,R_2R_2][t_1,t_2] & CF[R_2R_1,R_1R_2][t_1,t_2] & CF[R_2R_1,R_2R_2][t_1,t_2] \\
CF[R_1R_2,R_1R_1][t_1,t_2] & CF[R_1R_2,R_2R_1][t_1,t_2] & CF[R_2R_2,R_1R_1][t_1,t_2] & CF[R_2R_2,R_2R_1][t_1,t_2] \\
CF[R_1R_2,R_1R_2][t_1,t_2] & CF[R_1R_2,R_2R_2][t_1,t_2] & CF[R_2R_2,R_1R_2][t_1,t_2] & CF[R_2R_2,R_2R_2][t_1,t_2] \\
\end{pmatrix}
\end{eqnarray}
In particular case we can set $\hat{O}_{p1}(t_1)=\hat{I}_{p1}$ and $\hat{O}_{p2}(t_2)=\hat{I}_{p2}$. Setting both $\hat{O}_{p1}(t_1)$ and $\hat{O}_{p2}(t_2)$ to identity will give us mass flow between certain nodes of 2 Wannier qubit system.

We have
\begin{eqnarray}
CF[R_1R_1,R_1R_1][t_1,t_2]=\int_{x1 \in R_1}dx_1\int_{x1 \in R_1}dx_2\int_{x1 \in R_1}dx_3\int_{x1 \in R_1}dx_4|\psi(x_1,x_2,t_1)|^2|\psi(x_3,x_4,t_2)|^2= \nonumber \\
=prob(p1 \in R_1,p2 \in R_1,t_1)prob(p1 \in R_1,p2 \in R_1,t_2), \nonumber \\
CF[R_1R_1,R_2R_2][t_1,t_2]=\int_{x1 \in R_1}dx_1\int_{x1 \in R_1}dx_2\int_{x1 \in R_2}dx_3\int_{x1 \in R_2}dx_4|\psi(x_1,x_2,t_1)|^2|\psi(x_3,x_4,t_2)|^2= \nonumber \\
=prob(p1 \in R_1,p2 \in R_2,t_1)prob(p1 \in R_1,p2 \in R_2,t_2), \nonumber \\
CF[R_2R_2,R_1R_1][t_1,t_2]=\int_{x1 \in R_2}dx_1\int_{x1 \in R_2}dx_2\int_{x1 \in R_1}dx_3\int_{x1 \in R_1}dx_4|\psi(x_1,x_2,t_1)|^2|\psi(x_3,x_4,t_2)|^2= \nonumber \\
=prob(p2 \in R_2,p2 \in R_1,t_1)prob(p1 \in R_2,p2 \in R_1,t_2), \nonumber \\
CF[R_2R_2,R_2R_2][t_1,t_2]=\int_{x1 \in R_2}dx_1\int_{x1 \in R_2}dx_2\int_{x3 \in R_2}dx_3\int_{x4 \in R_2}dx_4|\psi(x_1,x_2,t_1)|^2|\psi(x_3,x_4,t_2)|^2= \nonumber \\
=prob(p1 \in R_2,p2 \in R_2,t_1)prob(p1 \in R_2,p2 \in R_2,t_2), \nonumber \\
\end{eqnarray}

When $\Delta t$ starts to increase all matrix terms start to oscillate and in case of stationary fields shall achieve certain oscillations periodicity T.

\begin{eqnarray}\hat{CFI}[p_1:R_k \rightarrow R_l,p_2:R_s \rightarrow R_w][t_1,t_2][\hat{O}_{p1}(t_1)][\hat{O}_{p2}(t_2)]=\hat{CF}[p_1:R_k \rightarrow R_l,p_2:R_s \rightarrow R_w][t_1,t_1+\Delta t]= \nonumber \\
\begin{pmatrix}
CF[R_1R_1,R_1R_1][t_1,t_2] & CF[R_1R_1,R_2R_1][t_1,t_2] & CF[R_2R_1,R_1R_1][t_1,t_2] & CF[R_2R_1,R_2R_1][t_1,t_2] \\
CF[R_1R_1,R_1R_2][t_1,t_2] & CF[R_1R_1,R_2R_2][t_1,t_2] & CF[R_2R_1,R_1R_2][t_1,t_2] & CF[R_2R_1,R_2R_2][t_1,t_2] \\
CF[R_1R_2,R_1R_1][t_1,t_2] & CF[R_1R_2,R_2R_1][t_1,t_2] & CF[R_2R_2,R_1R_1][t_1,t_2] & CF[R_2R_2,R_2R_1][t_1,t_2] \\
CF[R_1R_2,R_1R_2][t_1,t_2] & CF[R_1R_2,R_2R_2][t_1,t_2] & CF[R_2R_2,R_1R_2][t_1,t_2] & CF[R_2R_2,R_2R_2][t_1,t_2] \\
\end{pmatrix}.
\end{eqnarray}

This periodicity is not achieved in case of time-dependent external Wannier qubit system biasing fields (as electric or magnetic) that are not periodic.
\textbf{This correlation matrix mimics tight-binding Hamiltonian for 2 electrostatically coupled Wannier qubits. Yet it introduces anti-diagonal terms that are
non-zero and are correlated motion of two electrons in the same direction that is omitted in traditional tight-binding model. Hence validation of tight-binding model from
Schroedinger two coupled bodies formalism is always partial and is the simplification that is not fully accurate!!. The same reasoning can be conducted for 3 and N coupling
bodies as electrons.  }
\section{Conclusions}
The usage of hopping terms in tight-binding model was justified as by formula \ref{eqn:fundamental}. The prescription for exact computation of localized energy terms as $E_{p1}$ and $E_{p2}$ was given by formula \ref{given}. The origin of tight-binding model dissipation \cite{epidemic},\cite{dissipation} was identified in the framework of Schroedinger formalism. Various correlation functions using Schroedinger formalism and justifying the tight-binding model has been proposed.
The case of 2 Wannier qubit interaction was formulated in Schroedniger formalism using the straight and curvy semiconductor nanowires..
It consequently points to Hermitian and non-Hermitian Hamiltonian matrix formula. The effect or curvature of semiconductor quasi-one dimensional nanowires was expressed by proper equations of motion both in case of classical and quantum picture. It allows for modeling of quantum and classical SWAP gates using electron-electron interaction.
The effects of topology of open loop semiconductor nanowires can be studied by usage of Toeplitz matrix approach in different coordinate systems as by : Cartesian, cylindrical and spherical coordinates.
The conducted considerations allows for description of Wannier position based qubits with single and many electrons injected into source and drain of field effect transistor. They also give the base for modeling the quantum neural networks implemented by the chain of coupled quantum dots. The presented fundamental approach is useful in enhancement of tight-binding scheme as used in the design of quantum gates \cite{qgates}. The presented work is the extension of methodology given by \cite{Xu} as well as by \cite{Panos},\cite{Nbodies},\cite{Szafran}. The results obtained in Fig.3-14 for the case of Tanh Square cables shall be tested using Local Density of States observed in STM for different $\alpha$ coefficient. Various physical phenomena observed in condensed matter systems \cite{Spalek} can be simulated with the concept of quantum programmable matter demonstrated by position-dependent qubits controlled by electric signals \cite{QInternet},\cite{Jaynes}. Furthermore quantum machine learning can mimic any stochastic finite state machine by usage of tight-binding model as it was shown in \cite{epidemic}.


\begin{appendices}
\newpage

\section{Green functions in description of Wannier qubits} \label{App1}
In case of \textbf{propagator in time (but not in space)} the formula is given as
\setcounter{equation}{0}
\begin{eqnarray}
G_b(x_1,t_2,t_1)=(\hat{O}(x_1,t_2))^{-1}c \delta(t_2-t_1)=\frac{c \delta(t_2-t_1)}{\hat{O}(x_1,t_2)}=\frac{c \delta(t_2-t_1)}{i\hbar \frac{d}{dt_2}+\frac{\hbar^2}{2m}\frac{d^2}{dx_1^2}-V(x_1,t_2)+c}=\frac{c \delta(t_2-t_1)}{i\hbar \frac{d}{dt_2}-\hat{H}(x_1,t_2)+c}.
\end{eqnarray}
We notice that similar situation we have in case slightly different definition of propagators in time and in space, in space and in time given by $G_q$, $G_{aq}$ and $G_{bq}$ between finite points $(x_a,t_a)$ and $(x_b,t_b)$ as
\addtocounter{equation}{1}
\begin{eqnarray}
\int_{t_a}^{t_b} dt_1 \int_{x_a}^{x_b}dx_1 (G_q(x_b,t_b,x_1,t_1)\psi(x_1,t_1))=\psi(x_b,t_b),  \int_{x_a}^{x_b}dx_1 (G_{qa}(x_b,t,x_1,t)\psi(x_1,t))=\psi(x_b,t), \nonumber \\
\int_{t_a}^{t_b}dt_1 (G_{qb}(x,t_b,x,t_1)\psi(x,t_1))=\psi(x,t_b).
\end{eqnarray}
Consequently we obtain formulas for particle propagation in time or in space or both in time and in space in quasi-one dimensional world in the form as
\addtocounter{equation}{2}
\begin{eqnarray}
G_q(x_b,t_b,x_a,t_a)= \hat{O}(x_b,t_b)^{-1}[c \delta(x_b-x_a)\delta(t_b-t_a)]=\frac{c \delta(x_b-x_a)\delta(t_b-t_a)}{i\hbar \frac{d}{dt_b}+\frac{\hbar^2}{2m}\frac{d^2}{dx_b^2}-V(x_b,t_b)+c}=\frac{c \delta(x_b-x_a)\delta(t_b-t_a)}{i\hbar \frac{d}{dt_b}-\hat{H}(x_b,t_b)+c}, \nonumber \\
G_{qa}(x_b,t,x_a,t)= \hat{O}(x_b,t)^{-1}[c \delta(x_b-x_a)]=\frac{c \delta(x_b-x_a)}{i\hbar \frac{d}{dt}+\frac{\hbar^2}{2m}\frac{d^2}{dx_b^2}-V(x_b,t)+c}=\frac{c \delta(x_b-x_a)}{i\hbar \frac{d}{dt}-\hat{H}(x_b,t)+c}, \nonumber \\
G_{qb}(x,t_b,x,t_a)= \hat{O}(x,t_b)^{-1}[c \delta(t_b-t_a)]=\frac{c \delta(t_b-t_a)}{i\hbar \frac{d}{dt_b}+\frac{\hbar^2}{2m}\frac{d^2}{dx^2}-V(x,t_b)+c}=\frac{c \delta(t_b-t_a)}{i\hbar \frac{d}{dt_B}-\hat{H}(x,t_b)+c}.
\end{eqnarray}
It is quite straightforward to generalize results for 2 and 3 dimensional world so we have

\addtocounter{equation}{5}
\begin{eqnarray}
G_q(x_b,y_b,z_b,t_b,x_a,y_a,z_a,t_a)= \hat{O}(x_b,y_b,z_b,t_b)^{-1}[c \delta(x_b-x_a)\delta(y_b-y_a)\delta(z_b-z_a)\delta(t_b-t_a)]=\nonumber \\ =\frac{c \delta(x_b-x_a)\delta(y_b-y_a)\delta(z_b-z_a)\delta(t_b-t_a)}{i\hbar \frac{d}{dt_b}+\frac{\hbar^2}{2m}\frac{d^2}{dx_b^2}+\frac{\hbar^2}{2m}\frac{d^2}{dy_b^2}+\frac{\hbar^2}{2m}\frac{d^2}{dz_b^2}-V(x_b,y_b,z_b,t_b)+c} 
=\frac{c \delta(x_b-x_a)\delta(y_b-y_a)\delta(z_b-z_a)\delta(t_b-t_a)}{i\hbar \frac{d}{dt_b}-\hat{H}(x_b,y_b,z_b,t_b)+c}, \nonumber \\
G_{qa}(x_b,y_b,z_b,t,x_a,y_a,z_a,t)= \hat{O}(x_b,y_b,z_b,t)^{-1}[c \delta(x_b-x_a)\delta(y_b-y_a)\delta(z_b-z_a)]=\nonumber \\ =\frac{c \delta(x_b-x_a)\delta(y_b-y_a)\delta(z_b-z_a)}{i\hbar \frac{d}{dt}+\frac{\hbar^2}{2m}\frac{d^2}{dx_b^2}+\frac{\hbar^2}{2m}\frac{d^2}{dy_b^2}+\frac{\hbar^2}{2m}\frac{d^2}{dz_b^2}-V(x_b,y_b,z_b,t)+c} 
=\frac{c \delta(x_b-x_a)\delta(y_b-y_a)\delta(z_b-z_a)}{i\hbar \frac{d}{dt}-\hat{H}(x_b,y_b,z_b,t)+c}, \nonumber \\
G_{qb}(x,y,z,t_b,x_a,y_a,z_a,t_a)= \hat{O}(x,y,z,t_b)^{-1}[c \delta(t_b-t_a)]=\nonumber \\ =\frac{c \delta(t_b-t_a)}{i\hbar \frac{d}{dt_b}+\frac{\hbar^2}{2m}\frac{d^2}{dx_b^2}+\frac{\hbar^2}{2m}\frac{d^2}{dy_b^2}+\frac{\hbar^2}{2m}\frac{d^2}{dz_b^2}-V(x_b,y_b,z_b,t_b)+c} 
=\frac{c \delta(t_b-t_a)}{i\hbar \frac{d}{dt_b}-\hat{H}(x_b,y_b,z_b,t_b)+c}=\frac{\delta(t_b-t_a)}{i\hbar \frac{d}{dt_b}-\hat{H}(x_b,y_b,z_b,t_b)},
\end{eqnarray}
since $i\hbar \frac{d}{dt_b}-\hat{H}(x_b,y_b,z_b,t_b)=0$. Yet usage of non-zero c is stabilizing the numerical solutions since we can encounter dividing zero by zero etc.
Let us move from the case of 2 dimensional case of particle in effective potential into case of two interacting particles in 2 dimensional situation.
We have particle in 2 dimensional effective potential and thus equation of motion for operator $\hat{O}(x,y,t)$ is given as
\addtocounter{equation}{1}
\begin{eqnarray}
(i\hbar \frac{d}{dt}+\frac{\hbar^2}{2m}\frac{d^2}{dx^2} 
+\frac{\hbar^2}{2m}\frac{d^2}{dy^2}  \nonumber \\
-V(x,y,t)+
+c)\psi(x,y,t)=\hat{O}(x,y,t)\psi(x,y,t)=c \psi(x,y,t)
\end{eqnarray}
and in case of 2 interacting particles equation of motion for operator $\hat{O}(x_{p1},y_{p1},x_{p2},y_{p2},t)$ is given as
\addtocounter{equation}{3}

\begin{eqnarray}
[i\hbar \frac{d}{dt}+\frac{\hbar^2}{2m}(\frac{d^2}{dx_{p1}^2}+\frac{d^2}{dy_{p1}^2}+\frac{d^2}{dx_{p2}^2}+\frac{d^2}{dy_{p2}^2}) \nonumber \\
-V_{p2}(x_{p1},y_{p1},t)-V_{p2}(x_{p2},y_{p2},t)-V(x_{p1},y_{p1},x_{p2},y_{p2})+c] \psi(x_{p1},y_{p1},x_{p2},y_{p2},t) \nonumber \\
=\hat{O}(x_{p1},y_{p1},x_{p2},y_{p2},t)\psi(x_{p1},y_{p1},x_{p2},y_{p2},t) 
=c \psi(x_{p1},y_{p1},x_{p2},y_{p2},t). \nonumber \\
\end{eqnarray}
We have given definition of propagator for 2 interacting particles in 2 dimensions in time in the following way
\addtocounter{equation}{2}
\begin{eqnarray}
\int_{-\infty}^{+\infty} dt_1 \int_{-\infty}^{+\infty}dx_{Ap1}\int_{-\infty}^{+\infty}dy_{Ap1}\int_{-\infty}^{+\infty}dx_{Ap2}\int_{-\infty}^{+\infty}dy_{Ap2} (G(x_{Bp1},y_{Bp1},x_{Bp2},y_{Bp2},t_2,x_{Ap1},y_{Ap1},x_{Bp1},y_{Bp1},t_1)\times \nonumber \\ \times \psi(x_{Ap1},y_{Ap1},x_{Ap2},y_{Ap2},t_1))
=\psi(x_{Bp1},y_{Bp1},x_{Bp2},y_{Bp2},t_2)\nonumber \\
\end{eqnarray}
We obtain formula for 2 particle propagator in 2 dimensional world in time-dependent circumstances
\begin{eqnarray*}
G(x_{Bp1},y_{Bp1},x_{Bp2},y_{Bp2},t_2,x_{Ap1},y_{Ap1},x_{Bp1},y_{Bp1},t_1)= 
\nonumber \\
\frac{\delta(x_{Bp1}-x_{Ap1})\delta(y_{Bp1}-y_{Ap1})\delta(x_{Bp2}-x_{Ap2})\delta(y_{Bp2}-y_{Ap2})\delta(t_2-t_1)}{i\hbar \frac{d}{dt_2}+\frac{\hbar^2}{2m_{p1}}(\frac{d^2}{dx_{Bp1}^2}+\frac{d^2}{dy_{Bp1}^2})+\frac{\hbar^2}{2m_{p2}}(\frac{d^2}{dx_{Bp2}^2}+\frac{d^2}{dy_{Bp2}^2})-V(x_{Bp1},y_{Bp1},t_b)-V(x_{Bp2},y_{Bp2},t_b)-V(x_{Bp1},y_{Bp2},x_{Bp2},y_{Bp2},t_b)}\nonumber \\
=\frac{c\delta(x_{Bp1}-x_{Ap1})\delta(y_{Bp1}-y_{Ap1})\delta(x_{Bp2}-x_{Ap2})\delta(y_{Bp2}-y_{Ap2})\delta(t_2-t_1)}{i\hbar \frac{d}{dt_2}-\hat{H}(x_{Bp1},y_{Bp1},t_2)-\hat{H}(x_{Bp2},y_{Bp2},t_2)-\hat{H}_{int(p1-p2)}(x_{Bp1},y_{Bp1},x_{Bp2},y_{Bp2},t_2)+c}=. \nonumber \\
\frac{\delta(x_{Bp1}-x_{Ap1})\delta(y_{Bp1}-y_{Ap1})\delta(x_{Bp2}-x_{Ap2})\delta(y_{Bp2}-y_{Ap2})\delta(t_2-t_1)}{i\hbar \frac{d}{dt_2}-\hat{H}(x_{Bp1},y_{Bp1},t_2)-\hat{H}(x_{Bp2},y_{Bp2},t_2)-\hat{H}_{int(p1-p2)}(x_{Bp1},y_{Bp1},x_{Bp2},y_{Bp2},t_2)}.
\end{eqnarray*}
It will turn later that the most useful representation of propagator in space and in time is by Fourier transform. We start from propagator equation
\addtocounter{equation}{5}
\begin{eqnarray}
(i\hbar\frac{d}{dt_2}-\hat{H}(x_2,t_2))G(x_2,t_2,x_1,t_1) = (i\frac{d}{dt_2}+\frac{\hbar^2}{2m}(\frac{d^2}{dx_2^2})-V(x_2,t_2))G(x_2,t_2,x_1,t_1) = \delta(t_2-t_1)\delta(x_2-x_1), 
\end{eqnarray}
that can be converted by Fourier transform into more convenient form.
We use can bring picture of physical situation from position and time into momentum and frequency given in the way as
\addtocounter{equation}{1}
\begin{eqnarray}
\int_{-\infty}^{+\infty}d\omega_1 \frac{1}{\sqrt{2\pi}}\int_{-\infty}^{+\infty}\frac{1}{\sqrt{2\pi}}d\omega_2 e^{-i(\omega_2t_2-\omega_1t_1)}\hat{1}=\delta(t_2-t_1), \nonumber
\int_{-\infty}^{+\infty}\frac{1}{\sqrt{2\pi}}dk_1\int_{-\infty}^{+\infty}\frac{1}{\sqrt{2\pi}}dk_2 e^{-i(k_2 x_2-k_1x_1)}\hat{1}=\delta(x_2-x_1), \nonumber \\
G(x_2,t_2,x_1,t_1)=(\frac{1}{\sqrt{2\pi}})^4\int_{-\infty}^{+\infty}dk_2\int_{-\infty}^{+\infty}dk_1\int_{-\infty}^{+\infty}d\omega_2\int_{-\infty}^{+\infty}d\omega_1 e^{-i k_1 x_1} e^{-i k_2 x_2}e^{-i \omega_1 t_1}e^{-i \omega_2 x_2}G(k_2,\omega_2,k_1,\omega_1).
\end{eqnarray}
Using convolution scheme we have
\addtocounter{equation}{2}
\begin{eqnarray}
V(x_2,t_2)G(x_2,t_2,x_1,t_1)= \nonumber \\
=\frac{1}{(2\pi)^3}\int_{-\infty}^{+\infty} \int_{-\infty}^{+\infty} \int_{-\infty}^{+\infty} \int_{-\infty}^{+\infty} \int_{-\infty}^{+\infty} \int_{-\infty}^{+\infty}
dk_1dk_2d\omega_1d\omega_2d\omega_a dk_a e^{-jk_ax_2}e^{-jt_2\omega_a}e^{-jk_1x_1}e^{-jt_1\omega_1}\times \nonumber \\
\times V(k_a-k_2,\omega_a-\omega_2)G(k_2,\omega_2,k_1,\omega_1).
\end{eqnarray}
and it leads to the equation
\addtocounter{equation}{1}
\begin{eqnarray}
  (\hbar \omega_2+\frac{\hbar^2k_2^2}{2m})G(\omega_2,k_2,\omega_1,k_1)-\frac{1}{2\pi}\int_{-\infty}^{+\infty} \int_{-\infty}^{+\infty}d\omega_adk_aV(k_a-k_2,\omega_a-\omega_2)
  G(k_2,\omega_2,k_1,\omega_1)e^{-ix_2(k_a-k_2)}e^{-it_2(\omega_a-\omega_2)}=1. \nonumber \\
\end{eqnarray}
that finally implies position and momentum dependence of Green function for single particle in effective field in the form as
\addtocounter{equation}{1}
\begin{eqnarray}
G(k_2,\omega_2,k_1,\omega_1)=\frac{1}{
  (\hbar \omega_2+\frac{\hbar^2k_2^2}{2m})-\frac{1}{2\pi}\int_{-\infty}^{+\infty} \int_{-\infty}^{+\infty}d\omega_adk_aV(k_a-k_2,\omega_a-\omega_2)
  e^{-ix_2(k_a-k_2)}e^{-it_2(\omega_a-\omega_2)}}. \nonumber \\
\end{eqnarray}
and quite obviously we obtain
\addtocounter{equation}{2}
\begin{eqnarray}
G(x_2,t_2,x_1,t_1)=\nonumber \\
=\int_{-\infty}^{+\infty} \int_{-\infty}^{+\infty} \int_{-\infty}^{+\infty} \int_{-\infty}^{+\infty} \frac{dk_1dk_2 d\omega_1 d\omega_2 e^{i(k_1x_1+k_2x_2+k_3x_3+k_4x_4)}}{[
  (\hbar \omega_2+\frac{\hbar^2k_2^2}{2m})-\frac{1}{2\pi}\int_{-\infty}^{+\infty} \int_{-\infty}^{+\infty}d\omega_adk_aV(k_a-k_2,\omega_a-\omega_2)
  e^{-ix_2(k_a-k_2)}e^{-it_2(\omega_a-\omega_2)}](2\pi)^2}. \nonumber \\
\end{eqnarray}
Consequently we obtain
\addtocounter{equation}{1}
\begin{eqnarray*}
\psi(x_2,t_2)=
\nonumber \\
\int_{x_1}^{x_2} dx_1 \int_{t_1}^{t_2} dt_1 \int_{-\infty}^{+\infty} \int_{-\infty}^{+\infty} \int_{-\infty}^{+\infty} \int_{-\infty}^{+\infty} \frac{dk_1dk_2 d\omega_1 d\omega_2 e^{i(k_1x_1+k_2x_2+k_3x_3+k_4x_4)}\psi(x_1,t_1)\frac{1}{(2\pi)^2}}{[
  (\hbar \omega_2+\frac{\hbar^2k_2^2}{2m})-\frac{1}{2\pi}\int_{-\infty}^{+\infty} \int_{-\infty}^{+\infty}d\omega_adk_aV(k_a-k_2,\omega_a-\omega_2)
  e^{-ix_2(k_a-k_2)}e^{-it_2(\omega_a-\omega_2)}]}. 
\end{eqnarray*}
If we have position based qubit distributed between areas $R_1=(x_a,\frac{x_a+x_b}{2})$ and $R_2=(\frac{x_a+x_b}{2},x_2)$ we have the ratio of probabilities between electron presence in area $R_1$ denoted by $p_{R1}$ and area $R_2$ denoted by $p_{R2}$. 
Consequently we obtain the equation for Wannier qubit in time-dependent fields given as
\addtocounter{equation}{3}
\begin{eqnarray}
\frac{p_{R2}(t_2)}{p_{R1}(t_2)}=
\frac{\int_{\frac{x_a+x_b}{2}}^{x_b}|\psi(x_2,t_2)|^2dx_2}{\int_{x_a}^{\frac{x_a+x_b}{2}}|\psi(x_2,t_2)|^2dx_2}=
\nonumber \\
\frac{\int_{\frac{x_a+x_b}{2}}^{x_b}dx_2|\int_{x_1}^{x_2} dx_1 \int_{t_1}^{t_2} dt_1 \int_{-\infty}^{+\infty} \int_{-\infty}^{+\infty} \int_{-\infty}^{+\infty} \int_{-\infty}^{+\infty} \frac{dk_1dk_2 d\omega_1 d\omega_2 e^{i(k_1x_1+k_2x_2+k_3x_3+k_4x_4)}\psi(x_1,t_1)}{[
  (\hbar \omega_2+\frac{\hbar^2k_2^2}{2m})-\frac{1}{2\pi}\int_{-\infty}^{+\infty} \int_{-\infty}^{+\infty}d\omega_adk_aV(k_a-k_2,\omega_a-\omega_2)
  e^{-ix_2(k_a-k_2)}e^{-it_2(\omega_a-\omega_2)}]}|^2}{\int_{x_a}^{\frac{x_a+x_b}{2}}dx_2|\int_{x_1}^{x_2} dx_1 \int_{t_1}^{t_2} dt_1 \int_{-\infty}^{+\infty} \int_{-\infty}^{+\infty} \int_{-\infty}^{+\infty} \int_{-\infty}^{+\infty} \frac{dk_1dk_2 d\omega_1 d\omega_2 e^{i(k_1x_1+k_2x_2+k_3x_3+k_4x_4)}\psi(x_1,t_1)}{[
  (\hbar \omega_2+\frac{\hbar^2k_2^2}{2m})-\frac{1}{2\pi}\int_{-\infty}^{+\infty} \int_{-\infty}^{+\infty}d\omega_adk_aV(k_a-k_2,\omega_a-\omega_2)
  e^{-ix_2(k_a-k_2)}e^{-it_2(\omega_a-\omega_2)}]}|^2}. 
  \label{eqnq}
\end{eqnarray}
We have given dependence of quantum probabilities at given time instant $t_2$ on all time instants from time $t_1$ that is incorporated only on V(x,t) potential and initial state $\psi(x,t)$.

\section{Single particle correlation functions in Schroedinger formalism}

\small
\begin{eqnarray}
t_{s12}=t_{R_1 \rightarrow R_2}(t_1,t_2)=t_{R_1 \rightarrow R_2}(t_1,t_1+\Delta t) =\int_{x_a}^{\frac{x_a+x_b}{2}} dx_1 \int_{\frac{x_a+x_b}{2}}^{x_b} dx_2 \psi^{*}(x_1,t_1)\hat{H}(x_2,t_2)\psi(x_2,t_2)= \nonumber \\
t_{R_1 \rightarrow R_2}(t_1,t_2)=t_{R_1 \rightarrow R_2}(t_1,t_1+\Delta t) =\int_{x_a}^{\frac{x_a+x_b}{2}} dx_1 \int_{\frac{x_a+x_b}{2}}^{x_b} dx_2 \psi^{*}(x_1,t_1)\hat{H}(x_2,t_2)\int_{-\infty}^{+\infty}\int_{-\infty}^{+\infty}dx_3dt_3G(x_2,t_2,x_3,t_3)\psi(x_3,t_3)= \nonumber \\
=\int_{x_a}^{\frac{x_a+x_b}{2}} dx_1 \int_{\frac{x_a+x_b}{2}}^{x_b} dx_2 \psi^{*}(x_1,t_1)\hat{H}\psi(x_1+(x_2-x_1),t_1+(t_2-t_1))=\nonumber \\
=\int_{x_a}^{\frac{x_a+x_b}{2}} dx_1 \int_{\frac{x_a+x_b}{2}}^{x_b} dx_2 \psi^{*}(x_1,t_1)E(t_2)\psi(x_1+(x_2-x_1),t_1+(t_2-t_1))=\nonumber \\
=E(t_2)\int_{x_a}^{\frac{x_a+x_b}{2}} dx_1 \int_{\frac{x_a+x_b}{2}}^{x_b} dx_2 \psi^{*}(x_1,t_1)e^{\frac{i}{\hbar}(x_2-x_1)\hat{p}(x_1,t_1)}e^{-\frac{i}{\hbar}(t_2-t_1)\hat{H}(t_1)} 
\psi(x_1,t_1)
=\nonumber \\
=E(t_2)e^{-\frac{i}{\hbar}(t_2-t_1)E(t_1)}\int_{x_a}^{\frac{x_a+x_b}{2}} dx_1 \int_{\frac{x_a+x_b}{2}}^{x_b} dx_2 \psi^{*}(x_1,t_1)e^{\frac{i}{\hbar}(x_2-x_1)\hat{p}(x_1,t_1)} 
\psi(x_1,t_1)
=\nonumber \\
=E(t_2)e^{-\frac{i}{\hbar}(t_2-t_1)E(t_1)}\int_{x_a}^{\frac{x_a+x_b}{2}} dx_1 \int_{\frac{x_a+x_b}{2}-x_1}^{x_b-x_a} d \Delta x \psi^{*}(x_1,t_1)e^{\frac{i}{\hbar}\Delta x \hat{p}(x_1,t_1)} 
\psi(x_1,t_1)
=\nonumber \\
=E(t_2)e^{-\frac{i}{\hbar}(t_2-t_1)E(t_1)}[\int_{x_a}^{\frac{x_a+x_b}{2}} dx_1  [ \psi^{*}(x_1,t_1)\frac{\hbar}{i \hat{p}(x_1,t_1)}e^{\frac{i}{\hbar}\Delta x \hat{p}(x_1,t_1)}\psi(x_1,t_1)]_{\Delta x=\frac{x_a+x_b}{2}-x_1}^{\Delta x=x_b-x_a} 
=\nonumber \\
=E(t_2)e^{-\frac{i}{\hbar}(t_2-t_1)E(t_1)}[\int_{x_a}^{\frac{x_a+x_b}{2}} dx_1  [ \psi^{*}(x_1,t_1)\frac{\hbar}{i \hat{p}(x_1,t_1)}((e^{\frac{i}{\hbar}(x_b-x_a) \hat{p}(x_1,t_1)}-e^{\frac{i}{\hbar}(\frac{x_b+x_a}{2}-x_1) \hat{p}(x_1,t_1)})\psi(x_1,t_1))]= 
=\nonumber \\
=\frac{\hbar}{i}E(t_2)e^{-\frac{i}{\hbar}(t_2-t_1)E(t_1)}(<\frac{e^{\frac{i}{\hbar}(x_b-x_a) \hat{p}(x,t_1)}}{\hat{p}(x,t_1)}>_{x\in R_1}-<\frac{e^{\frac{\hbar}{i}(\frac{x_b+x_a}{2}-x)p(x,t_1)}}{\hat{p}(x,t_1)}>_{x\in R_1})= \nonumber \\
=\frac{\hbar}{i}E(t_2)e^{-\frac{i}{\hbar}(t_2-t_1)E(t_1)}(<\frac{Cos(\frac{i}{\hbar}(x_b-x_a) \hat{p}(x,t_1))-Cos(\frac{\hbar}{i}(\frac{x_b+x_a}{2}-x)p(x,t_1))}{\hat{p}(x,t_1)}>_{x\in R_1} \nonumber \\
+i<\frac{Sin(\frac{i}{\hbar}(x_b-x_a) \hat{p}(x,t_1))-Sin(\frac{\hbar}{i}(\frac{x_b+x_a}{2}-x)p(x,t_1))}{\hat{p}(x,t_1)}>_{x\in R_1})=
\nonumber \\
=\frac{\hbar}{i}E(t_2)e^{-\frac{i}{\hbar}(t_2-t_1)E(t_1)}(<2\frac{Sin(\frac{i}{\hbar}(\frac{3}{4}x_b-\frac{1}{4}x_a-\frac{1}{2}x) \hat{p}(x,t_1))
Sin(\frac{\hbar}{i}(\frac{3}{4}x_a-\frac{1}{4}x_b-x)p(x,t_1))}{\hat{p}(x,t_1)}>_{x\in R_1} \nonumber \\
-i2<\frac{Sin(\frac{i}{\hbar}(\frac{x_b}{4}-\frac{3}{4}x_a+\frac{x}{2}) \hat{p}(x,t_1))Cos(\frac{\hbar}{i}(\frac{3}{4}x_b-\frac{1}{4}x_a-\frac{x}{2})p(x,t_1))}{\hat{p}(x,t_1)}>_{x\in R_1})= \nonumber \\
=\int_{x_a}^{\frac{x_a+x_b}{2}} dx_1 \int_{\frac{x_a+x_b}{2}}^{x_b} dx_2 \psi^{*}(x_1,t_1)\sum_nE_n(t_2)c_n(t_2)\psi(x_1+(x_2-x_1),t_1+(t_2-t_1))=\nonumber \\
=\int_{x_a}^{\frac{x_a+x_b}{2}} dx_1 \int_{\frac{x_a+x_b}{2}}^{x_b} dx_2 \psi^{*}(x_1,t_1)\sum_n c_n(t_2) E_n(t_2)e^{-(t_2-t_1) \frac{i}{\hbar}E_n(t_1)}(e^{(x_2-x_1) \frac{i}{\hbar}\frac{\hbar}{i}\frac{d}{dx_1}}\psi_n(x_1,t_1))= \nonumber \\
=\int_{x_a}^{\frac{x_a+x_b}{2}} dx_1 \int_{\frac{x_a+x_b}{2}}^{x_b} dx_2 \psi^{*}(x_1,t_1)\sum_n c_n(t_2) E_n(t_2)e^{-(t_2-t_1) \frac{i}{\hbar}E_n(t_1)}(e^{(x_2-x_1) \frac{d}{dx_1}}\psi_n(x_1,t_1))= \nonumber \\
=\sum_n c_n(t_2) E_n(t_2)e^{-(t_2-t_1) \frac{i}{\hbar}E_n(t_1)}\int_{x_a}^{\frac{x_a+x_b}{2}} dx_1 \int_{\frac{x_b-x_a}{2}-x_1}^{x_b-x_a} d\Delta x \psi^{*}(x_1,t_1)(e^{\Delta x \frac{d}{dx_1}}\psi_n(x_1,t_1))= \nonumber \\
=\sum_n c_n(t_2) E_n(t_2)e^{-(t_2-t_1) \frac{i}{\hbar}E_n(t_1)}\int_{x_a}^{\frac{x_a+x_b}{2}} dx_1  [ \psi^{*}(x_1,t_1)((\frac{1}{\frac{d}{dx_1}}e^{\Delta x \frac{d}{dx_1}})\psi_n(x_1,t_1))]_{\Delta x=\frac{x_b-x_a}{2}-x_1}^{\Delta x=x_b-x_a}= \nonumber \\
=\sum_n c_n(t_2) E_n(t_2)e^{-(t_2-t_1) \frac{i}{\hbar}E_n(t_1)}\int_{x_a}^{\frac{x_a+x_b}{2}} dx_1 [\psi^{*}(x_1,t_1)(\frac{\psi_n(x_1,t_1)}{(\frac{d}{dx_1}\psi_n(x_1,t_1))}[e^{\Delta x \frac{d}{dx_1}}\psi_n(x_1,t_1)])]_{\Delta x=\frac{x_b-x_a}{2}-x_1}^{\Delta x=x_b-x_a}= \nonumber \\
=\sum_{n=0}^{+\infty} \sum_{k=0}^{+\infty} c_{E_n(t_2)}c_{E_k(t_1)}^{*} E_n(t_2)e^{-(t_2-t_1) \frac{i}{\hbar}E_n(t_1)}\int_{x_a}^{\frac{x_a+x_b}{2}} dx_1 [\psi_k^{*}(x_1,t_1)(\frac{\psi_n(x_1,t_1)}{\frac{d}{dx_1}\psi_n(x_1,t_1)}e^{\Delta x \frac{d}{dx_1} }\psi_n(x_1,t_1))]_{\Delta x=\frac{x_b-x_a}{2}-x_1}^{\Delta x=x_b-x_a}= \nonumber \\
=\sum_{n=0}^{+\infty} \sum_{k=0}^{+\infty} c_{E_n(t_2)}c_{E_k(t_1)}^{*} E_n(t_2)e^{-(t_2-t_1) \frac{i}{\hbar}E_n(t_1)}\int_{x_a}^{\frac{x_a+x_b}{2}} dx_1 [\psi_k^{*}(x_1,t_1)(\frac{\psi_n(x_1,t_1)}{\frac{d}{dx_1}\psi_n(x_1,t_1)}(e^{(x_b-x_a)\frac{d}{dx_1}}-e^{(\frac{x_b+x_a}{2}-x_1)\frac{d}{dx_1}})\psi_n(x_1,t_1))]= \nonumber \\
=\sum_{n=0}^{+\infty} \sum_{k=0}^{+\infty} c_{E_n(t_2)}c_{E_k(t_1)}^{*} E_n(t_2)e^{-(t_2-t_1) \frac{i}{\hbar}E_n(t_1)} \times \nonumber \\
\times\int_{x_a}^{\frac{x_a+x_b}{2}} dx_1 [\psi_k^{*}(x_1,t_1)(\frac{\psi_n(x_1,t_1)}{\frac{d}{dx_1}\psi_n(x_1,t_1)}(e^{(x_b-x_a)\frac{d}{dx_1}}-e^{(\frac{x_b+x_a}{2}-x_1)\frac{d}{dx_1}})\psi_n(x_1,t_1))].
\end{eqnarray}

Doing in the same fashion we can obtain the time dependent value of operator $E_{p1}$ in the form as
\begin{eqnarray}
E_{p1}=t_{R_1 \rightarrow R_1}(t_1,t_2)=t_{R_1 \rightarrow R_1}(t_1,t_1+\Delta t) =\int_{x_a}^{\frac{x_a+x_b}{2}} dx_1 \int_{x_a}^{\frac{x_a+x_b}{2}} dx_2 \psi^{*}(x_1,t_1)\hat{H}(x_2,t_2)\psi(x_2,t_2)= \nonumber \\
=\int_{x_a}^{\frac{x_a+x_b}{2}} dx_1 \int_{x_a}^{\frac{x_a+x_b}{2}} dx_2 \psi^{*}(x_1,t_1)\hat{H}(x_2,t_2)\int_{-\infty}^{+\infty}\int_{-\infty}^{+\infty}dx_3dt_3G(x_2,t_2,x_3,t_3)\psi(x_3,t_3)= \nonumber \\
=\int_{x_a}^{\frac{x_a+x_b}{2}} dx_1 \int_{x_a}^{\frac{x_a+x_b}{2}} dx_2 \psi^{*}(x_1,t_1)\hat{H}(x_2,t_2)\int_{t_1}^{t_2}\int_{x_1}^{x_2}dx_3dt_3G_q(x_2,t_2,x_3,t_3)\psi(x_3,t_3)= \nonumber \\
\int_{x_a}^{\frac{x_a+x_b}{2}} dx_1 \int_{x_a}^{\frac{x_a+x_b}{2}} dx_2 \psi^{*}(x_1,t_1)\hat{H}(x_2,t_2)\psi(x_1+(x_2-x_1),t_1+(t_2-t_1))=\nonumber \\
=\int_{x_a}^{\frac{x_a+x_b}{2}} dx_1 \int_{x_a}^{\frac{x_a+x_b}{2}} dx_2 \psi^{*}(x_1,t_1)E(t_2)((e^{(x_2-x_1) \frac{i}{\hbar}\hat{p}(x_1,t_1)}e^{-(t_2-t_1) (\frac{i}{\hbar}\hat{H}(t_1)}\psi(x_1,t_1))))=\nonumber \\
=E(t_2)e^{-(t_2-t_1) \frac{i}{\hbar}E(t_1)}\int_{x_a}^{\frac{x_a+x_b}{2}} dx_1 \int_{x_a}^{\frac{x_a+x_b}{2}} dx_2 \psi^{*}(x_1,t_1)((e^{(x_2-x_1) \frac{i}{\hbar}\hat{p}(x_1,t_1)}\psi(x_1,t_1)))=\nonumber \\
=\int_{x_a}^{\frac{x_a+x_b}{2}} dx_1 \int_{-x_1+x_a}^{\frac{x_a+x_b}{2}-x_1} d\Delta x \psi^{*}(x_1,t_1) E(t_2)e^{-(t_2-t_1) \frac{i}{\hbar}E(t_1)}(e^{\Delta x \frac{i}{\hbar}\hat{p}(x_1,t_1)}\psi(x_1,t_1))= \nonumber \\
=\int_{x_a}^{\frac{x_a+x_b}{2}} dx_1 \psi^{*}(x_1,t_1) E(t_2)e^{-(t_2-t_1) \frac{i}{\hbar}E(t_1)}[\frac{\hbar}{i\hat{p}(x_1,t_1)}e^{\Delta x \frac{i}{\hbar}\hat{p}(x_1,t_1)}\psi(x_1,t_1)]_{\Delta x=-x_1+x_a}^{\Delta x=\frac{x_a+x_b}{2}-x_1}= \nonumber \\
=\int_{x_a}^{\frac{x_a+x_b}{2}} dx_1 \psi^{*}(x_1,t_1) E(t_2)e^{-(t_2-t_1) \frac{i}{\hbar}E(t_1)}[\frac{\hbar}{i\hat{p}(x_1,t_1)}(e^{(\frac{x_a+x_b}{2}-x_1) \frac{i}{\hbar}\hat{p}(x_1,t_1)}-e^{(-x_1+x_a) \frac{i}{\hbar}\hat{p}(x_1,t_1)})\psi(x_1,t_1)]= \nonumber \\
=\int_{x_a}^{\frac{x_a+x_b}{2}} dx_1 \psi^{*}(x_1,t_1) E(t_2)e^{-(t_2-t_1) \frac{i}{\hbar}E(t_1)}[\frac{\psi(x_1,t_1)}{\frac{d}{dx_1}\psi(x_1,t_1)}(e^{(\frac{x_a+x_b}{2}-x_1) \frac{d}{dx_1}}-e^{(-x_1+x_a) \frac{d}{dx_1}})\psi(x_1,t_1)]= \nonumber \\
=\int_{x_a}^{\frac{x_a+x_b}{2}} dx_1 \psi^{*}(x_1,t_1) E(t_2)e^{-(t_2-t_1) \frac{i}{\hbar}E(t_1)}[\frac{\psi(x_1,t_1)}{\frac{d}{dx_1}\psi(x_1,t_1)}(e^{(\frac{x_a+x_b}{2}-x_1) \frac{\frac{d}{dx_1}\psi(x_1,t_1)}{\psi(x_1,t_1)}}-e^{(-x_1+x_a)(\frac{\frac{d}{dx_1}\psi(x_1,t_1)}{\psi(x_1,t_1)})})\psi(x_1,t_1)]= \nonumber \\
=E_{p1}(t_1,t_1+\Delta t)=
E(t_2)e^{-\Delta t \frac{i}{\hbar}E(t_1)}\langle [\frac{\hbar}{i\hat{p}(t_1)}(e^{(\frac{x_a+x_b}{2}-x) \frac{i}{\hbar}\hat{p}(t_1)}-e^{(-x+x_a) \frac{i}{\hbar}\hat{p}(t_1)})]
\rangle_{x \in R_1}= \nonumber \\
\sum_n c_n(t_2)\int_{x_a}^{\frac{x_a+x_b}{2}} dx_1 \psi^{*}(x_1,t_1) E_n(t_2)e^{-(t_2-t_1) \frac{i}{\hbar}E_n(t_1)}[\frac{\hbar}{i\hat{p}_n(x_1,t_1)}(e^{(\frac{x_a+x_b}{2}-x_1) \frac{i}{\hbar}\hat{p}_n(x_1,t_1)}-e^{(-x_1+x_a) \frac{i}{\hbar}\hat{p}_n(x_1,t_1)})\psi_n(x_1,t_1)]= \nonumber \\
=\sum_n c_n(t_2)\int_{x_a}^{\frac{x_a+x_b}{2}} dx_1 \psi^{*}(x_1,t_1) E_n(t_2)e^{-(t_2-t_1) \frac{i}{\hbar}E(t_1)}[\frac{1}{\frac{d}{dx_1}}(e^{(\frac{x_a+x_b}{2}-x_1) \frac{d}{dx_1}}-e^{(-x_1+x_a) \frac{d}{dx_1}})\psi_n(x_1,t_1)]= \nonumber \\
\sum_n c_n(t_2)\int_{x_a}^{\frac{x_a+x_b}{2}} dx_1 \psi^{*}(x_1,t_1) E_n(t_2)e^{-(t_2-t_1) \frac{i}{\hbar}E_k(t_1)}[\frac{\psi_n(x_1,t_1)}{\frac{d}{dx_1}\psi_n(x_1,t_1)}(e^{(\frac{x_a+x_b}{2}-x_1) \frac{d}{dx_1}}-e^{(-x_1+x_a) \frac{d}{dx_1}})\psi_n(x_1,t_1)]= \nonumber \\
\sum_{k,n=0}^{k=+\infty}c_n(t_2)c_k^{*}(t_1)\int_{x_a}^{\frac{x_a+x_b}{2}} dx_1 \psi_k^{*}(x_1,t_1) E_n(t_2)e^{-(t_2-t_1) \frac{i}{\hbar}E_n(t_1)}[\frac{\psi_n(x_1,t_1)}{\frac{d}{dx_1}\psi_n(x_1,t_1)}(e^{(\frac{x_a+x_b}{2}-x_1) \frac{d}{dx_1}}-e^{(-x_1+x_a) \frac{d}{dx_1}})\psi_n(x_1,t_1)]= \nonumber \\
=\sum_{n,k=0}^{n,k=+\infty} \sqrt{p_{E_n}(t_2)p_{E_k}(t_1)}e^{i(\phi_{E_n}(t_2)-\phi_{E_n}(t_1))}\times \nonumber \\ \times \int_{x_a}^{\frac{x_a+x_b}{2}} dx_1 \psi_k^{*}(x_1,t_1) E_n(t_2)e^{-(t_2-t_1) \frac{i}{\hbar}E(t_1)}[\frac{\psi_n(x_1,t_1)}{\frac{d}{dx_1}\psi_n(x_1,t_1)}(e^{(\frac{x_a+x_b}{2}-x_1) \frac{d}{dx_1}}-e^{(-x_1+x_a) \frac{d}{dx_1}})\psi_n(x_1,t_1)]
= \nonumber \\ \hbar E(t_2)e^{-\Delta t \frac{i}{\hbar}E(t_1)}
( \langle
 \frac{Cos((\frac{x_a+x_b}{2}-x)\hat{p}(t_1))-Cos((-x+x_a)\hat{p}(t_1))}{i\hat{p}(t_1)}
\rangle_{R_1} 
+i \langle
 \frac{Sin((\frac{x_a+x_b}{2}-x)\hat{p}(t_1))-Sin((-x+x_a)\hat{p}(t_1))}{i\hat{p}(t_1)}
\rangle_{R_1} ) \nonumber \\
=2\hbar E(t_2)e^{-\Delta t \frac{i}{\hbar}E(t_1)}
( \langle
 \frac{Sin((\frac{x_b+3x_a}{4}-x)\hat{p}(t_1))Sin((x+\frac{x_a+x_b}{4})\hat{p}(t_1))}{i\hat{p}(t_1)}
\rangle 
+i \langle
 \frac{Cos((\frac{x_b+3x_a}{4}-x)\hat{p}(t_1))Sin((+\frac{x_b-x_a}{4}-x)\hat{p}(t_1))}{i\hat{p}(t_1)}
\rangle ) \nonumber \\
=\hbar E(t_2)e^{-\Delta t \frac{i}{\hbar}E(t_1)} \Bigg[ e^{iArcTan(\frac{\frac{Cos((\frac{x_b+3x_a}{4}-x)\hat{p}(t_1))Sin((+\frac{x_b-x_a}{4}-x)\hat{p}(t_1))}{i\hat{p}(t_1)}}{\langle
 \frac{Sin((\frac{x_b+3x_a}{4}-x)\hat{p}(t_1))Sin((x+\frac{x_a=x_b}{4})\hat{p}(t_1))}{i\hat{p}(t_1)}|\langle [\frac{\hbar}{i\hat{p}(t_1)}(e^{(\frac{x_a+x_b}{2}-x) \frac{i}{\hbar}\hat{p}(t_1)}-e^{(-x+x_a) \frac{i}{\hbar}\hat{p}(t_1)})]
\rangle
\rangle})} \Bigg] \times \nonumber \\ \times |\langle [\frac{\hbar}{i\hat{p}(t_1)}(e^{(\frac{x_a+x_b}{2}-x) \frac{i}{\hbar}\hat{p}(t_1)}-e^{(-x+x_a) \frac{i}{\hbar}\hat{p}(t_1)})]
\rangle|
\label{given}
\end{eqnarray}

It is quite obvious to obtain
\begin{eqnarray}
E_{p2}=t_{R_2 \rightarrow R_2}(t_1,t_2)=t_{R_2 \rightarrow R_2}(t_1,t_1+\Delta t) =\int_{\frac{x_a+x_b}{2}}^{x_b} dx_1 \int_{\frac{x_a+x_b}{2}}^{x_b} dx_2 \psi^{*}(x_1,t_1)\hat{H}(x_2,t_2)\psi(x_2,t_2)= \nonumber \\
=\int_{\frac{x_a+x_b}{2}}^{x_b} dx_1 \int_{\frac{x_a+x_b}{2}}^{x_b} dx_2 \psi^{*}(x_1,t_1)\hat{H}(x_2,t_2)\int_{-\infty}^{+\infty}\int_{-\infty}^{+\infty}dx_3dt_3G(x_2,t_2,x_3,t_3)\psi(x_3,t_3)= \nonumber \\
=\int_{\frac{x_a+x_b}{2}}^{x_b} dx_1 \int_{\frac{x_a+x_b}{2}}^{x_b} dx_2 \psi^{*}(x_1,t_1)\hat{H}(x_2,t_2)\int_{t_1}^{t_2}\int_{x_1}^{x_2}dx_3dt_3G_q(x_2,t_2,x_3,t_3)\psi(x_3,t_3)= \nonumber \\
\int_{\frac{x_a+x_b}{2}}^{x_b} dx_1 \int_{\frac{x_a+x_b}{2}}^{x_b} dx_2 \psi^{*}(x_1,t_1)\hat{H}(x_2,t_2)\psi(x_1+(x_2-x_1),t_1+(t_2-t_1))=\nonumber \\
=\int_{\frac{x_a+x_b}{2}}^{x_b} dx_1 \int_{\frac{x_a+x_b}{2}}^{x_b} dx_2 \psi^{*}(x_1,t_1)E(t_2)e^{(x_2-x_1) \frac{i}{\hbar}\hat{p}(x_1,t_1)}\psi(x_1,t_1+(t_2-t_1))=\nonumber \\
=\int_{\frac{x_a+x_b}{2}}^{x_b} dx_1 \int_{\frac{x_a+x_b}{2}}^{x_b} dx_2 \psi^{*}(x_1,t_1)E(t_2)((e^{(x_2-x_1) \frac{i}{\hbar}\hat{p}(x_1,t_1)}e^{-(t_2-t_1) (\frac{i}{\hbar}\hat{H}(t_1)}\psi(x_1,t_1))))=\nonumber \\
=E(t_2)e^{-(t_2-t_1) \frac{i}{\hbar}E(t_1)}\int_{\frac{x_a+x_b}{2}}^{x_b} dx_1 \int_{\frac{x_a+x_b}{2}}^{x_b} dx_2 \psi^{*}(x_1,t_1)((e^{(x_2-x_1) \frac{i}{\hbar}\hat{p}(x_1,t_1)}\psi(x_1,t_1)))=\nonumber \\
=\int_{\frac{x_a+x_b}{2}}^{x_b} dx_1 \int_{-x_1+\frac{x_b+x_a}{2}}^{x_b-x_1} d\Delta x \psi^{*}(x_1,t_1) E(t_2)e^{-(t_2-t_1) \frac{i}{\hbar}E(t_1)}(e^{\Delta x \frac{i}{\hbar}\hat{p}(x_1,t_1)}\psi(x_1,t_1))= \nonumber \\
=\int_{\frac{x_a+x_b}{2}}^{x_b} dx_1 \psi^{*}(x_1,t_1) E(t_2)e^{-(t_2-t_1) \frac{i}{\hbar}E(t_1)}[\frac{\hbar}{i\hat{p}(x_1,t_1)}e^{\Delta x \frac{i}{\hbar}\hat{p}(x_1,t_1)}\psi(x_1,t_1)]_{\Delta x=-x_1+\frac{x_a+x_b}{2}}^{\Delta x=x_b-x_1} \nonumber \\
=\int_{\frac{x_a+x_b}{2}}^{x_b} dx_1 \psi^{*}(x_1,t_1) E(t_2)e^{-(t_2-t_1) \frac{i}{\hbar}E(t_1)}[\frac{\hbar}{i\hat{p}(x_1,t_1)}((e^{(x_b-x_1) \frac{i}{\hbar}\hat{p}(x_1,t_1)}-e^{(-x_1+\frac{x_a+x_b}{2}) \frac{i}{\hbar}\hat{p}(x_1,t_1)})\psi(x_1,t_1)]=\nonumber \\
\langle E(t_2)e^{-(t_2-t_1) \frac{i}{\hbar}E(t_1)}\frac{\hbar}{i\hat{p}(t_1)}e^{(x_b-x) \frac{i}{\hbar}\hat{p}(t_1)}\rangle_{x \in R_2}-\langle E(t_2)e^{-(t_2-t_1) \frac{i}{\hbar}E(t_1)}\frac{\hbar}{i\hat{p}(t_1)}e^{(x_b-x) \frac{i}{\hbar}\hat{p}(t_1)}e^{(-x+\frac{x_a+x_b}{2}) \frac{i}{\hbar}\hat{p}(x_1,t_1)})\rangle_{x \in R_2 }=\nonumber \\
=\int_{\frac{x_a+x_b}{2}}^{x_b} dx_1 \psi^{*}(x_1,t_1) E(t_2)e^{-(t_2-t_1) \frac{i}{\hbar}E(t_1)}[\frac{1}{\frac{d}{dx_1}}((e^{(x_b-x_1) \frac{d}{dx_1}}-e^{(-x_1+\frac{x_a+x_b}{2})\frac{d}{dx_1}})\psi(x_1,t_1)]=\nonumber \\
\sum_{n,m=0}^{+\infty} c_m^{*}(t_1)c_n(t_2)\int_{\frac{x_a+x_b}{2}}^{x_b} dx_1 \psi_m^{*}(x_1,t_1) E_n(t_2)e^{-(t_2-t_1) \frac{i}{\hbar}E_n(t_1)}[\frac{\psi_n(x_1,t_1)}{\frac{d}{dx_1}\psi_n(x_1,t_1)}(e^{(x_b-x_1) \frac{d}{dx_1}}-e^{(-x_1+\frac{x_a+x_b}{2})\frac{d}{dx_1}})\psi_n(x_1,t_1)] \nonumber \\
=\sum_{n=0}^{+\infty}\sum_{m=0}^{+\infty}\sqrt{p_m^{*}(t_1)p_n(t_2)}e^{i(\phi_n(t_2)-\phi_n(t_1))}E_n(t_2)e^{-(t_2-t_1) \frac{i}{\hbar}E_n(t_1)} \times \nonumber \\
\times
\int_{\frac{x_a+x_b}{2}}^{x_b} dx_1 \psi_m^{*}(x_1,t_1)[\frac{\psi_n(x_1,t_1)}{\frac{d}{dx_1}\psi_n(x_1,t_1)}(e^{(x_b-x_1) \frac{d}{dx_1}}-e^{(-x_1+\frac{x_a+x_b}{2})\frac{d}{dx_1}})\psi_n(x_1,t_1)]. \nonumber \\
\end{eqnarray}
Here we have assumed that Hamiltonian can be time-dependent and that $p_n$ and $p_m$ are probabilities of occupancy of n-th and m-th eigenergy denoted by $E_n$ and $E_m$, while $\psi(x,t)=\sum_{n=0}^{+\infty}c_n(t)\psi_n(x,t)=$ \nonumber \\ $=\sum_{n=0}^{+\infty}e^{i\phi_n(t)}\sqrt{p_n(t)}\psi_n(x,t)$ and
$\hat{H}_t\psi_n(x,t)=E_n(t)\psi_n(x,t)$ with $\sum_{n=0}^{+\infty}p_n(t)=1$.

\section{Two-particle correlation functions in Schroedinger formalism} \label{TwoPCorrelation}
\begin{eqnarray}
t_{s12,p1-p2,AntiCorr2}=t_{R_{2,p1} \rightarrow R_{1,p1},R_{1,p2} \rightarrow R_{2,p2}}(t_1,t_2) 
=t_{R_{2,p1} \rightarrow R_{1,p1},R_{1,p2} \rightarrow R_{2,p2}}(t_1,t_1+\Delta t)= \nonumber \\ =\int_{\frac{x_a+x_b}{2}}^{x_b} dx_{1,p1} \int_{x_a}^{\frac{x_a+x_b}{2}} dx_{2,p1}
\int_{x_a}^{\frac{x_a+x_b}{2}}dx_{1,p2} \int_{\frac{x_a+x_b}{2}}^{x_b} dx_{2,p2}\psi^{*}(x_{1,p1},x_{1,p2},t_1)\hat{H}(x_{2,p1},x_{2,p2},t_2)\psi(x_{2,p1},x_{2,p2},t_2), \nonumber \\
\end{eqnarray}
\begin{eqnarray}
t_{s12,p1-p2,Corr1}=t_{R_{1,p1} \rightarrow R_{2,p1},R_{1,p2} \rightarrow R_{2,p2}}(t_1,t_2) 
=t_{R_{1,p1} \rightarrow R_{2,p1},R_{1,p2} \rightarrow R_{2,p2}}(t_1,t_1+\Delta t)= \nonumber \\ =\int_{x_a}^{\frac{x_a+x_b}{2}} dx_{1,p1} \int_{\frac{x_a+x_b}{2}}^{x_b} dx_{2,p1}
\int_{x_a}^{\frac{x_a+x_b}{2}} dx_{1,p2} \int_{\frac{x_a+x_b}{2}}^{x_b}  dx_{2,p2}\psi^{*}(x_{1,p1},x_{1,p2},t_1)\hat{H}(x_{2,p1},x_{2,p2},t_2)\psi(x_{2,p1},x_{2,p2},t_2)= \nonumber \\
\end{eqnarray}
\begin{eqnarray}
t_{s12,p1-p2,Corr2}=t_{R_{2,p1} \rightarrow R_{1,p1},R_{2,p2} \rightarrow R_{1,p2}}(t_1,t_2) 
=t_{R_{2,p1} \rightarrow R_{1,p1},R_{2,p2} \rightarrow R_{1,p1}}(t_1,t_1+\Delta t)= \nonumber \\ =\int_{\frac{x_a+x_b}{2}}^{x_b}dx_{1,p1} \int_{x_a}^{\frac{x_a+x_b}{2}} dx_{2,p1}
\int_{\frac{x_a+x_b}{2}}^{x_b} dx_{1,p2} \int_{x_a}^{\frac{x_a+x_b}{2}}   dx_{2,p2}\psi^{*}(x_{1,p1},x_{1,p2},t_1)\hat{H}(x_{2,p1},x_{2,p2},t_2)\psi(x_{2,p1},x_{2,p2},t_2)= \nonumber \\
\end{eqnarray}

\begin{eqnarray}
E_{[p1A+p1B],p1-p2}=t_{R_{1,p1} \rightarrow R_{1,p1},R_{1,p2} \rightarrow R_{1,p2}}(t_1,t_2) 
=t_{R_{1,p1} \rightarrow R_{1,p1},R_{1,p2} \rightarrow R_{1,p1}}(t_1,t_1+\Delta t)= \nonumber \\ =\int_{x_a}^{\frac{x_a+x_b}{2}}dx_{1,p1} \int_{x_a}^{\frac{x_a+x_b}{2}} dx_{2,p1}
\int_{x_a}^{\frac{x_a+x_b}{2}} dx_{1,p2} \int_{x_a}^{\frac{x_a+x_b}{2}}   dx_{2,p2}\psi^{*}(x_{1,p1},x_{1,p2},t_1)\hat{H}(x_{2,p1},x_{2,p2},t_2)\psi(x_{2,p1},x_{2,p2},t_2),  \nonumber \\
\end{eqnarray}

\begin{eqnarray}
E_{[p2A+p2B],p1-p2}=t_{R_{1,p1} \rightarrow R_{1,p1},R_{1,p2} \rightarrow R_{1,p2}}(t_1,t_2) 
=t_{R_{1,p1} \rightarrow R_{1,p1},R_{1,p2} \rightarrow R_{1,p1}}(t_1,t_1+\Delta t)= \nonumber \\ =\int_{x_a}^{\frac{x_a+x_b}{2}}dx_{1,p1} \int_{x_a}^{\frac{x_a+x_b}{2}} dx_{2,p1}
\int_{x_a}^{\frac{x_a+x_b}{2}} dx_{1,p2} \int_{x_a}^{\frac{x_a+x_b}{2}}   dx_{2,p2}\psi^{*}(x_{1,p1},x_{1,p2},t_1)\hat{H}(x_{2,p1},x_{2,p2},t_2)\psi(x_{2,p1},x_{2,p2},t_2),  \nonumber \\
\end{eqnarray}

\begin{eqnarray}
E_{[p1A+p2B],p1-p2}=t_{R_{1,p1} \rightarrow R_{1,p1},R_{2,p2} \rightarrow R_{2,p2}}(t_1,t_2) 
=t_{R_{1,p1} \rightarrow R_{1,p1},R_{2,p2} \rightarrow R_{2,p1}}(t_1,t_1+\Delta t)= \nonumber \\ =\int_{x_a}^{\frac{x_a+x_b}{2}}dx_{1,p1} \int_{x_a}^{\frac{x_a+x_b}{2}} dx_{2,p1}
\int_{\frac{x_a+x_b}{2}}^{x_b} dx_{1,p2} \int_{\frac{x_a+x_b}{2}}^{x_b}   dx_{2,p2}\psi^{*}(x_{1,p1},x_{1,p2},t_1)\hat{H}(x_{2,p1},x_{2,p2},t_2)\psi(x_{2,p1},x_{2,p2},t_2),  \nonumber \\
\end{eqnarray}

\begin{eqnarray}
E_{[p2A+p1B],p1-p2}(t_1,t_2)=t_{R_{2,p1} \rightarrow R_{2,p1},R_{1,p2} \rightarrow R_{1,p2}}(t_1,t_2) 
=t_{R_{2,p1} \rightarrow R_{2,p1},R_{1,p2} \rightarrow R_{1,p2}}(t_1,t_1+\Delta t)= \nonumber \\ =\int_{\frac{x_a+x_b}{2}}^{x_b} dx_{1,p1} \int_{x_a}^{\frac{x_a+x_b}{2}} dx_{1,p1}
\int_{\frac{x_a+x_b}{2}}^{x_b} dx_{1,p2} \int_{\frac{x_a+x_b}{2}}^{x_b}   dx_{2,p2}\psi^{*}(x_{1,p1},x_{1,p2},t_1)\hat{H}(x_{2,p1},x_{2,p2},t_2)\psi(x_{2,p1},x_{2,p2},t_2),  \nonumber \\
\end{eqnarray}

\begin{eqnarray}
t_{p1:R1 \rightarrow R2,p2:R1}(t_1,t_2)=t_{R_{1,p1} \rightarrow R_{2,p1},R_{1,p2} \rightarrow R_{1,p2}}(t_1,t_2) 
=t_{R_{1,p1} \rightarrow R_{2,p1},R_{1,p2} \rightarrow R_{1,p2}}(t_1,t_1+\Delta t)= \nonumber \\ =\int_{x_a}^{\frac{x_a+x_b}{2}} dx_{1,p1} \int_{\frac{x_a+x_b}{2}}^{x_b} dx_{2,p1}
\int_{x_a}^{\frac{x_a+x_b}{2}} dx_{1,p2} \int_{x_a}^{\frac{x_a+x_b}{2}}   dx_{2,p2}\psi^{*}(x_{1,p1},x_{1,p2},t_1)\hat{H}(x_{2,p1},x_{2,p2},t_2)\psi(x_{2,p1},x_{2,p2},t_2),  \nonumber \\
\end{eqnarray}

\begin{eqnarray}
t_{p1:R1 \rightarrow R2,p2:R2}(t_1,t_2)=t_{R_{1,p1} \rightarrow R_{2,p1},R_{2,p2} \rightarrow R_{2,p2}}(t_1,t_2) 
=t_{R_{1,p1} \rightarrow R_{2,p1},R_{2,p2} \rightarrow R_{2,p2}}(t_1,t_1+\Delta t)= \nonumber \\ =\int_{x_a}^{\frac{x_a+x_b}{2}} dx_{1,p1} \int_{\frac{x_a+x_b}{2}}^{x_b} dx_{2,p1}
\int_{\frac{x_a+x_b}{2}}^{x_b} dx_{1,p2} \int_{\frac{x_a+x_b}{2}}^{x_b}   dx_{2,p2}\psi^{*}(x_{1,p1},x_{1,p2},t_1)\hat{H}(x_{2,p1},x_{2,p2},t_2)\psi(x_{2,p1},x_{2,p2},t_2),  \nonumber \\
\end{eqnarray}

\begin{eqnarray}
t_{p1:R1,p2:R1 \rightarrow R2}(t_1,t_2)=t_{R_{1,p1} \rightarrow R_{1,p1},R_{1,p2} \rightarrow R_{2,p2}}(t_1,t_2) 
=t_{R_{1,p1} \rightarrow R_{1,p1},R_{1,p2} \rightarrow R_{2,p2}}(t_1,t_1+\Delta t)= \nonumber \\ =\int_{x_a}^{\frac{x_a+x_b}{2}} dx_{1,p1} \int_{x_a}^{\frac{x_a+x_b}{2}} dx_{2,p1}
\int_{x_a}^{\frac{x_a+x_b}{2}} dx_{1,p2} \int_{\frac{x_a+x_b}{2}}^{x_b} dx_{2,p2}\psi^{*}(x_{1,p1},x_{1,p2},t_1)\hat{H}(x_{2,p1},x_{2,p2},t_2)\psi(x_{2,p1},x_{2,p2},t_2),  \nonumber \\
\end{eqnarray}

\begin{eqnarray}
t_{p1:R1,p2:R2 \rightarrow R1}(t_1,t_2)=t_{R_{1,p1} \rightarrow R_{1,p1},R_{2,p2} \rightarrow R_{1,p2}}(t_1,t_2) 
=t_{R_{1,p1} \rightarrow R_{1,p1},R_{2,p2} \rightarrow R_{1,p2}}(t_1,t_1+\Delta t)= \nonumber \\ =\int_{x_a}^{\frac{x_a+x_b}{2}} dx_{1,p1} \int_{x_a}^{\frac{x_a+x_b}{2}} dx_{2,p1}
\int_{\frac{x_a+x_b}{2}}^{x_b} dx_{1,p2} \int_{x_a}^{\frac{x_a+x_b}{2}} dx_{2,p2}\psi^{*}(x_{1,p1},x_{1,p2},t_1)\hat{H}(x_{2,p1},x_{2,p2},t_2)\psi(x_{2,p1},x_{2,p2},t_2),  \nonumber \\
\end{eqnarray}

\begin{eqnarray}
t_{p1:R2,p2:R1 \rightarrow R2}(t_1,t_2)=t_{R_{2,p1} \rightarrow R_{2,p1},R_{1,p2} \rightarrow R_{2,p2}}(t_1,t_2) 
=t_{R_{2,p1} \rightarrow R_{2,p1},R_{1,p2} \rightarrow R_{2,p2}}(t_1,t_1+\Delta t)= \nonumber \\ =\int_{\frac{x_a+x_b}{2}}^{x_b} dx_{1,p1} \int_{\frac{x_a+x_b}{2}}^{x_b} dx_{2,p1}
\int_{x_a}^{\frac{x_a+x_b}{2}} dx_{1,p2} \int_{\frac{x_a+x_b}{2}}^{x_b} dx_{2,p2}\psi^{*}(x_{1,p1},x_{1,p2},t_1)\hat{H}(x_{2,p1},x_{2,p2},t_2)\psi(x_{2,p1},x_{2,p2},t_2),  \nonumber \\
\end{eqnarray}

\begin{eqnarray}
t_{p1:R2,p2:R2 \rightarrow R1}(t_1,t_2)=t_{R_{2,p1} \rightarrow R_{2,p1},R_{2,p2} \rightarrow R_{1,p2}}(t_1,t_2) 
=t_{R_{2,p1} \rightarrow R_{2,p1},R_{2,p2} \rightarrow R_{1,p2}}(t_1,t_1+\Delta t)= \nonumber \\ =\int_{\frac{x_a+x_b}{2}}^{x_b} dx_{1,p1} \int_{\frac{x_a+x_b}{2}}^{x_b} dx_{2,p1}
\int_{\frac{x_a+x_b}{2}}^{x_b} dx_{1,p2} \int_{x_a}^{\frac{x_a+x_b}{2}} dx_{2,p2}\psi^{*}(x_{1,p1},x_{1,p2},t_1)\hat{H}(x_{2,p1},x_{2,p2},t_2)\psi(x_{2,p1},x_{2,p2},t_2),  \nonumber \\
\end{eqnarray}

\section{Open loops confining single electron in spherical coordinates} \label{spherical}

\normalsize
\begin{eqnarray}
-\frac{\hbar^2}{2m} \nabla^2 \psi(r,\phi, \Theta)=-\frac{\hbar^2}{2m}(\frac{1}{r^2}\frac{d}{dr}[r^2 \frac{d}{dr}]+\frac{1}{r^2}\frac{1}{Sin(\Theta)}\frac{d}{d\Theta}[ Sin(\Theta)\frac{d}{d\Theta}]+\frac{1}{r^2 Sin(\Theta)^2}\frac{d^2}{d\phi^2})\psi(r,\phi, \Theta)= \nonumber \\
=(E-V(r,\Theta,\phi))\psi(r,\Theta, \phi).
\end{eqnarray}
We have
\begin{eqnarray}
-\frac{\hbar^2}{2m}(\frac{d^2}{dr^2}+\frac{2}{r}\frac{d}{dr}+\frac{1}{r^2}\frac{d^2}{d\Theta^2}+\frac{1}{r^2}\frac{Cos(\Theta)}{Sin(\Theta)}[ \frac{d}{d\Theta}]+\frac{1}{r^2 Sin(\Theta)^2}\frac{d^2}{d\phi^2})\psi(r,\phi, \Theta)= \nonumber \\ =(E-V(r,\Theta,\phi))\psi(r,\Theta, \phi).
\end{eqnarray}
We have given $r(s)$, $\Theta(s)$ and $\phi(s)$. We assume than nanowire is parametrized by s real value.
We have $\frac{d}{dr}=\frac{ds}{dr}\frac{d}{ds}=\frac{1}{r'(s)}\frac{d}{ds}$, $\frac{d}{d\Theta}=\frac{ds}{d\Theta}\frac{d}{ds}=\frac{1}{\Theta'(s)}\frac{d}{ds}$ and $\frac{d}{d\phi}=\frac{ds}{d\phi}\frac{d}{ds}=\frac{1}{\phi'(s)}\frac{d}{ds}$ and consequently we have
$\frac{d^2}{dr^2}=\frac{ds}{dr}\frac{d}{ds}(\frac{ds}{dr}\frac{d}{ds})=\frac{1}{r'(s)}\frac{d}{ds}(\frac{1}{r'(s)}\frac{d}{ds})=\frac{1}{(r'(s))^2}\frac{d^2}{ds^2}-\frac{r''(s)}{(r'(s))^3}\frac{d}{ds}$, $\frac{d^2}{d\Theta^2}=\frac{ds}{d\Theta}\frac{d}{ds}(\frac{ds}{d\Theta}\frac{d}{ds})=\frac{1}{\Theta'(s)}\frac{d}{ds}(\frac{1}{\Theta'(s)}\frac{d}{ds})=\frac{1}{(\Theta'(s))^2}\frac{d^2}{ds^2}-\frac{\Theta''(s)}{(\Theta'(s))^3}\frac{d}{ds}$ and $\frac{d^2}{d\phi^2}=\frac{ds}{d\phi}\frac{d}{ds}(\frac{ds}{d\phi}\frac{d}{ds})=\frac{1}{(\phi'(s))^2}\frac{d^2}{ds^2}-\frac{\phi''(s)}{(\phi'(s))^3}\frac{d}{ds}$. We obtain
\begin{eqnarray}
-\frac{\hbar^2}{2m}(\frac{1}{(r'(s))^2}\frac{d^2}{ds^2}-\frac{r''(s)}{(r'(s))^3}\frac{d}{ds}+\frac{2}{r}(\frac{1}{r'(s)}\frac{d}{ds})+\frac{1}{r^2}(\frac{1}{(\Theta'(s))^2}\frac{d^2}{ds^2}-\frac{\Theta''(s)}{(\Theta'(s))^3}\frac{d}{ds})+\frac{1}{r^2}\frac{Cos(\Theta)}{Sin(\Theta)}\frac{1}{\Theta'(s)}\frac{d}{ds} \nonumber \\
+\frac{1}{r^2 Sin(\Theta)^2}(\frac{1}{(\phi'(s))^2}\frac{d^2}{ds^2}-\frac{\phi''(s)}{(\phi'(s))^3}\frac{d}{ds}))\psi(r,\phi, \Theta)=\nonumber \\ =(E-V(r,\Theta,\phi))\psi(r,\Theta, \phi).
\end{eqnarray}
Finally we obtain
\begin{eqnarray}
-\frac{\hbar^2}{2m}([\frac{1}{(r'(s))^2}+\frac{1}{r^2}(\frac{1}{(\Theta'(s))^2}+\frac{1}{r^2 Sin(\Theta)^2}(\frac{1}{(\phi'(s))^2}))]\frac{d^2}{ds^2}
+[-\frac{r''(s)}{(r'(s))^3}+\frac{2}{r}(\frac{1}{r'(s)})-\frac{1}{r^2(s)}\frac{\Theta''(s)}{(\Theta'(s))^3}+ \nonumber \\
+\frac{1}{r^2}\frac{Cos(\Theta)}{Sin(\Theta)}\frac{1}{\Theta'(s)}-\frac{1}{r^2 Sin(\Theta)^2}\frac{\phi''(s)}{(\phi'(s))^3}]\frac{d}{ds}
)\psi(s)=\nonumber \\ =(E-V(s))\psi(s),
\end{eqnarray}
that we can write in compact form as
\begin{eqnarray}
\Bigg[-\frac{\hbar^2}{2m}\Bigg[(f(s)\frac{d^2}{ds^2}-g(s)\frac{d}{ds})\Bigg]+V(s)\Bigg] \psi(s)=E \psi(s),
\end{eqnarray}
where
\begin{eqnarray}
f(s)=[\frac{1}{(r'(s))^2}+\frac{1}{r^2}(\frac{1}{(\Theta'(s))^2}+\frac{1}{r^2 Sin(\Theta)^2}(\frac{1}{(\phi'(s))^2}))], \nonumber \\
g(s)=[-\frac{r''(s)}{(r'(s))^3}+\frac{2}{r}(\frac{1}{r'(s)})-\frac{1}{r^2(s)}\frac{\Theta''(s)}{(\Theta'(s))^3}+\frac{1}{r^2}\frac{Cos(\Theta)}{Sin(\Theta)}\frac{1}{\Theta'(s)}-\frac{1}{r^2 Sin(\Theta)^2}\frac{\phi''(s)}{(\phi'(s))^3}].
\end{eqnarray}
\section{Open curvy loops confining single electron in cylindrical coordinates in Schroedinger formalism} \label{cylindrical}

In cylindrical coordinates we have
\begin{eqnarray*}
 [-\frac{\hbar^2}{2m}[\frac{d^2}{dr^2}+\frac{1}{r}\frac{d}{dr}+\frac{d^2}{dz^2}]+V(r,\phi,z)]  \psi(r(\phi),\phi,z)=E \psi(r(\phi),\phi,z)
\end{eqnarray*}
and we limit considerations to one plane so we have
\begin{eqnarray*}
 [-\frac{\hbar^2}{2m}[\frac{d^2}{dr^2}+\frac{1}{r}\frac{d}{dr}]+V(r,\phi)]  \psi(r(\phi),\phi)=E \psi(r(\phi),\phi)
\end{eqnarray*}
where $r=r(\phi)$, so we can write
\begin{eqnarray}
\frac{d}{dr}=\frac{d\phi}{dr}\frac{d}{d\phi}=\frac{1}{r'(\phi)}\frac{d}{d\phi}, \frac{d^2}{dr^2}=\frac{1}{r'(\phi)}\frac{d}{d\phi}[\frac{1}{r'(\phi)}\frac{d}{d\phi}]=\frac{1}{(r'(\phi))^2}\frac{d^2}{d\phi^2}-\frac{r''(\phi)}{(r'(\phi))^3}\frac{d}{d\phi}
\end{eqnarray}
Finally we obtain
\begin{eqnarray*}
 [-\frac{\hbar^2}{2m}[\frac{1}{(r'(\phi))^2}\frac{d^2}{d\phi^2}-\frac{r''(\phi)}{(r'(\phi))^3}\frac{d}{d\phi}+\frac{1}{r}\frac{1}{r'(\phi)}\frac{d}{d\phi}]+V(r,\phi)]  \psi(\phi)=E \psi(\phi)
\end{eqnarray*}
that can be simplified into
\begin{eqnarray*}
 [-\frac{\hbar^2}{2m}[\frac{1}{(r'(\phi))^2}\frac{d^2}{d\phi^2}+[-\frac{r''(\phi)}{(r'(\phi))^3}+\frac{1}{r}\frac{1}{r'(\phi)}]\frac{d}{d\phi}]+V(\phi)]  \psi(\phi)=E \psi(\phi)
\end{eqnarray*}
and that we can write in compact form as
\begin{eqnarray}
\Bigg[-\frac{\hbar^2}{2m}\Bigg[(f(s)\frac{d^2}{ds^2}-g(s)\frac{d}{ds})\Bigg]+V(s)\Bigg] \psi(s)=E \psi(s),
\end{eqnarray}
where
\begin{eqnarray}
f(s)=[\frac{1}{(r'(\phi))^2}],g(s)=-\frac{r''(\phi)}{(r'(\phi))^3}+\frac{1}{r}\frac{1}{r'(\phi)}.
\end{eqnarray}
Let us now express any open loop in cylindrical coordinates so we have $r(s), \phi(s),z(s)$. In such case
we have $\frac{d}{dr}=\frac{ds}{dr}\frac{d}{ds}=\frac{1}{r'(s)}\frac{d}{ds}$, $\frac{d}{d\phi}=\frac{ds}{d\phi}\frac{d}{ds}=\frac{1}{\phi'(s)}\frac{d}{ds}$ and $\frac{d}{dz}=\frac{ds}{dz}\frac{d}{ds}=\frac{1}{z'(s)}\frac{d}{ds}$ and consequently we have
$\frac{d^2}{dr^2}=\frac{ds}{dr}\frac{d}{ds}(\frac{ds}{dr}\frac{d}{ds})=\frac{1}{r'(s)}\frac{d}{ds}(\frac{1}{r'(s)}\frac{d}{ds})=\frac{1}{(r'(s))^2}\frac{d^2}{ds^2}-\frac{r''(s)}{(r'(s))^3}\frac{d}{ds}$, $\frac{d^2}{d\phi^2}=\frac{ds}{d\phi}\frac{d}{ds}(\frac{ds}{d\phi}\frac{d}{ds})=\frac{1}{\phi'(s)}\frac{d}{ds}(\frac{1}{\phi'(s)}\frac{d}{ds})=\frac{1}{(\phi'(s))^2}\frac{d^2}{ds^2}-\frac{\phi''(s)}{(\phi'(s))^3}\frac{d}{ds}$ and $\frac{d^2}{dz^2}=\frac{ds}{dz}\frac{d}{ds}(\frac{ds}{dz}\frac{d}{ds})=\frac{1}{(z'(s))^2}\frac{d^2}{ds^2}-\frac{z''(s)}{(z'(s))^3}\frac{d}{ds}$.
Consequently we obtain
\begin{eqnarray*}
 [-\frac{\hbar^2}{2m}[\frac{1}{(r'(s))^2}\frac{d^2}{ds^2}-\frac{r''(s)}{(r'(s))^3}\frac{d}{ds}+\frac{1}{r}(\frac{1}{r'(s)}\frac{d}{ds})+(\frac{1}{(z'(s))^2}\frac{d^2}{ds^2}-\frac{z''(s)}{(z'(s))^3}\frac{d}{ds})]+V(r,\phi,z)]  \psi(r(\phi),\phi,z)=E \psi(r(\phi),\phi,z)
\end{eqnarray*}
and we can write
\begin{eqnarray*}
 [-\frac{\hbar^2}{2m}([\frac{1}{(r'(s))^2}+\frac{1}{(z'(s))^2}]\frac{d^2}{ds^2}+[\frac{1}{r}\frac{1}{r'(s)}-\frac{r''(s)}{(r'(s))^3}-\frac{z''(s)}{(z'(s))^3}]\frac{d}{ds})]+V(r,\phi,z)]  \psi(r(\phi),\phi,z)=E \psi(r(\phi),\phi,z),
\end{eqnarray*}
that can be summarized by equation
\begin{eqnarray}
\Bigg[-\frac{\hbar^2}{2m}\Bigg[(f(s)\frac{d^2}{ds^2}-g(s)\frac{d}{ds})\Bigg]+V(s)\Bigg] \psi(s)=E \psi(s),
\end{eqnarray}
where $f(s)=[\frac{1}{(r'(s))^2}+\frac{1}{(z'(s))^2}]$ and $g(s)=[\frac{1}{r}\frac{1}{r'(s)}-\frac{r''(s)}{(r'(s))^3}-\frac{z''(s)}{(z'(s))^3}]$.

\begin{landscape}
\section{Matrix representation of Schroedinger equation in curvy cylindrical coordinates}
We express the Hamiltonian of deformed nanowire for single electron in effective potential with use of Toeplitz matrix approach.
Using the cylindrical coordinated we obtain
\begin{eqnarray}
\hat{H}_{1, N \times N}= -\frac{\hbar^2}{2m (\Delta\phi)^2}
\begin{pmatrix}
-2\frac{1}{(r'(\phi_1))^2}-\frac{2m(\Delta\phi)^2}{\hbar^2}V(\phi_1) & \frac{1}{(r'(\phi_2))^2}  &  0 & 0 \\
\frac{1}{(r'(\phi_2))^2} & -2\frac{1}{(r'(\phi_2))^2}-\frac{2m(\Delta\phi)^2}{\hbar^2}V(\phi_2) &  \frac{1}{(r'(\phi_3))^2} & 0 \\
0  & \frac{1}{(r'(\phi_3))^2}  & -2\frac{1}{(r'(\phi_3))^2}-\frac{2m(\Delta\phi)^2}{\hbar^2}V(\phi_3) & \frac{1}{(r'(\phi_4))^2} \\
0  & 0   & \frac{1}{(r'(\phi_4))^2}   & -2\frac{1}{(r'(\phi_4))^2}-\frac{2m(\Delta\phi)^2}{\hbar^2}V(\phi_4)
\end{pmatrix}
\end{eqnarray}
and
\begin{eqnarray}
\hat{H}_{2, N \times N}= -\frac{1}{2}\frac{\hbar^2}{2m(\Delta\phi)}
\begin{pmatrix}
-[-\frac{r''(\phi)}{(r'(\phi_1))^3}+\frac{1}{r(\phi_1)}\frac{1}{r'(\phi_1)}] & [-\frac{r''(\phi_2)}{(r'(\phi_2))^3}+\frac{1}{r}\frac{1}{r'(\phi_2)}]  &   0 & 0 \\
0 & -[-\frac{r''(\phi_2)}{(r'(\phi_2))^3}+\frac{1}{r(\phi_2)}\frac{1}{r'(\phi_2)}] &  [-\frac{r''(\phi_3)}{(r'(\phi_3))^3}+\frac{1}{r(\phi_3)}\frac{1}{r'(\phi_3)}] & 0 \\
0  & 0  & 0 & [-\frac{r''(\phi_4)}{(r'(\phi_4))^3}+\frac{1}{r(\phi_4)}\frac{1}{r'(\phi_4)}] \\
\end{pmatrix}
\end{eqnarray}
and
\begin{eqnarray}
\hat{H}_{3, N \times N}= -\frac{1}{2}\frac{\hbar^2}{2m(\Delta\phi)}
\begin{pmatrix}
[-\frac{r''(\phi)}{(r'(\phi_1))^3}+\frac{1}{r(\phi_1)}\frac{1}{r'(\phi_1)}] & 0  &   0 & 0 \\
-[-\frac{r''(\phi_2)}{(r'(\phi_2))^3}+\frac{1}{r}\frac{1}{r'(\phi_2)}] & [-\frac{r''(\phi_2)}{(r'(\phi_2))^3}+\frac{1}{r(\phi_2)}\frac{1}{r'(\phi_2)}] &  0 & 0 \\
0  & -[-\frac{r''(\phi_3)}{(r'(\phi_3))^3}+\frac{1}{r(\phi_3)}\frac{1}{r'(\phi_3)}]  & [-\frac{r''(\phi_3)}{(r'(\phi_3))^3}+\frac{1}{r(\phi_3)}\frac{1}{r'(\phi_3)}] & 0 \\
0  & 0   & -[-\frac{r''(\phi_4)}{(r'(\phi_4))^3}+\frac{1}{r(\phi_4)}\frac{1}{r'(\phi_4)}]   & [-\frac{r''(\phi_4)}{(r'(\phi_4))^3}+\frac{1}{r(\phi_4)}\frac{1}{r'(\phi_4)}]
\end{pmatrix}.
\end{eqnarray}
we can obtain antysymmetric component of Hamiltonian given as
\begin{eqnarray}
\hat{H}_{2, N \times N}+\hat{H}_{3, N \times N}= -\frac{\hbar^2}{4m(\Delta\phi)}
\begin{pmatrix}
0 & [-\frac{r''(\phi_2)}{(r'(\phi_2))^3}+\frac{1}{r(\phi_2)}\frac{1}{r'(\phi_2)}]  &   0 & 0 \\
-[-\frac{r''(\phi_2)}{(r'(\phi_2))^3}+\frac{1}{r}\frac{1}{r'(\phi_2)}] & 0 &  [-\frac{r''(\phi_3)}{(r'(\phi_3))^3}+\frac{1}{r(\phi_3)}\frac{1}{r'(\phi_3)}] & 0 \\
0  & -[-\frac{r''(\phi_3)}{(r'(\phi_3))^3}+\frac{1}{r(\phi_3)}\frac{1}{r'(\phi_3)}]  & 0 & [-\frac{r''(\phi_4)}{(r'(\phi_4))^3}+\frac{1}{r(\phi_4)}\frac{1}{r'(\phi_4)}] \\
0  & 0   & -[-\frac{r''(\phi_4)}{(r'(\phi_4))^3}+\frac{1}{r(\phi_4)}\frac{1}{r'(\phi_4)}]   & 0
\end{pmatrix}.
\end{eqnarray}
We have finally $\hat{H}_{1, N \times N}+\hat{H}_{2, N \times N}+\hat{H}_{3, N \times N}$
\tiny
\begin{eqnarray}
= -\frac{\hbar^2}{2m (\Delta\phi)^2}
\begin{pmatrix}
-2\frac{1}{(r'(\phi_1))^2}-\frac{2m(\Delta\phi)^2}{\hbar^2}V(\phi_1) & \frac{1}{(r'(\phi_2))^2}+[-\frac{r''(\phi_2)}{(r'(\phi_2))^3}+\frac{1}{r(\phi_2)}\frac{1}{r'(\phi_2)}]\frac{\Delta \phi}{2}  &  0 & 0 \\
\frac{1}{(r'(\phi_2))^2}-[-\frac{r''(\phi_2)}{(r'(\phi_2))^3}+\frac{1}{r(\phi_2)}\frac{1}{r'(\phi_2)}]\frac{\Delta \phi}{2} & -2\frac{1}{(r'(\phi_2))^2}-\frac{2m(\Delta\phi)^2}{\hbar^2}V(\phi_2) &  \frac{1}{(r'(\phi_3))^2}+\frac{1}{r(\phi_3)}\frac{1}{r'(\phi_3)}]\frac{\Delta \phi}{3} & 0 \\
0  & \frac{1}{(r'(\phi_3))^2}  & -2\frac{1}{(r'(\phi_3))^2}-\frac{2m(\Delta\phi)^2}{\hbar^2}V(\phi_3) & \frac{1}{(r'(\phi_4))^2}+\frac{r''(\phi_4)}{(r'(\phi_4))^3}+\frac{1}{r(\phi_4)}\frac{1}{r'(\phi_4)}]\frac{\Delta \phi}{2} \\
0  & 0   & \frac{1}{(r'(\phi_4))^2}-[\frac{r''(\phi_4)}{(r'(\phi_4))^3}+\frac{1}{r(\phi_4)}\frac{1}{r'(\phi_4)}]\frac{\Delta \phi}{2}   & -2\frac{1}{(r'(\phi_4))^2}-\frac{2m(\Delta\phi)^2}{\hbar^2}V(\phi_4)
\end{pmatrix}
\end{eqnarray}

\end{landscape}

\end{appendices}

\end{document}